\documentclass[]{aa}  

\usepackage{graphicx}
\usepackage{txfonts}
\usepackage{rotating}
\usepackage{longtable}
\usepackage{multicol}
\usepackage{lscape}
\usepackage[maxfloats=25]{morefloats}
\usepackage{amsmath}
\usepackage{xfrac}
\usepackage[switch]{lineno}
\usepackage{booktabs,caption,fixltx2e}
\usepackage[flushleft]{threeparttable}
\usepackage{placeins}
\usepackage{float}

\begin{document}

   \title{Location of $\gamma$-ray emission and magnetic field strengths in OJ\,287}


   \author{J.A. Hodgson
          \inst{1,10}
          \and
          T.P. Krichbaum\inst{1}
           \and
          A.P. Marscher\inst{2}
          \and 
          S.G. Jorstad\inst{2,3}  \and B. Rani\inst{1} \and I. Marti-Vidal\inst{1,4} \and U. Bach\inst{1} \and S. Sanchez\inst{5} \and M. Bremer\inst{6} \and M. Lindqvist\inst{4} \and M. Uunila\inst{7} \and J. Kallunki\inst{7} \and P. Vicente\inst{8} \and L. Fuhrmann\inst{1} \and E. Angelakis\inst{1} \and V. Karamanavis\inst{1} \and I. Myserlis\inst{1} \and I. Nestoras\inst{1} \and C. Chidiac\inst{1} \and A. Sievers\inst{5} \and M. Gurwell\inst{9} \and J. A. Zensus\inst{1}
          }
\institute{International Center for Radio Astronomy Research, Perth, Western Australia}
\institute{Max-Planck-Institut f\"{u}r Radioastronomie, Auf dem H\"{u}gel 69, 53121, Bonn, Germany 
\\ \email{jhodgson@kasi.re.kr}
\and Institute for Astrophysical Research, Boston University, 725 Commonwealth Avenue, Boston, MA 02215
\and Astronomical Institute, St. Petersburg State University, Universitetskij Pr. 28, Petrodvorets, 
198504 St. Petersburg, Russia
\and Dept. of Earth and Space Sciences, Chalmers Univ. of Technology, Onsala
Space Observatory, SE-43992 Onsala, Sweden
\and Institut de Radio Astronomie Millim\'etrique, Avenida Divina Pastora 7, Local 20, 18012 Granada, Spain
\and Institut de Radio Astronomie Millim\'etrique, 300 rue de la Piscine, Domaine Universitaire, 38406 Saint Martin d'Heres, France
\and Aalto University Mets\"{a}hovi Radio Observatory, Mets\"{a}hovintie 114, FIN-02540 Kylm\"{a}l\"{a}, Finland
\and Observatorio de Yebes (IGN), Apartado 148, 19080, Yebes, Spain.
\and Harvard–Smithsonian Center for Astrophysics, Cambridge, MA 02138, USA
\and Korea Astronomy and Space Institute, 776 Daedeokdae-ro, Yuseong-gu, Daejeon 34055, Korea
}


   \date{Received ; accepted }

 
  \abstract
   {The $\gamma$-ray BL Lac object OJ\,287 is known to exhibit inner-parsec ``jet-wobbling'', high degrees of variability at all wavelengths and quasi-stationary features including an apparent ($\approx 100^{\circ}$) position angle change in projection on the sky plane.}
   {Sub-$50$ micro-arcsecond resolution 86\,GHz observations with the global mm-VLBI array (GMVA) supplement ongoing multi-frequency VLBI blazar monitoring at lower frequencies. Using these maps together with cm/mm total intensity and $\gamma$-ray observations from \emph{Fermi}/LAT from 2008-2014, we aimed to determine the location of $\gamma$-ray emission and to explain the inner-mas structural changes. }
   {Observations with the GMVA offer approximately double the angular resolution compared with 43\,GHz VLBA observations and allow us to observe above the synchrotron self-absorption peak frequency. Fermi-LAT $\gamma$-ray data were reduced and analysed. The jet was spectrally decomposed at multiple locations along the jet. From this we could derive estimates of the magnetic field using equipartition and synchrotron self-absorption arguments. How the field decreases down the jet allowed an estimate of the distance to the jet apex and an estimate of the magnetic field strength at the jet apex and in the broad line region. Combined with accurate kinematics we attempt to locate the site of $\gamma$-ray activity, radio flares and spectral changes.}
   {Strong $\gamma$-ray flares appeared to originate from either the ``core'' region, a downstream stationary feature, or both, with $\gamma$-ray activity significantly correlated with radio flaring in the downstream quasi-stationary feature. Magnetic field estimates were determined at multiple locations along the jet, with the magnetic field found to be $\geq$\,1.6\,G in the ``core'' and $\leq$\,0.4\,G in the downstream quasi-stationary feature. We therefore found upper limits on the location of the VLBI ``core'' as $\lesssim$\,6.0\,pc from the jet apex and determined an upper limit on the magnetic field near the jet base of the order of thousands of Gauss.  }
   {}


   \maketitle
%

\section{Introduction}

Radio loud active galactic nuclei (AGN) feature highly energised relativistic jets which are likely produced by the conversion of gravitational energy around a central super-massive black hole (SMBH) \citep{BZ77,BP82}. Blazars are a subclass of AGN with the jet direction being nearly parallel to our line of sight \citep{Urry95}. This causes relativistic effects including apparent superluminal motion, reduction of variability timescales and the apparent quasi-periodic changes of inner jet orientation that we refer to as jet ``wobbling''. This ``wobbling'' is thought to to be caused either by geometric effects \citep[e.g.,][]{jor05,bach05} or due to binary black hole procession \citep[e.g.,][]{binary_ref}. The BL Lac object OJ\,287 \cite[$z$=0.306,][]{z_ref} is a well studied blazar, harbouring a SMBH with widely varying mass estimates of $~4 \times 10^{8}-~1.8 \times 10^{10}$\,M$_{\odot}$ and exhibiting quasi-periodic flaring that has been suggested as due to a binary black hole system \citep{mass_ref,liu2002,binary_ref_2}. \\

\begin{table*}[th]
\caption{Overview of VLBI observations of OJ\,287}
\begin{center}
\begin{tabular}{ccccccc}
\hline \hline
Epoch & Frequency  & Participating Stations & Beam  & Position Angle & Recording rate & Polarisation \\ 
      &   [GHz]    &          & [maj:min mas] &   [$^{\circ}$]   &   [Mbit/s]     &        \\ \hline
2008.77 & 86.2 & All & 0.211 ; 0.047 & -9.3  & 512 & Dual$^{3}$ \\
2009.36 & 86.2 & All & 0.219 ; 0.051 & -2.5  & 512  & Dual$^{3}$ \\
2009.77 & 86.2 & All & 0.221 ; 0.045 & -2.6  & 512  & Dual$^{3}$ \\
2010.35 & 86.2 & All & 0.269 ; 0.056 & -2.7  & 512  & Dual$^{3}$ \\
2011.36 & 86.2 & All & 0.245 ; 0.047 & -5.6  & 512  & Dual$^{3}$ \\
2011.76 & 86.2 & All & 0.255 ; 0.078 &  3.2 & 512  & Dual$^{3}$ \\
2012.38 & 86.2 & All$^{2}$ & 0.230 ; 0.063 &  22.0 & 512  & Dual$^{3,4}$ \\
2007.45-2013.57 & 43.13 & VLBA & 0.351 ; 0.145$^{1}$ & -2.9   & 512  & Dual \\
2008.70-2012.39 & 15.36 & VLBA & 0.891 ; 0.379$^{1}$ & -5.5   & 512  & Dual \\ \hline
\multicolumn{6}{l}{\textsuperscript{}\footnotesize{$^{1}$ Beam sizes are indicative only and will depend on uv coverage. $^{2}$ Yebes participated. }} \\
\multicolumn{6}{l}{\textsuperscript{}\footnotesize{$^{3}$ Onsala only supported LCP. $^{4}$ Yebes only supported LCP.}}
\end{tabular}
\end{center}
\label{epochs2}
\end{table*}

The jet kinematics, light curves and polarisation properties of OJ\,287 have been recently studied by \citet{agudo11,agudo12}, with the position angle (PA) of the jet axis appearing to change by $\approx$\,$100^\circ$ between 2004 and 2006. Gamma-ray emission was suggested to be correlated with mm-radio flaring and placed at least 14 pc away from the central engine, largely in agreement with spectral energy distribution (SED) modelling by \citet{kushwaha13}. Recent flaring activity in 2011-2012 was analysed by \citet{sawada15} using 22\,GHz VLBI maps. They  interpreted the $\gamma$-ray flaring as possibly being a new jet component that is unresolved at 22\,GHz. \\

Very long baseline interferometry (VLBI) has long been used to provide high angular resolution images of blazars, with angular resolutions of $\approx$0.1--0.2\,milli-arcseconds (mas) achievable with 43 GHz VLBI. Visible structural variations can occur within weeks to months, requiring high cadence monitoring. Such monitoring programs include the Monitoring Of Jets in AGN with VLBA Experiments (MOJAVE) program at 15\, GHz \citep{MOJAVE_ref} and the VLBA-BU-BLAZAR 43 GHz blazar monitoring program \citep{mar08}. A subset of these sources has been observed approximately semi-yearly at 86\,GHz using the Global mm-VLBI Array (GMVA) from 2008 until now. It has been proposed by \citet{DalyMarscher88}, \citet{Marscher08Rev}, \citet{Cawthorn13} and others that the mm-wave ``core'' in VLBI images could be the first of a series of recollimation shocks produced when a jet becomes under-pressured compared to its surrounding medium. At lower frequencies, the ``core'' is usually assumed to be a $\tau=1$ surface where the radiation begins to be self absorbed. Global 3\,mm VLBI using the GMVA allows the imaging of regions above the self-absorption turnover frequency with angular resolution approaching $\approx$40\,$\mu$as. \\

Since the launch of EGRET, on board the Compton Gamma Ray Observatory \citep{egret}, $\gamma$-rays have been known to be produced in AGN, but the site of their production remains elusive \citep{fichtel94,jor01}. With the launch in 2008 of the \emph{Fermi Gamma-ray Space Telescope} and the Large Area Telescope instrument on board (\emph{Fermi}), long-term $\gamma$-ray light curves of many ($>$\,$10^{3}$) AGN have been observed \citep{abdo10,2fgl}. Sites for $\gamma$-ray production are proposed to be either within the Broad Line Region (BLR) close to the central engine \citep{blandford95} or further along the jet in recollimation shocks or other jet features \citep{mar14}. The observational evidence is conflicting with some favouring the former scenario \citep[e.g.,][]{tav10,rani13b} and others the latter \citep[e.g.,][]{jor10,rani13a,fuhrmann14}. While TeV emission is not detected from all blazars, it is interesting to note that the comparable source with a similar redshift 0716+714, exhibits TeV emission, while OJ\,287 does not \citep{tash08,anderhub09,wang13}.\\

Here, we aimed to further test this scenario, using recent 15 and 43 GHz data and the semi-annual GMVA observations at 86\,GHz to derive kinematic properties and estimate magnetic field strengths in individual VLBI components. In Section 2, we present the data obtained and the methods to reduce and analyse the data. In Section 3, we present our observational results and Section 4 contains a deeper analysis using a spectral decomposition to compute magnetic field strengths at multiple locations in the jet. We then devise a method to estimate the location of jet features relative to the jet apex. In Section 5, we present the interpretation of these results and discuss them in the context of prevailing theories. In Section 6, we present our conclusions and outlook for the future. Dates throughout the paper are presented in decimal years. A linear scale of 4.64 pc/mas and a luminosity distance $D_{L}$ of 1.63 Gpc, at the source redshift of z=0.306 was adopted with standard cosmological parameters of $\Omega_{m} = 0.302$, $\Omega_{\lambda} = 0.698$ and $H_{0} = 68$ km s$^{-1}$ Mpc$^{-1}$ \citep{cosmo,spe07}.  \\

\section{Observations and data analysis}

\begin{table*}%
\caption{Overview of stations used in global 3 mm VLBI observations}
\begin{center}
\begin{tabular}{llccc}
\hline \hline
Station & Country & Effective Diameter & Typical SEFD$^{1}$  & Polarisation\\ 
        &         &      [m]           &     [Jy]      &   \\ \hline
Mets\"{a}hovi & Finland & 14 & 17500 & Dual\\
Onsala & Sweden & 20 & 5500 & LCP\\ 
Effelsberg & Germany & 80 & 1500 & Dual\\ 
Plateau de Bure & France & 34 & 500 & Dual\\ 
Pico Veleta & Spain & 30 & 700 & Dual\\ 
Yebes & Spain & 40 & 1700 & LCP \\ 
VLBA (x8) & United States & 25 & 2000 & Dual\\ \hline
\multicolumn{5}{l}{\textsuperscript{}\footnotesize{$^{1}$ System equivalent flux density}}
\end{tabular}
\end{center}
\label{stations}
\end{table*}

\subsection{GMVA observations}

The GMVA combines the eight 3\,mm receiver equipped stations of the VLBA and up to six European observatories, including Effelsberg, Onsala, Mets\"{a}hovi, Pico Veleta, Plateau de Bure and since 2012, Yebes \citep{hodgson14}. Data were recorded at 512 Mbit/s, with eight 8 MHz channels, in dual polarisation. Onsala and Yebes observed in left circular polarisation (LCP) only. For these stations, LCP was assumed to be equal to RCP and hence Stokes I. Scans of approximately 7 minutes every 15 minutes were recorded with pointing and calibration performed on European stations in the gaps between scans. A summary of observations is given in Table \ref{epochs2} and a summary of participating stations is given in Table \ref{stations}. Between 2008 and 2012, observations were taken approximately every six months, except in 2010 where only one observation was made. Data were correlated using the DiFX correlator after $\sim$2011.4 and earlier with the Mark IV correlator at the Max-Planck-Institut f\"{u}r Radioastronomie in Bonn, Germany \citep{deller07,deller11}.\\

Data were fringe-fitted and calibrated using standard procedures in the Astronomical Image Processing System (AIPS) for high frequency VLBI data reduction \cite[e.g.,][]{jor05} with extended procedures written in ParselTongue as described by \citet{marti12}. Within AIPS, amplitudes were corrected for system temperatures, sky opacity and gain-elevation. Phase calibration was performed on the brightest sources and scans within the experiment and fringe-fitting was performed averaged over all intermediate frequencies (IFs) in order to increase SNR. Relative flux density accuracy of VLBI measurements as compared against F-GAMMA (section 2.3.2) and VLA/EVLA flux densities are within 5-10\%. An example 3\,mm VLBI map of the source is presented with a near-in-time 7\,mm map in Fig. \ref{3mm121}. The remaining epochs are also presented with near-in-time 7\,mm maps in Fig. \ref{3mm082} to Fig. \ref{3mm112}.

\subsection{VLBA observations at 15 and 43 GHz}

\begin{figure*}
\centering
\includegraphics[width=0.8\linewidth]{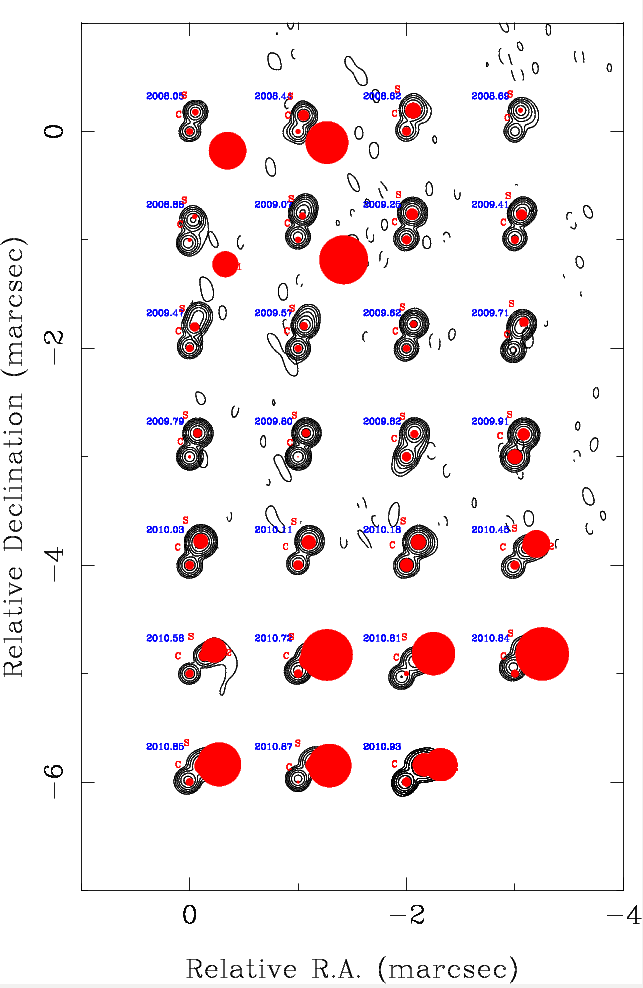}
\caption{7\,mm maps until 2010, super-resolved with a circular 0.10\,mas beam. }
\label{all7mm1}
\end{figure*}

\begin{figure*}
\centering
\includegraphics[width=\linewidth]{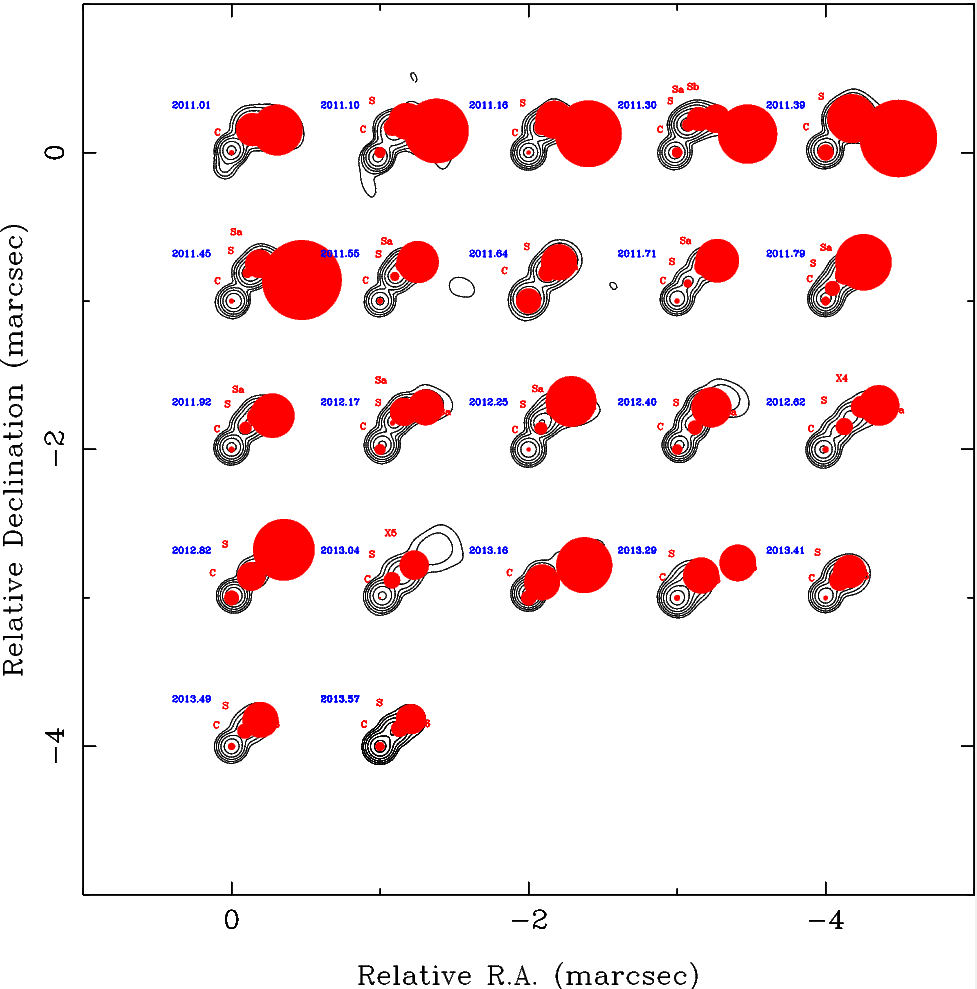}
\caption{7\,mm maps from 2011 onwards, super-resolved with a circular 0.10\,mas beam. }
\label{all7mm2}
\end{figure*}

In total, 72 observations of OJ287 were obtained approximately monthly as part of a VLBA-BU-BLAZAR 43\,GHz VLBA monitoring program of $ \gamma $-ray bright blazars \citep{mar08}, with increased cadence during the flaring events of August 2007, October 2009 and November 2011. The data are publicly available from the VLBA-BU-BLAZAR program website and were re-imaged by us for this work. Reduction was performed according to \citet{jor05}. 15\,GHz VLBI images were obtained as part of the MOJAVE monitoring program with data reduction and errors described in \citet{MOJAVE_ref}. 15\,GHz MOJAVE data were primarily used to provide VLBI flux measurements at 15\,GHz. For spectral index determination, we selected seven epochs of near-simultaneous MOJAVE VLBI and VLBA-BU-BLAZAR data. Data for both the BU Blazar Monitoring Program and the MOJAVE program were correlated using DiFX at the National Radio Astronomy Observatory (NRAO) Array Operations Centre in Soccorro, New Mexico \citep{deller07,deller11}. A typical 7\,mm map is shown in Fig. \ref{eg7mm}. A full sequence of super-resolved 7\,mm maps are shown in Fig. \ref{all7mm1} and Fig. \ref{all7mm2}. Because the observations were not truly simultaneous with the 3\,mm GMVA observations, when analyses were performed combining near-in-time epochs, the observation date is displayed with a tilde (e.g., $\sim$2009.4).  \\

\subsection{Imaging and model-fitting}

Images were produced in DIFMAP using the CLEAN algorithm \citep{clean}. In addition to this, amplitude and phase self-calibration was also performed \citep{selfcal,difmap}. In order to parametrise images for analysis, circular Gaussian components were fitted to the visibility data. This allowed us to represent the location, size and flux densities of distinct regions of the jet. Errors were independently estimated for several test epochs using the techniques described by \citet{mueller14} and \citet{frank12} with a conservative 10$\sigma$ error on separations and 5$\sigma$ errors on component sizes assumed \citep[e.g.,][]{bia15}. We found that errors on the sizes and core separation are consistently $\sim$20\% of the beam for unresolved components and 15\% of the full-width half maximum (FWHM) for resolved components. Similarly, errors on fitted fluxes (excluding possible systematic errors) were consistently $\sim$10\% and errors in the PA were $\sim$5$^{\circ}$. These values have been adopted throughout the paper. Errors were then propagated by producing simulated variables with Gaussian distributions 1000 times. When performing a model-fit in DIFMAP, a reduced $\chi$-squared value is calculated as an estimate of the goodness-of-fit. Occasionally, model-fitting a Gaussian component would fit the FWHM of the component to a delta or almost delta component, probably indicating unresolved structure. In this situation, the FWHM of the component was forced to an approximation of 1/5th of the beam. \\

In VLBI experiments such as this, absolute phase information is lost due to applying phase self-calibration during imaging, requiring maps to be aligned in order to perform analysis. As these are not phase referencing experiments \cite[e.g.,][]{ros05}, there are two methods to align maps. (1) is to use a two-dimensional cross-correlation on optically thin emission under the assumption that optically thin emission is co-spatial at all frequencies \cite[e.g.,][]{fromm13_3}. This approach was attempted but did not yield meaningful results for these data, possibly due to resolved transverse jet structure or differences in position due to differences in observation dates (with up to two weeks difference) or a combination of the above. We hence used a second approach (2) to align based on model-fitted components and morphological similarities  \cite[e.g.,][]{kadler04}. We employed this method here, aligning on the southern-most permanent feature (component ``C'') in all epochs. \\

\subsection{Long-term total intensity lightcurves}

Long-term total intensity light-curves from late 2008 until 2014.1 were obtained from $ \gamma $-ray to cm wavelengths at E$>$219\,MeV, 350 GHz (0.87 mm), 225 GHz (1.3 mm), 86.24 GHz (3 mm) and 43 GHz (7 mm).  \\

\subsubsection{Gamma-ray lightcurves}
We made use of 100\,MeV--300\,GeV data of the source from 2008.6 to
2014.2, which was observed in survey mode by the {\it Fermi}-LAT \citep{Atw09}.
Photons in the `source' event class were selected for the analysis. We analysed the LAT data using the standard ScienceTools\footnote{ScienceTools can be downloaded from http://fermi.gsfc.nasa.gov/ssc/data/analysis/documentation/} (software version v9.32.5) 
and instrument response function P7REP\_SOURCE. We analysed a region of interest of 10$^{\circ}$ in radius, centred at the position of the $\gamma$-ray 
source associated with OJ\,287 using a maximum-likelihood algorithm \citep{mattox96}.

\begin{figure}
\includegraphics[width=\linewidth]{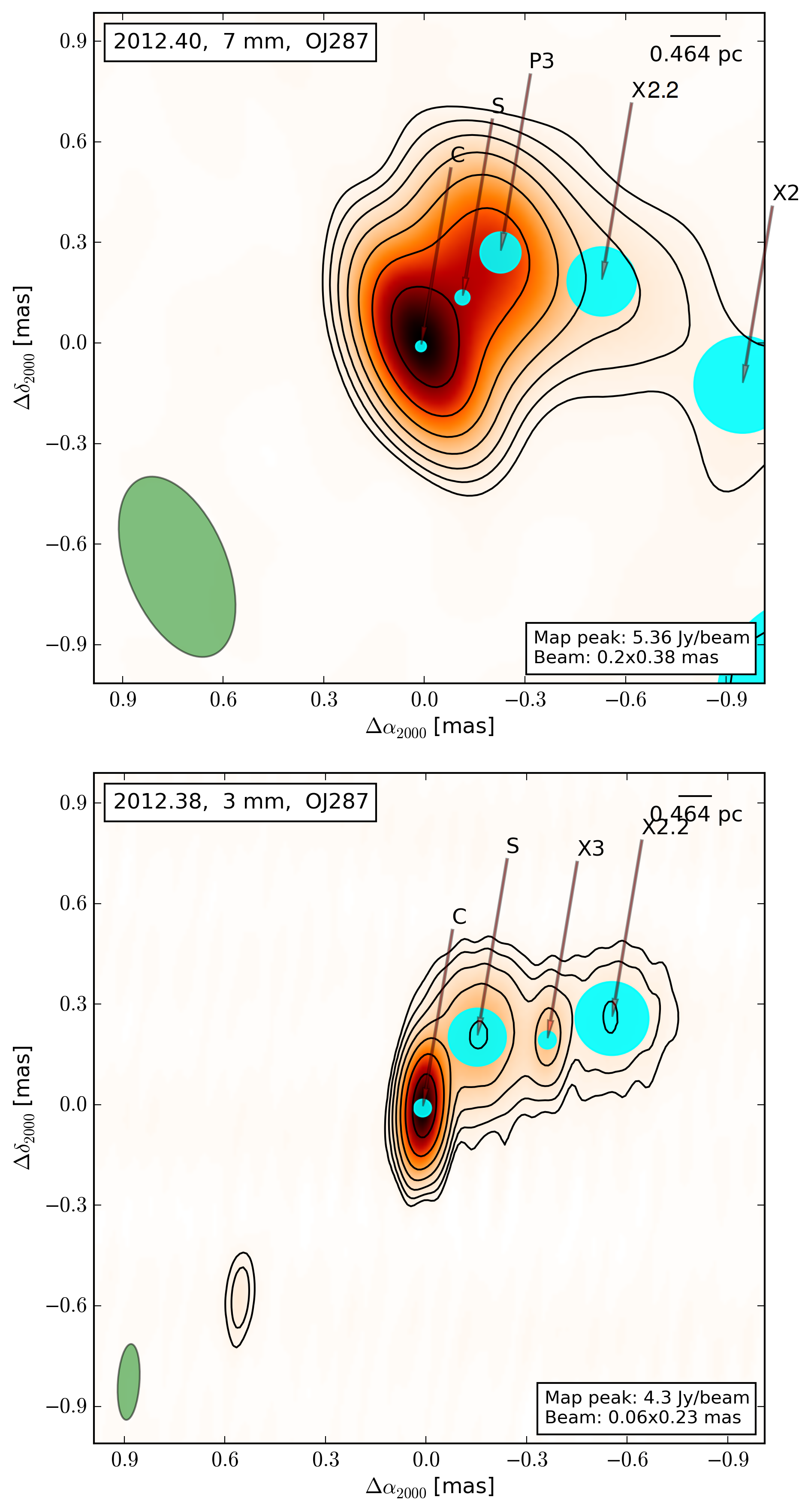}
\caption{Near-in-time 3\,mm and 7\,mm VLBI observations of OJ\,287 in $\sim$2012.4. Contours: -1, 1, 2, 4, 8, 16, 32, 64\% of peak flux density. Beam size and peak flux density are shown in the bottom right corner. The restoring beam ellipse is shown in bottom left corner. The position and size of Gaussian model-fit components is represented as filled blue circles within the map. Labels represent the identification of the VLBI component. Other epochs are shown in the appendix. }
\label{3mm121}
\end{figure}

In the unbinned likelihood analysis\footnote{http://fermi.gsfc.nasa.gov/ssc/data/analysis/scitools/likelihood\_tutorial.html}, we included all the 23 sources of the 2FGL catalogue \citep{2fgl_} within 10$^{\circ}$ and the recommended Galactic diffuse background (gll{\_}iem{\_}v05.fits) and the isotropic background (iso{\_}source{\_}v05.txt) emission components. \\
We compute photon flux light curves above the ``decorrelation energy'' ($E_{0}$), \citep{lott12}, which
minimises the correlations between integrated photon flux and photon index. Over the course of the past 6
years of observations, we obtain $E_{0}$ = 219 MeV.  We generated the constant uncertainty (15\%) light curve
above $E_{0}$ through the adaptive  binning method following \citet{lott12}. The adaptive binned light
curve is produced by modelling the spectra  by a simple power law ($ N(E) = N_{0}E^{-\Gamma}$, $N_{0}$ : prefactor and $\Gamma$ : power law index). 
The estimated systematic uncertainty on the flux is 10\% at 100 MeV, 5\% at 500 MeV, and 20\% at 10 GeV \citep{ackermann12} which are comparatively 
less dominant than the statistical errors. The OJ\,287 $\gamma$-ray light curve (top) and spectral index (bottom) are presented in Panel 1 of Fig. \ref{radio_lcs}.

\subsubsection{Radio lightcurves}

The cm/mm radio light curves of OJ\,287 have been obtained within the framework of a {\sl Fermi}-LAT related monitoring program of $\gamma$-ray blazars \citep[F-GAMMA program, for reduction details, see:][]{fuhrmann07, manolis08, fuhrmann14}. The millimetre observations are closely coordinated with the more general flux monitoring conducted by IRAM, and data from both programs are included in this paper. The overall frequency range spans from 2.64\,GHz to 142\,GHz using the Effelsberg 100-m and IRAM 30-m telescopes. The 225 GHz (1.3\,mm) and 345\,GHz (0.87\,mm) flux density data was obtained at the Submillimeter Array (SMA) near the summit of Mauna Kea (Hawaii).  OJ287 (J0854+201) is included in an ongoing monitoring program at the SMA to determine the flux densities of compact extragalactic radio sources that could be used as calibrators at mm wavelengths \citep{gurwell07}.  Observations of available potential calibrators are from time to time taken for 3 to 5 minutes, and the measured source signal strength calibrated against known standards, typically solar system objects (Titan, Uranus, Neptune, or Callisto).  Data from this program are updated regularly and are available at the SMA website. These Light curves are shown in Panel 2 of Fig. \ref{radio_lcs}. 
\section{Results}
\subsection{Morphology}
\begin{figure}
\includegraphics[width=\linewidth]{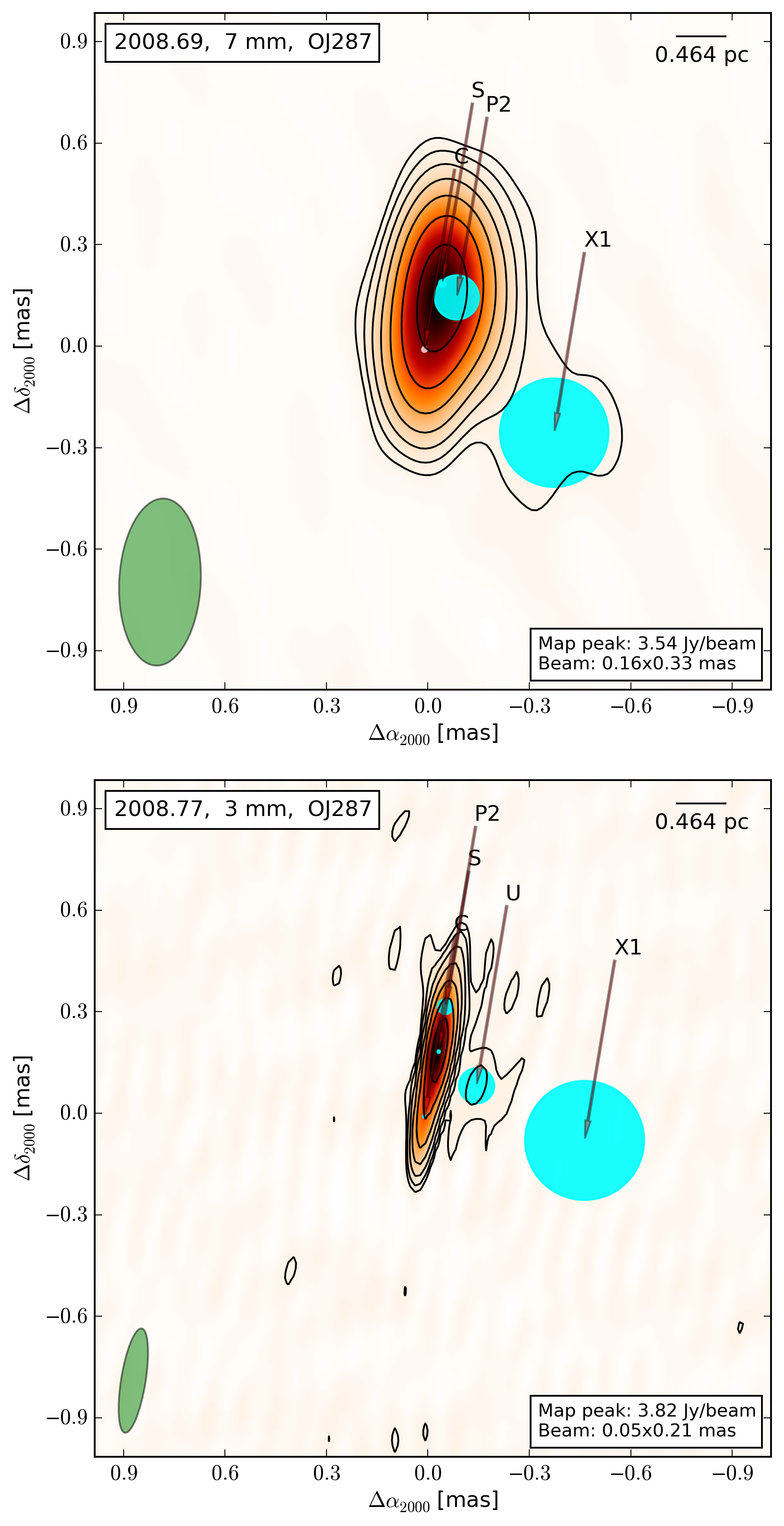}
\caption{Near-in-time 3\,mm and 7\,mm VLBI observations of OJ\,287 in $\sim$2008.7. Contours: -1,-0.5, 0.5, 1, 2, 4, 8, 16, 32, 64\% of peak flux density. Other features are described in Fig. \ref{3mm121}. }
\label{3mm082}
\end{figure}

\begin{figure}
\includegraphics[width=\linewidth]{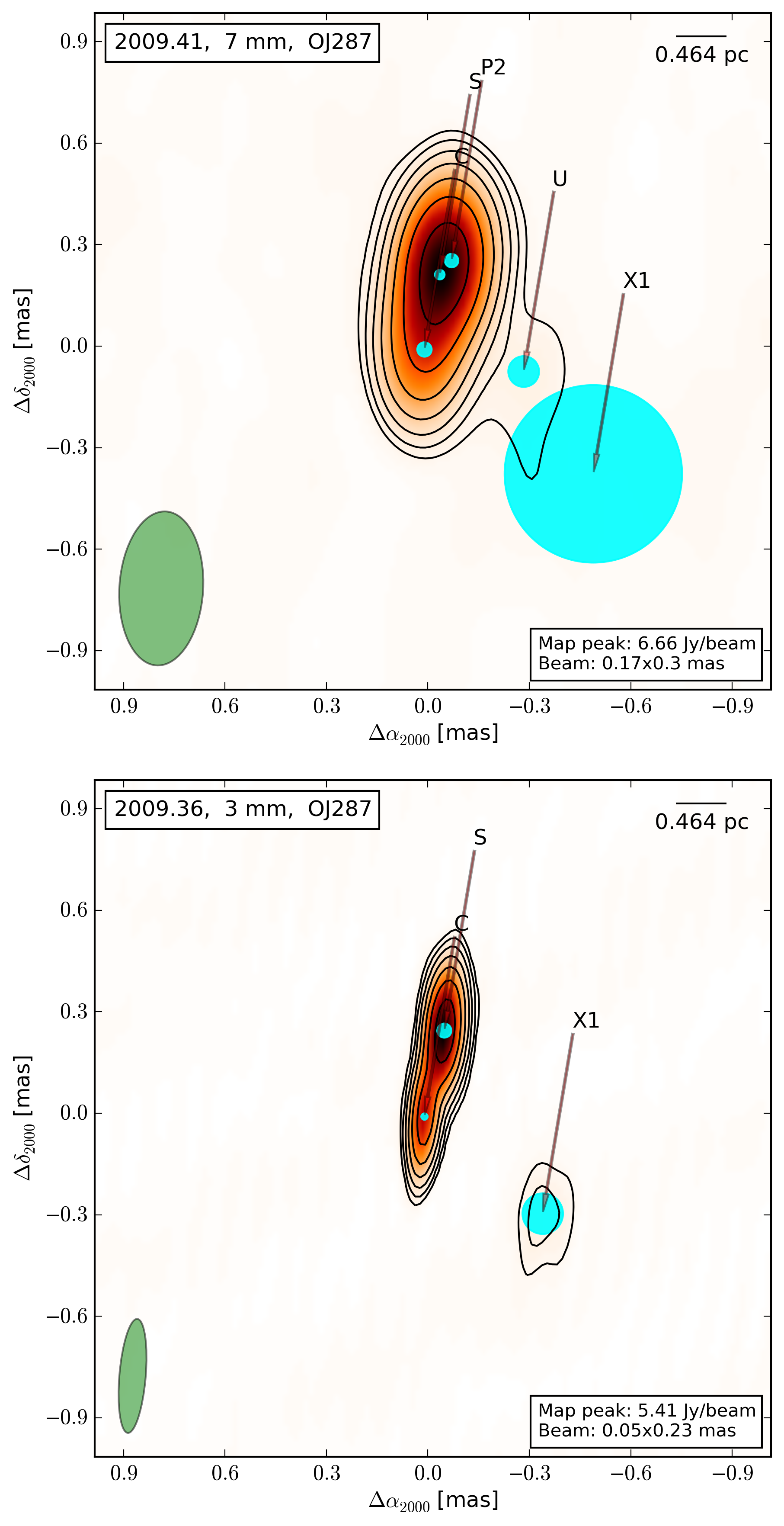}
\caption{Near-in-time 3\,mm and 7\,mm VLBI observations of OJ\,287 in $\sim$2009.4. Contours: -1, 1, 2, 4, 8, 16, 32, 64\% of peak flux density. Other features are described in Fig. \ref{3mm121}. }
\label{3mm091}
\end{figure}

\begin{figure}
\includegraphics[width=\linewidth]{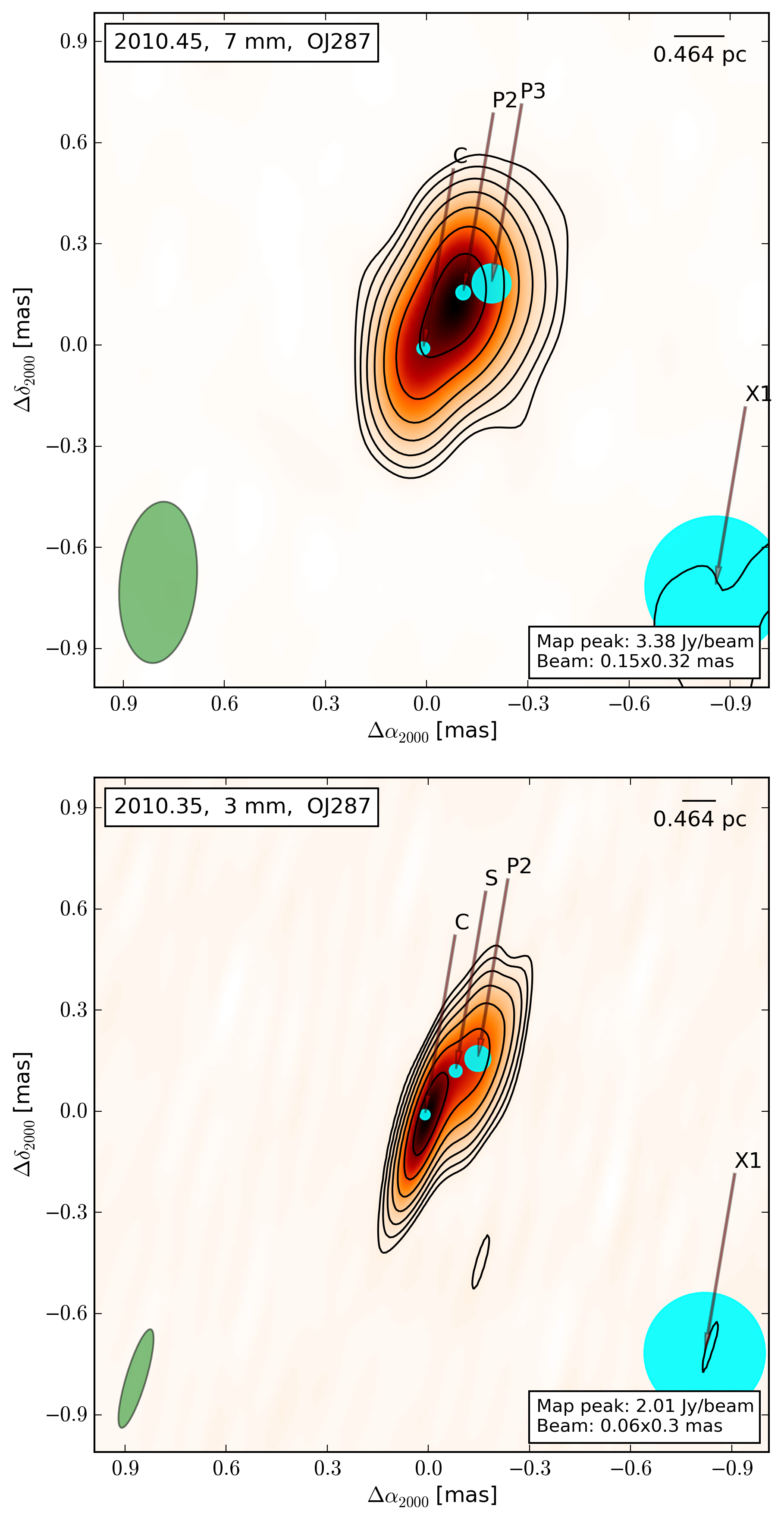}
\caption{Near-in-time 3\,mm and 7\,mm VLBI observations of OJ\,287 in $\sim$2010.4. Contours: -1, 1, 2, 4, 8, 16, 32, 64\% of peak flux density. Other features are described in Fig. \ref{3mm121}.}
\label{3mm101}
\end{figure}

\begin{figure}
\includegraphics[width=\linewidth]{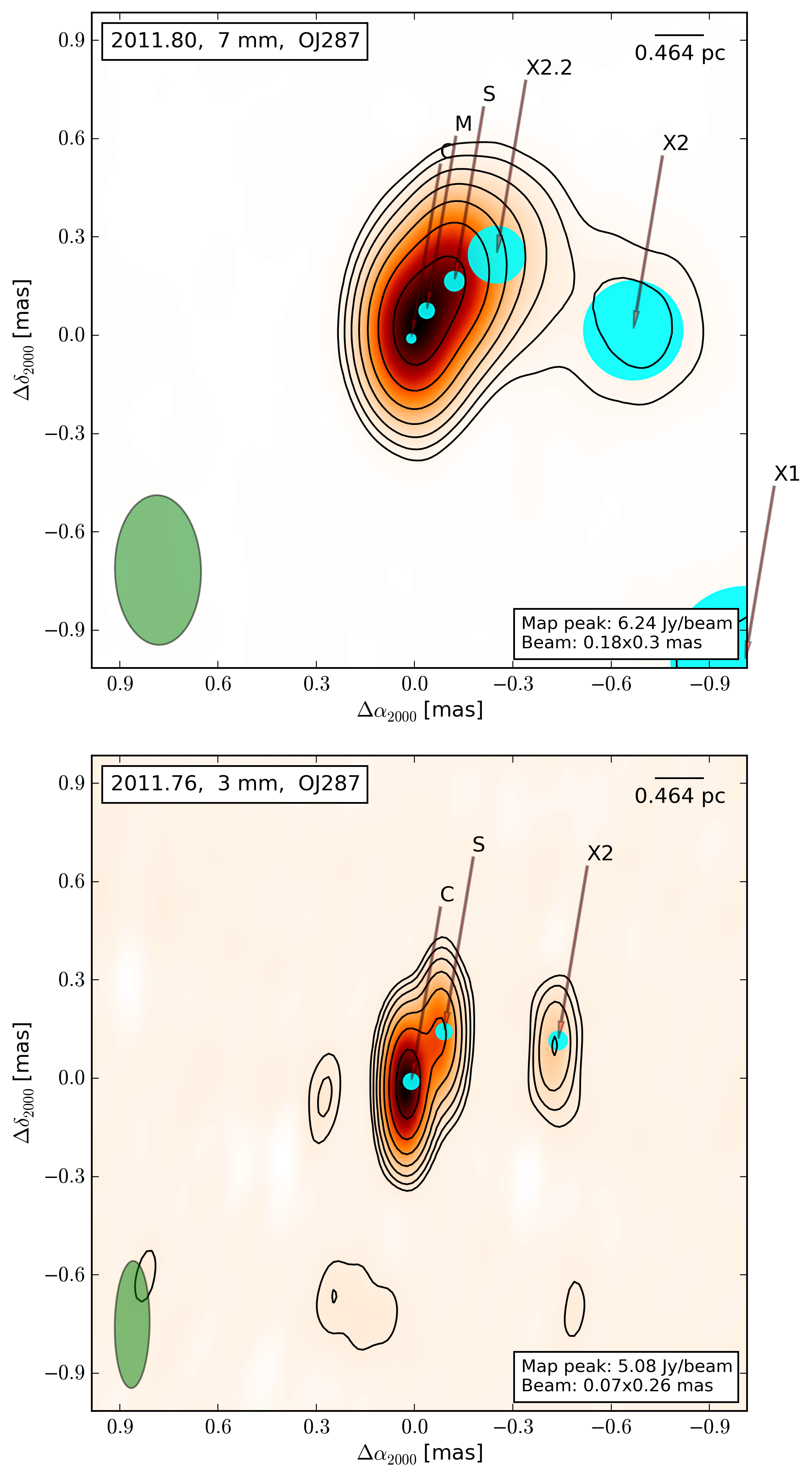}
\caption{Near-in-time 3\,mm and 7\,mm VLBI observations of OJ\,287 in $\sim$2011.8. Contours: -1, 1, 2, 4, 8, 16, 32, 64\% of peak flux density. Other features are described in Fig. \ref{3mm121}.}
\label{3mm112}
\end{figure}

In Fig. \ref{3mm121} and in Figs. \ref{3mm082}--\ref{3mm112}, 7\,mm maps are shown in the top panels, with near-in-time GMVA observations of OJ\,287 displayed in the bottom panels. In all epochs, both 3\,mm and 7\,mm VLBI maps show broadly consistent morphology, with only minor differences in structure. In all epochs, two bright quasi-stationary features were detected approximately 0.2\,mas from each other in a roughly north-south orientation. Moving jet components appear at the northern feature and then move in a westerly or south-westerly direction. \\

\subsection{Stationary features and ``core'' identification}
\begin{figure}
\includegraphics[width=\linewidth]{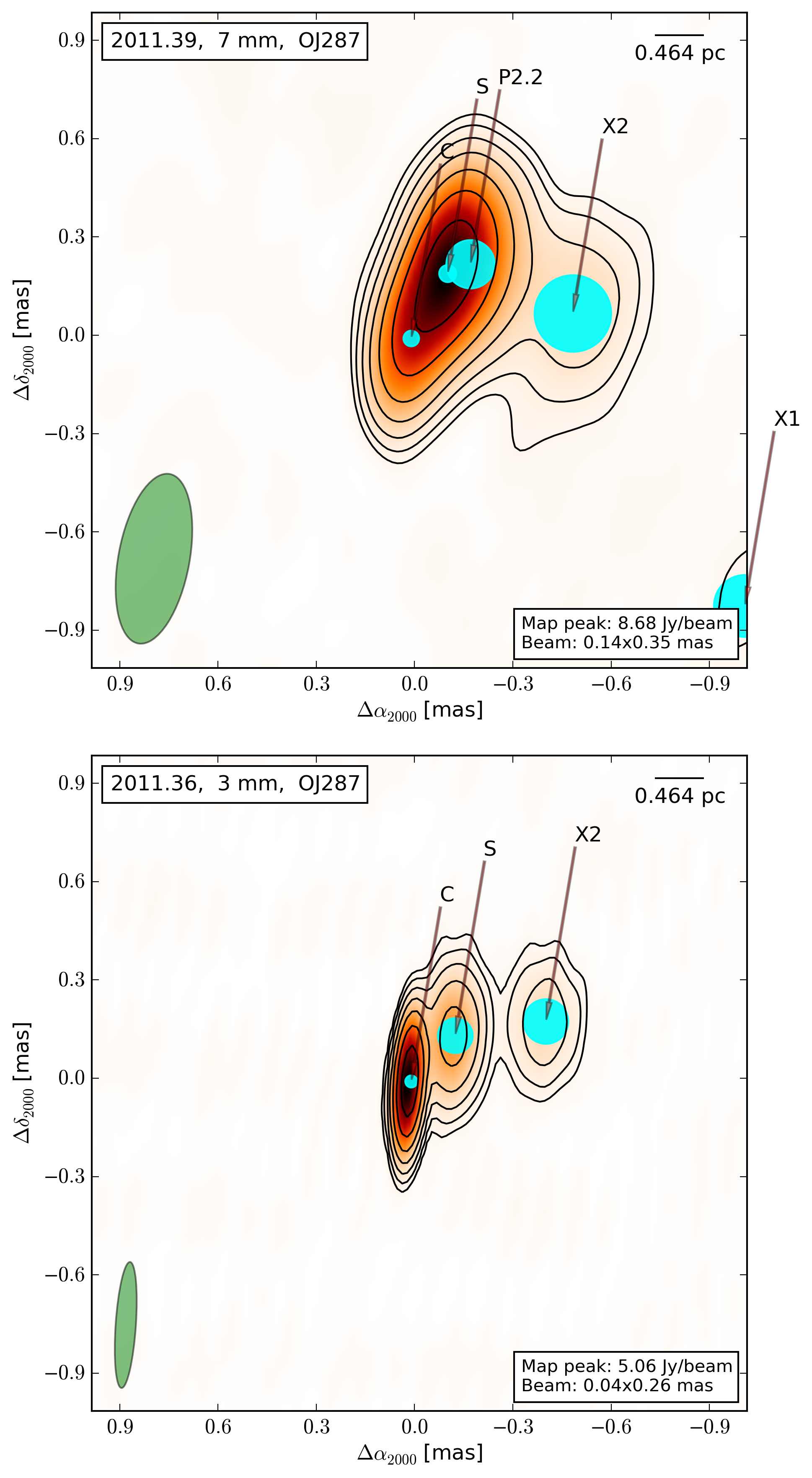}
\caption{Near-in-time 3\,mm and 7\,mm VLBI observations of OJ\,287 in $\sim$2011.4. Contours: -1, 1, 2, 4, 8, 16, 32, 64\% of peak flux density. Other features are described in Fig. \ref{3mm121}.}
\label{3mm111}
\end{figure}

To perform accurate kinematics, a common point of reference must be defined, typically taken to be the most upstream visible component or
VLBI ``core'' \cite[e.g.,][]{jor05}. The ``core'' is typically identified on the basis of i) morphology, ii) a smaller size, higher flux 
densities and correspondingly higher brightness temperatures than downstream components, iii) an optically thick (inverted) spectrum, and iv)
a stronger variability of the flux density. On average the southernmost stationary feature was smaller and brighter with a higher brightness 
temperature than the north-west component, although on some occasions the reverse situation was true. 
\begin{center}
\begin{figure*}[htbp]
\includegraphics[width=\linewidth]{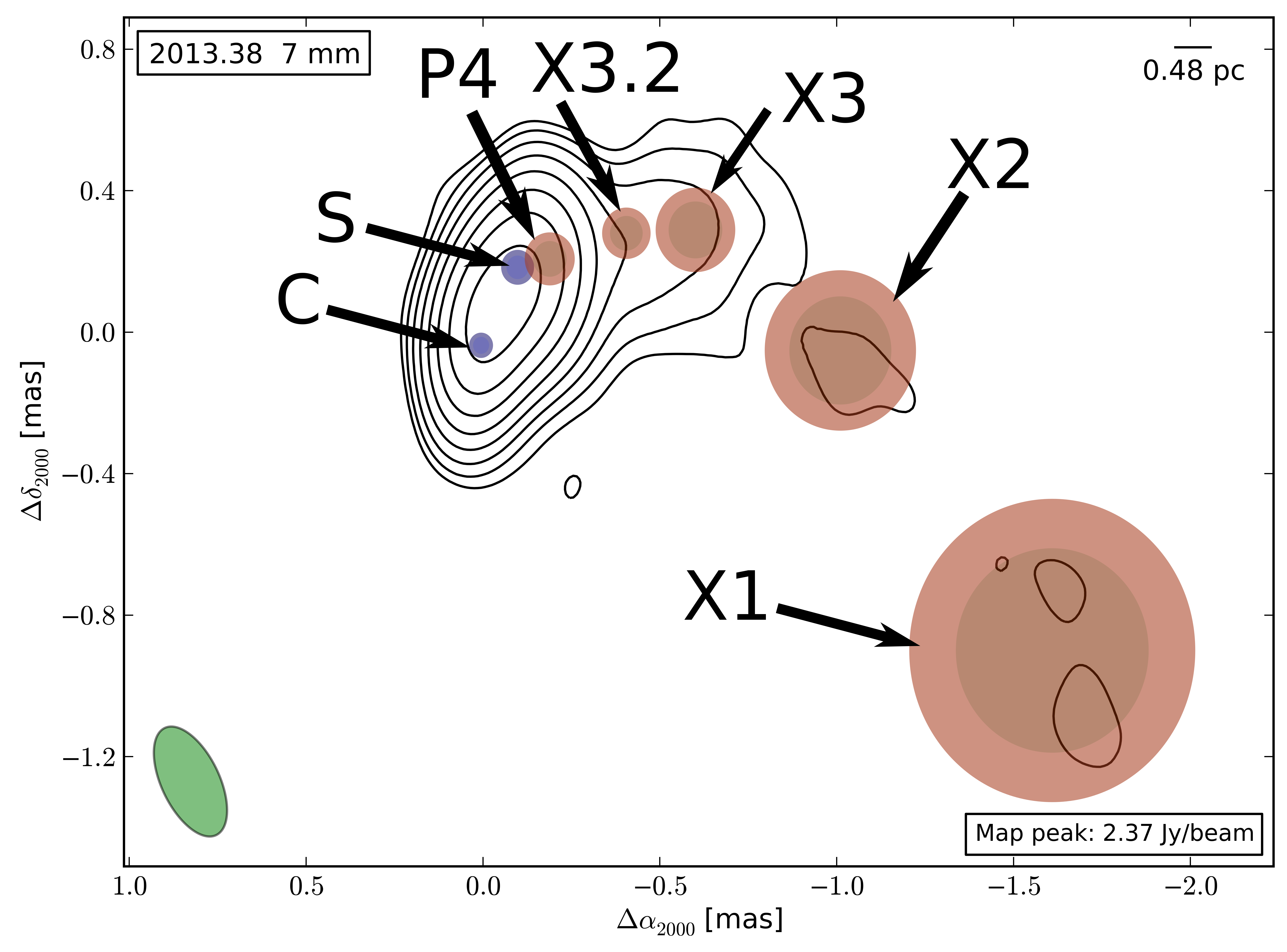}
\caption{A recent example 7\,mm map showing all fitted components and the shape of the jet. Contours: -1, -0.5, 0.5, 1, 2, 4, 8, 16, 32, 64\%. Other features are described in Fig. \ref{3mm082}.}
\label{eg7mm}
\end{figure*}
\end{center}
Both components exhibit both negative and
positive spectral indices at different times, with high degrees of variability. However, the southern-most component shows on average a more 
inverted spectrum (see Section \ref{sec:decomp} for more details). Travelling components are always first identified in the maps near the northern 
component (e.g., Fig. \ref{3mm111}), moving away from the southern component. Features detected between the northern and southern components cannot, 
however, be conclusively identified with ejected components. We cannot exclude the possibility that these components travel towards the southern component
from the northern component, implying that the southern component could be a counter-jet. We consider this unlikely, since the source has historically exhibited 
highly superluminal motion and any counter-jet would be highly Doppler de-boosted \citep{jor05}. This leads us to identify the southern component as the ``core''
(labelled $C$), consistent with the ``core'' identification of \citet{agudo11}. We refer to the northern component as ``the quasi-stationary feature'' (labelled $S$)
as its proper motion was-0.004 $ \pm $ 1.01\,mas/year and consistent with stationarity. Component $S$ could be seen as early as late 2004, but its current location $\sim$0.2 mas
from the core was not established until late 2008, after which it was persistent in all later epochs \citep{agudo12}. \\

\subsection{Moving Component Identification and Kinematics}

\begin{table*}
\centering
\begin{threeparttable}%
\caption{Table of fitted components and derived properties}
\label{comps2}
\begin{tabular}{lcccccc}
\hline \hline
                            & $X1_{43}$      & $X2_{43}$       & $X2.2_{43}$      & $X3_{43}$     & $X3.2_{43}$ \\ \hline
$\mu$ [mas/yr]           & 0.27 $\pm$ 0.10 & 0.50 $\pm$ 0.24 & 0.40 $\pm$ 0.24 & 0.45 $\pm$ 0.13 & 0.50 $\pm$ 0.24  \\ 
$\beta_{\text{app}}$ [c]    & 3.8 $\pm$ 1.5 & 7.3 $\pm$ 3.5 & 5.8 $\pm$ 3.5 & 6.6 $\pm$ 1.9 & 7.3 $\pm$ 3.5 \\ 
Av. PA [$^{\circ}$]     & -127.4 $\pm$ 0.1 & -89.5 $\pm$ 0.5 &  -56.6 $\pm$ 1.1 & -51.5  $\pm$ 1.8 & -63.4 $\pm$ 1.4 \\
$\theta_{\text{crit}}$  [$^{\circ}$] & 15.2 $\pm$ 3.4 & 7.8 $\pm$ 2.8 & 9.8 $\pm$ 2.0 & 8.7 $\pm$ 2.5 & 7.9 $\pm$ 1.5  \\ 
$\delta_{\text{VLBI}}$  & 3.8 $\pm$ 0.4 & 7.9 $\pm$ 2.7 & 5.8 $\pm$ 0.7 & 7.4 $\pm$ 1.7 & 7.6 $\pm$ 2.5  \\ 
$\Gamma_{\text{min}}$   & 3.9 $\pm$ 0.2   &  7.4 $\pm$ 0.2 & 5.9 $\pm$ 0.3 & 6.6 $\pm$ 1.7 & 7.4 $\pm$ 1.2 \\
$C_{0}$            & 2007.89 $\pm$ 0.34 & 2010.06 $\pm$ 0.19 & 2010.85 $\pm$ 0.26 & 2011.97 $\pm$ 0.22 & 2012.75 $\pm$ 0.28   \\ 
$S_{0}$           & 2008.69 $\pm$ 0.34 & 2010.35 $\pm$ 0.19 & 2011.10 $\pm$ 0.26 & 2012.17 $\pm$ 0.22 & 2013.04 $\pm$ 0.28 \\
\hline
\end{tabular}
\begin{tablenotes}
\small
\item $\mu$: component proper speed [mas/yr]; $\beta_{\text{app}}$: Apparent super-luminal motion [c]; $\theta_{\text{crit}}$: critical angle; $\delta_{\text{VLBI}}$: Doppler factor from component speeds; $\Gamma_{\text{min}}$: minimum Lorentz factor from component speeds; $C_{0}$: estimated component passage time past C; $S_{0}$: estimated component passage time for S.
\end{tablenotes}
\end{threeparttable}
\end{table*}

\begin{figure*}[!htb]
\centering
\includegraphics[width=\linewidth]{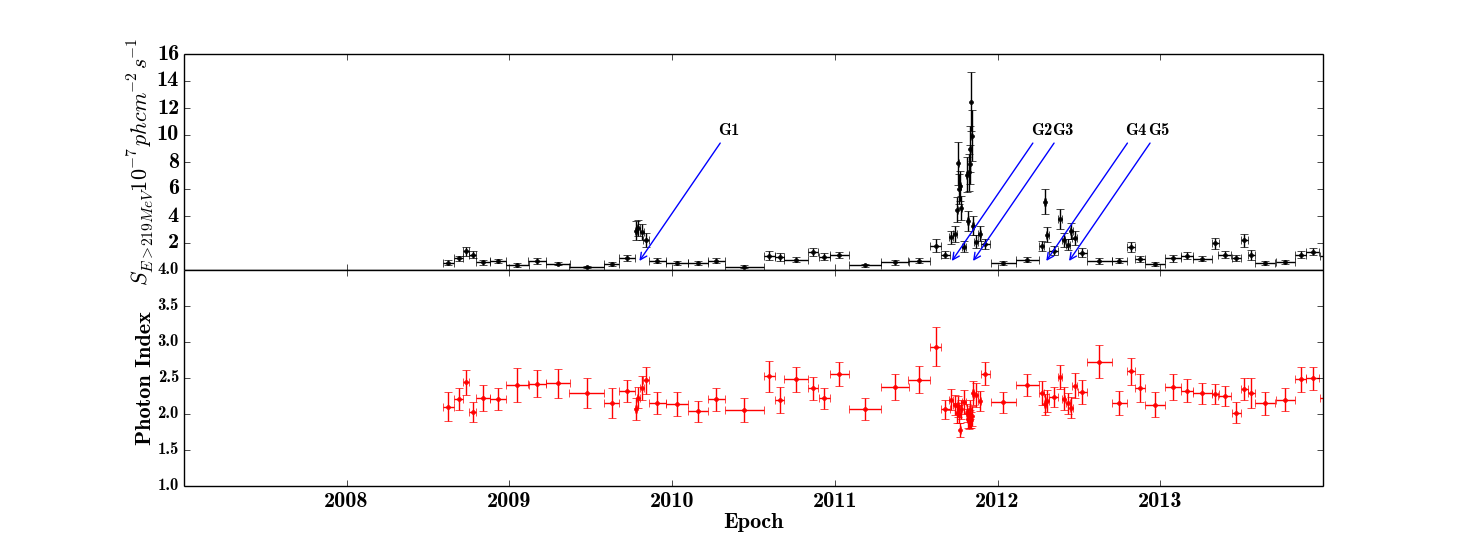}
\includegraphics[width=\linewidth]{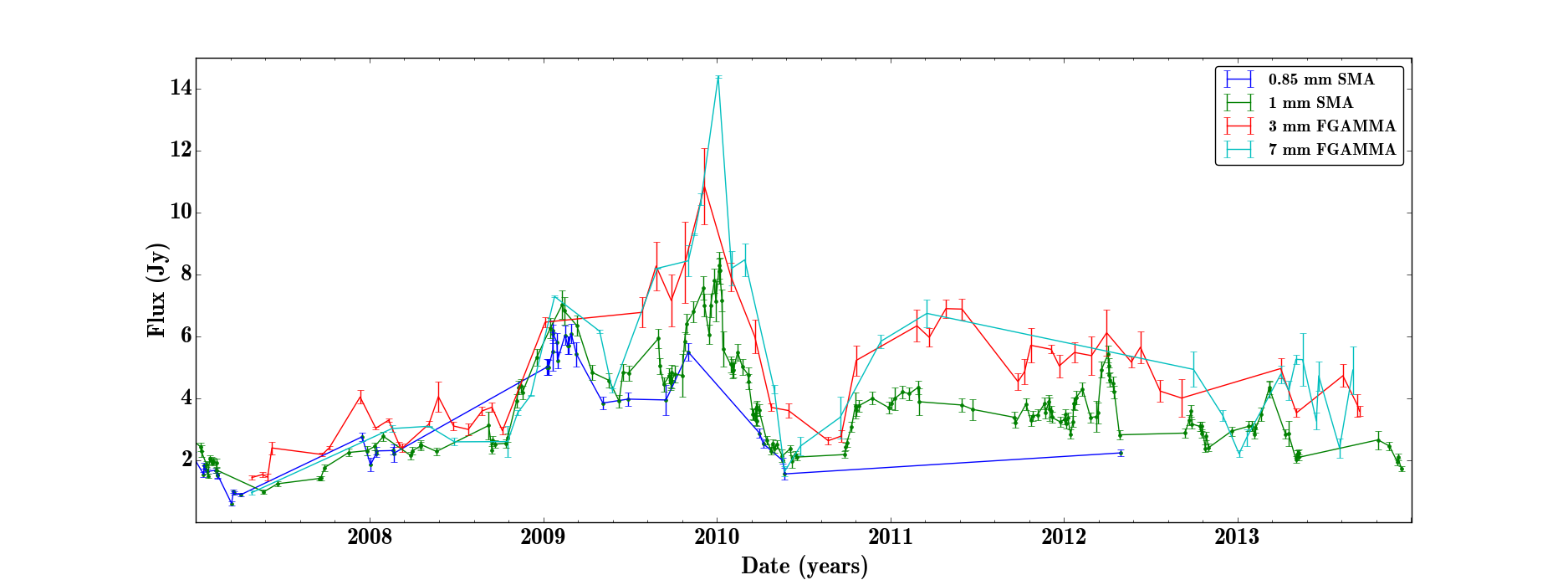}
\includegraphics[width=0.95\linewidth]{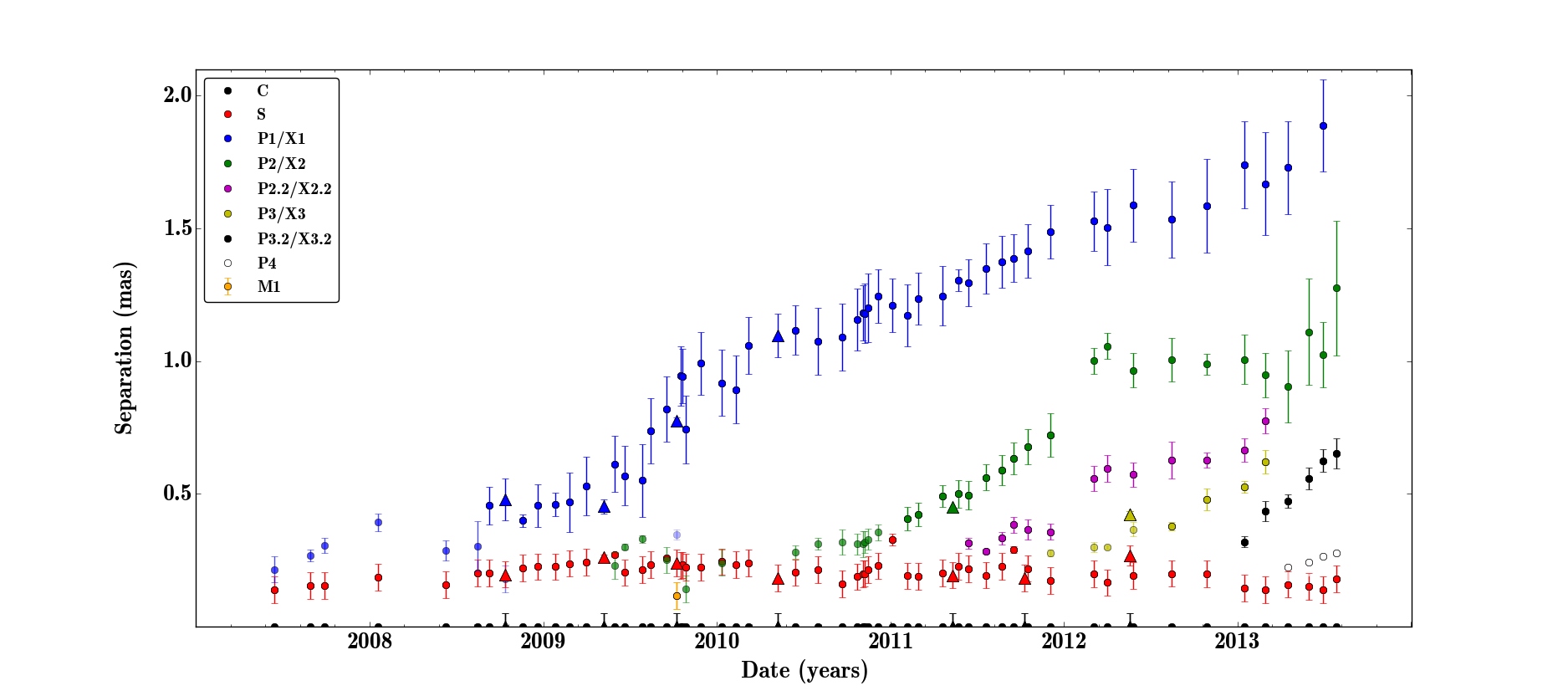}
\caption{Panel 1: 100\,MeV--300\,GeV Gamma-ray photon flux (top) and photon index (bottom). Gamma-ray flares are labelled G1-G5. Panel 2: Total intensity mm-wave
light-curves from 2007 until 2013. 7\,mm (cyan); 3\,mm (red); 1\,mm (green); 0.85\,mm (blue). Panel 3: ``Core'' separation plot of components in OJ\,287. Circular
symbols denote 43\,GHz data points. Triangle symbols denote 86\,GHz data points. }
\label{radio_lcs}
\end{figure*}

A summary of the kinematic properties, including the component proper motions of OJ\,287 is given in Table \ref{comps2}. 
Radial displacement from the ``core'' as a function of time is shown in panel 3 of Fig. \ref{radio_lcs}. If a component was simultaneously detected 
at 3 and 7\,mm, a cross-identification between the frequencies is implied. well-defined moving components (over a beam-size away from
the quasi-stationary feature) are labelled $Xn$ or $Xn.2$. When a moving component is less than a beam-size away from a stationary feature 
($\approx$0.15 mas at 7\,mm and $\approx$0.07 mas at 3\,mm), it is difficult to discriminate between the quasi-stationary feature and the 
moving feature and therefore we label it $P$. Their fluxes are then summed for the purposes of flare identification. When a component
was identified between C and S, it was labelled $M$ and could possibly be associated with a later ejected moving features. If a component cannot be 
easily cross-identified or identified with either a moving or stationary feature, it is labelled U.   \\ 
To aid in consistent component identification, 7\,mm model-fits use the previous epoch's best model as a starting model for the current epoch. 
These 7\,mm model-fits aid the model-fitting and cross-identification of 3\,mm maps. Discriminating between component denotations is based on 
positional and kinematic properties. We could exclude that components have been occasionally mis-identified, although the effect on derived properties should be small.  \\

Based on setting the southernmost stationary component as the ``core'' as the reference point, we could derive kinematics of components 
relative to this point. The reference was labelled $C$ and the derived kinematics are shown in panel 3 of Fig. \ref{radio_lcs}. The kinematics were 
derived only from components further than one beam-size away from component S. Apparent component speeds vary between $\beta_{\text{app}}$\,$\approx$4-7\,c, 
slower than previously reported values \citep{jor01,jor05,agudo12}.  Over the course of observations, there appears to have been three component ejections past 
the stationary feature, labelled X1, X2 and X3, with two additional fainter ejections labelled X2.2 and X3.2. Additionally, there was a component detected that was ejected before the observation period and is labelled X0. In $\sim$2009.8, the component located between C and S, 
labelled M, could be associated with the component X2, seen in later epochs. During a period of increased flux density in S during late 2009 to early 2010, 
additional structure could be detected (labelled P2), but there was no component ejection detected. There was possibly a new component ejection in the most recent data
($\approx$\,2013.3), but we cannot conclusively identify this without more data. We tentatively identify it as P4, although it could be a trailing component. Proper motion 
calculations are made only using epochs where the component was identified with an X feature in more than five epochs.  A recent 7\,mm map showing all fitted components is 
shown in Fig. \ref{eg7mm}. \\

\begin{figure*}[htbp]
\includegraphics[width=\linewidth]{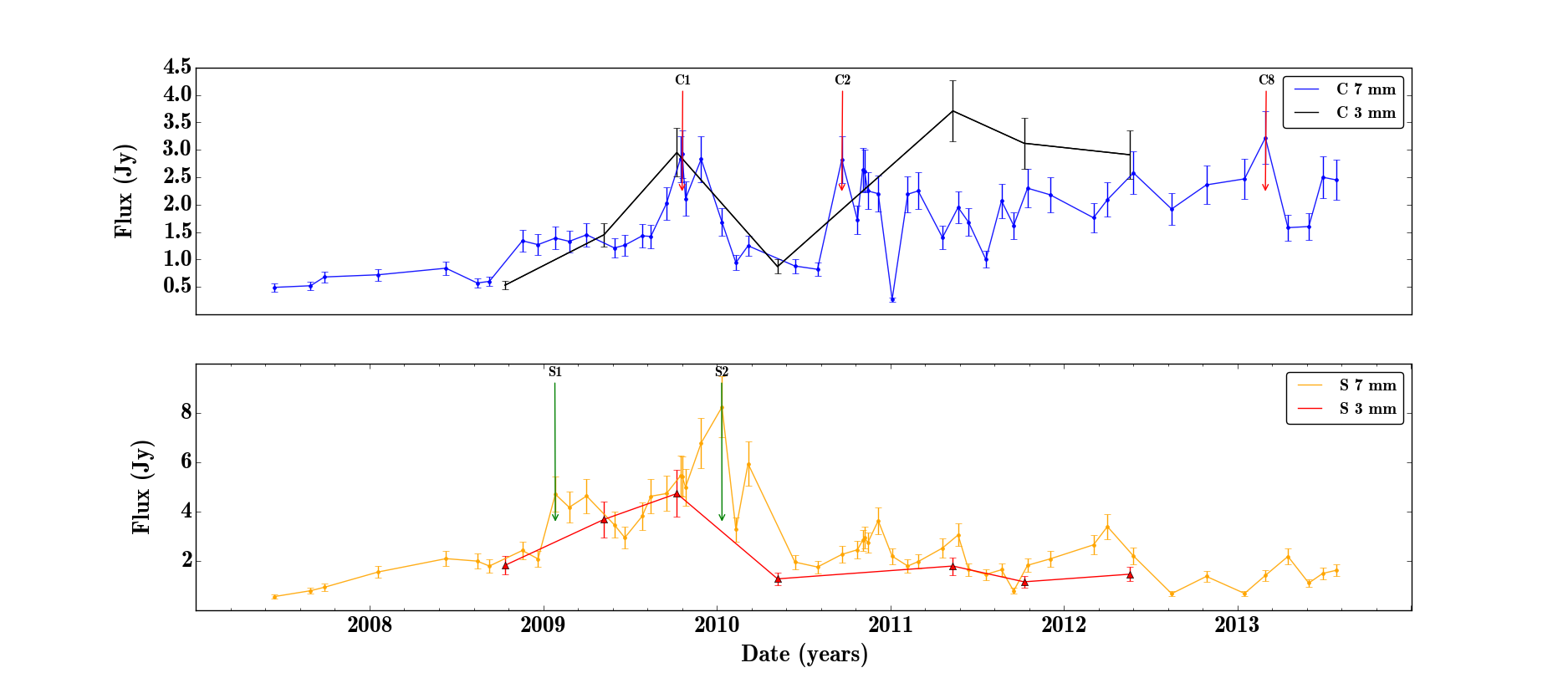}
\caption{Decomposed light-curves from 3\,mm and 7\,mm VLBI model-fits. Top panel: 
VLBI light-curves from C, with significant detected flares labelled in red.
Bottom panel: VLBI light-curves from S, with significant detected flares denoted with green arrows.   }
\label{vlbi_lcs}
\end{figure*}
\subsection{Position Angle and Trajectories}
\label{sec:PA}
\begin{center}
\begin{figure*}
\includegraphics[width=1.0\linewidth]{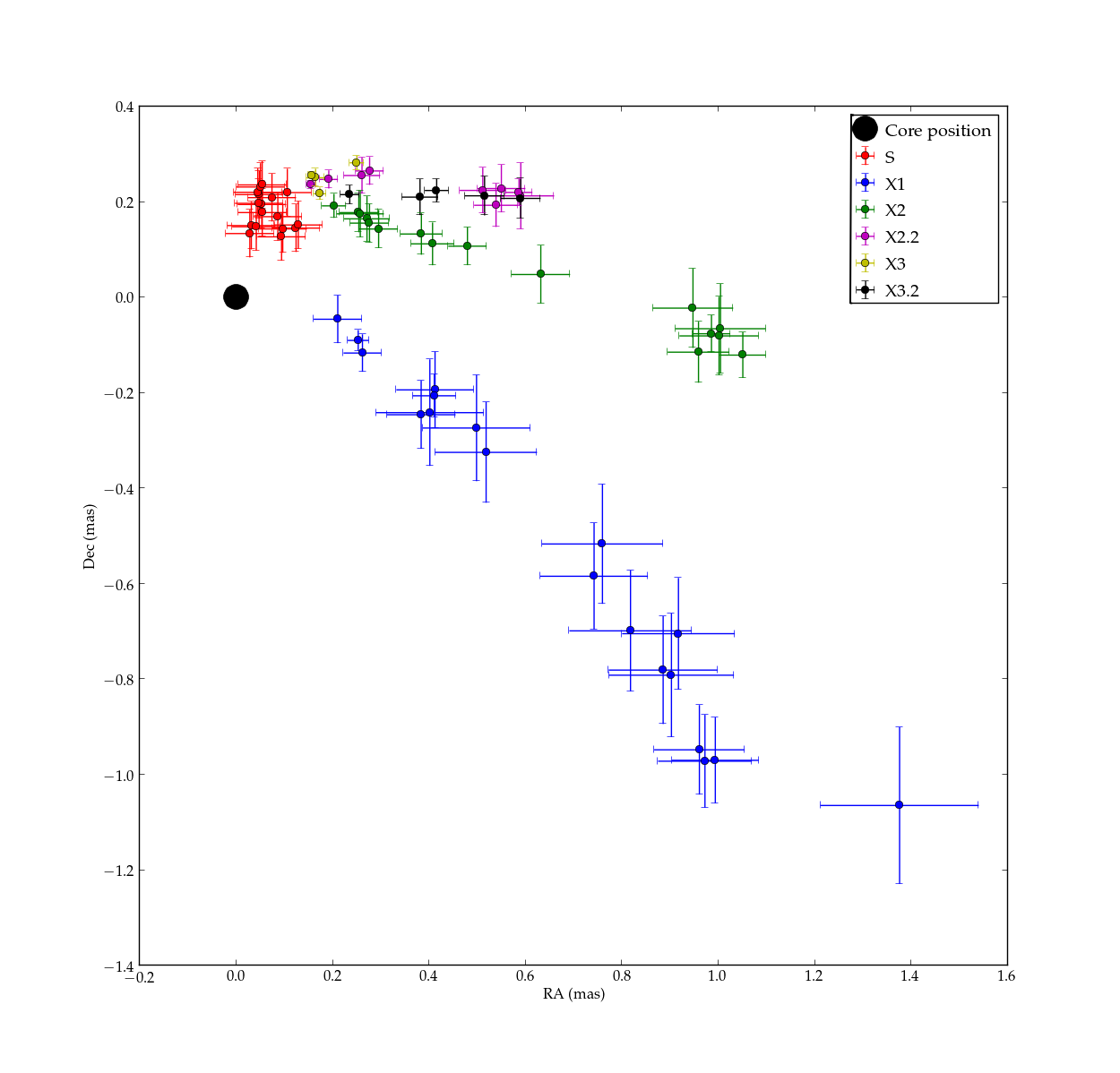}
\caption{Component S and components X1-X3.2. Components X2 and later are travelling westerly or south-westerly from PA -40$^{\circ}$ to PA -100$^{\circ}$. Component X1 travels south-westerly with a PA of approximately -120$^{\circ}$.  }
\label{xy}
\end{figure*}
\end{center}

\begin{figure*}
\includegraphics[width=\linewidth]{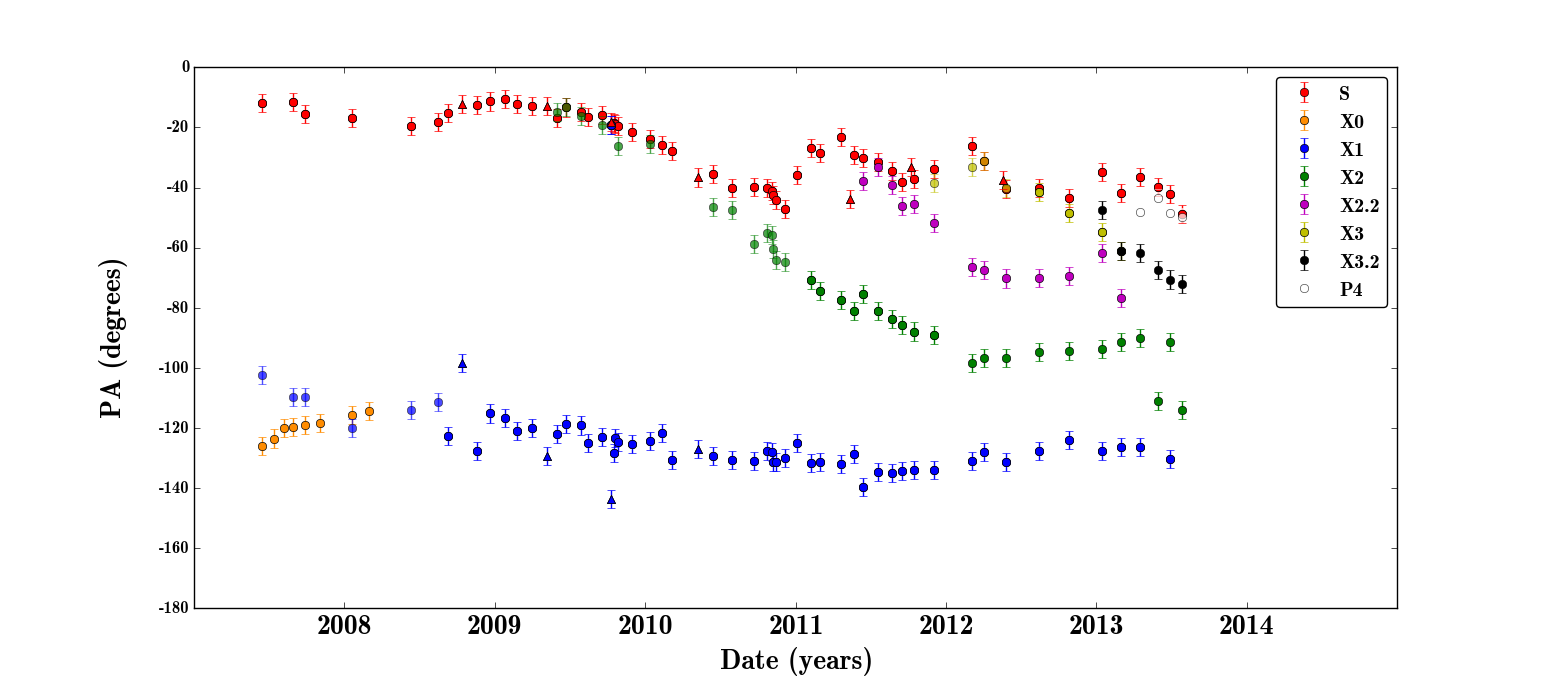}
\caption{Evolution of the PA of all components with time. A lighter coloured symbol denotes a P component, rather than an X component.  }
\label{PA}
\end{figure*}

Figure \ref{xy} shows component trajectories relative to the ``core'' for all components and for all images, while Fig. \ref{PA} shows the relative PA. 
Components $X1$ and $X0$ move in the previous jet direction. Components $X2$ and later have only slightly different trajectories, but are ejected at very different PAs,
and very different trajectories when compared with components X0 and X1. The PA of the stationary feature changes from $\sim -10^{\circ}$ in early 2009 to $\sim -40^{\circ}$ in
mid 2010, coinciding with the ejection of component $X2$. The PA then fluctuates by approximately $\pm20^{\circ}$ up to the most recent epochs. The 3\,mm PA in $S$ was consistent 
with 7\,mm observations except in $\sim$2011.4 where a $\sim$20$^{\circ}$ discrepancy was observed. Components $X2$ and later are all co-spatial with $S$ when ejected, but component
$X1$ was not, having been ejected $\approx$-100$^{\circ}$ away from the stationary feature. \\

\subsection{Flaring}
\label{longLC}

\subsubsection{Gamma-ray Flaring}
\label{gamma-LC}

\begin{table}%
\centering
\begin{threeparttable}
\centering
\caption{Details of $\gamma$-ray flares}
\begin{tabular}{lcccc}
\hline \hline
ID &  $\Delta T_{\gamma}$ & $T_{\gamma,max}$ & $S_{\gamma,max}$ & Strength \\ 
   &    [yr]              &                  &  [$\times10^{-7}$ ph/\,cm$^{-2}$\,s$^{-1}$] &   \\   \hline
G1 & 0.024                & 2009.79          & $3.09 \pm 0.58$  & 2.1 \\
G2 & 0.058                & 2011.76          & $7.90 \pm 1.60$  & 5.4 \\
G3 & 0.030                & 2011.84          & $12.48 \pm 2.16$ & 8.5 \\
G4 & 0.036                & 2012.29          & $5.04 \pm 0.92$  & 3.4 \\
G5 & 0.036                & 2012.38          & $3.78 \pm 0.76$  & 2.6 \\ \hline
\end{tabular}
\label{gamma_flares}
\begin{tablenotes}
\small
\item $\Delta T_{\gamma}$: Duration of flare, in years, taken as the difference between surrounding quiescent data points; $T_{\gamma,max}$ data of flare peak; $S_{\gamma,max}$ $\gamma$-ray flux density at flare peak; Strength was the ratio of the quiescent level and the flare peak. 
\end{tablenotes}
\end{threeparttable}
\end{table}

The $\gamma$-ray light curve is plotted in panel 1 of Fig. \ref{radio_lcs}, with flare peaks listed in Table \ref{gamma_flares}. To determine the significance of  $\gamma$-ray flares, we estimate the quiescent level taking the average of the 20 minimum data points. A flare was then defined as when the $\gamma$-ray flux was at least two standard deviations above this level. Flare duration was defined as the period of consecutive data points determined to be above the quiescent level. The flare peak was then the highest $\gamma$-ray flux measured during this period. Strength was defined as the ratio of the quiescent level and the flare peak.

Five significant $\gamma$-ray flares were found and labelled G1-G5. The observed $\gamma$-ray photon index is plotted in the bottom of Panel 1 of Fig. \ref{radio_lcs}, but exhibits no statistically significant fluctuations due to the relatively large uncertainties associated with the individual data points, remaining nearly constant at 2.3$\pm$0.2.

\subsubsection{Radio Flaring}
\label{sec:vlbi_radio}

\begin{table}%
\centering
\begin{threeparttable}
\centering
\caption{Flux density significance limits.}
\begin{tabular}{lccc}
\hline \hline
Component &  1\,$\sigma$ & 2\,$\sigma$ & 3\,$\sigma$ \\ 
   &    [Jy]             &     [Jy]             &  [Jy]    \\   \hline
C & 1.9               & 2.8           & 3.8   \\
S & 1.9              & 4.2           & 13.2   \\
\hline
\end{tabular}
\label{flare_sig}
\begin{tablenotes}
\small
\item The significance thresholds for detected flux densities as determined from 5000 simulated light curves. 
\end{tablenotes}
\end{threeparttable}
\end{table}

\begin{table}%
\centering
\begin{threeparttable}
\centering
\caption{Details of radio flares in C}
\begin{tabular}{lcccc}
\hline \hline
ID &  $\Delta T_{C}$ & $T_{C,max}$ & $S_{C,max}$ & Significance \\ 
   &    [yr]              &                  &  [Jy] &   \\   \hline
C0 & -                  & 2008.88           & $1.34 \pm 0.14$   & - \\
C1 & 0.41               & 2009.80           & $2.92 \pm 0.15$   & 2\,$\sigma$ \\
C2 & 0.23               & 2010.72           & $2.82 \pm 0.14$   & 2\,$\sigma$ \\
C3 & 0.20               & 2010.84           & $2.64 \pm 0.13$   & 1\,$\sigma$ \\
C4 & 0.29               & 2011.16           & $2.25 \pm 0.11$   & 1\,$\sigma$ \\
C5 & 0.16               & 2011.64           & $2.01 \pm 0.10$   & 1\,$\sigma$ \\
C6 & 0.46               & 2011.79           & $2.30 \pm 0.11$   & 1\,$\sigma$ \\
C7 & 0.45               & 2012.40           & $2.58 \pm 0.13$   & 1\,$\sigma$ \\
C8 & 0.67               & 2013.16           & $3.22 \pm 0.16$   & 2\,$\sigma$ \\ \hline
\end{tabular}
\label{radio_flaresC}
\begin{tablenotes}
\small
\item $\Delta T_{C}$: Duration of flare, in years, taken as the difference between surrounding quiescent data points; $T_{C,max}$ data of flare peak; $S_{C,max}$ radio flux density at flare peak; Significance denotes if the flare was a 1\,$\sigma$ or 2\,$\sigma$ detection. 
\end{tablenotes}
\end{threeparttable}
\end{table}

\begin{table}%
\centering
\begin{threeparttable}
\centering
\caption{Details of radio flares in S}
\begin{tabular}{lcccc}
\hline \hline
ID &  $\Delta T_{S}$ & $T_{S,max}$ & $S_{S,max}$ & Significance \\ 
   &    [yr]              &                  &  [Jy] &   \\   \hline
S1 & 0.57               & 2009.07           & $4.70 \pm 0.47$   & 2\,$\sigma$ \\
S2 & 1.76               & 2010.03           & $8.24 \pm 0.82$   & 2\,$\sigma$ \\
S3 & 0.29               & 2010.93           & $3.19 \pm 0.32$   & 1\,$\sigma$ \\
S4 & 0.29               & 2011.39           & $3.05 \pm 0.31$   & 1\,$\sigma$ \\
S5 & 0.91               & 2012.25           & $3.39 \pm 0.34$   & 1\,$\sigma$ \\
S6 & 0.25               & 2013.29           & $2.18 \pm 0.21$   & 1\,$\sigma$ \\\hline
\end{tabular}
\label{radio_flaresS}
\begin{tablenotes}
\small
\item $\Delta T_{S}$: Duration of flare, in years, taken as the difference between surrounding quiescent data points; $T_{S,max}$ data of flare peak; $S_{S,max}$ radio flux density at flare peak; Significance denotes if the flare was a 1\,$\sigma$ or 2\,$\sigma$ detection. 
\end{tablenotes}
\end{threeparttable}
\end{table}

Total intensity light-curves from 7 to 0.85\,mm are presented in Fig. \ref{radio_lcs}. The VLBI flux decomposition is shown in Fig. \ref{vlbi_lcs}. We see that the total flux density $S_{\text{tot}}$ was dominated by the ``core'' and the quasi-stationary feature, with the sum of $S_{C}$ and $S_{S}$ nearly equal to the total flux density at all times. The flares were identified solely from VLBI flux decomposition of components C and S at 7\,mm. Flare significance could not be determined in a similar way to the $\gamma$-ray light curves and hence, to determine the significance, light-curves were simulated 5000 times using the DELCgen simulation code, which was adapted from \citet{simul}. From this, the 1\,$\sigma$, 2\,$\sigma$ and 3\,$\sigma$ flux density confidence levels were determined (see: Table \ref{flare_sig}). We found no 3\,$\sigma$ detections in either C or S. Flare detections at the 2\,$\sigma$ level were found in C (C1,C2 and C8) and also in S (S1 and S2). All other flares were detected only at the 1\,$\sigma$ level and could be due to statistical fluctuations, particularly in C. We have taken the 2\,$\sigma$ level as the threshold of significant flaring activity. All detected flares are tabulated in Table \ref{radio_flaresC} for C and Table \ref{radio_flaresS} for S. \\

\subsubsection{Spectral decomposition}
\label{sec:decomp}

\begin{figure*}
\centering
\includegraphics[width=0.8\linewidth]{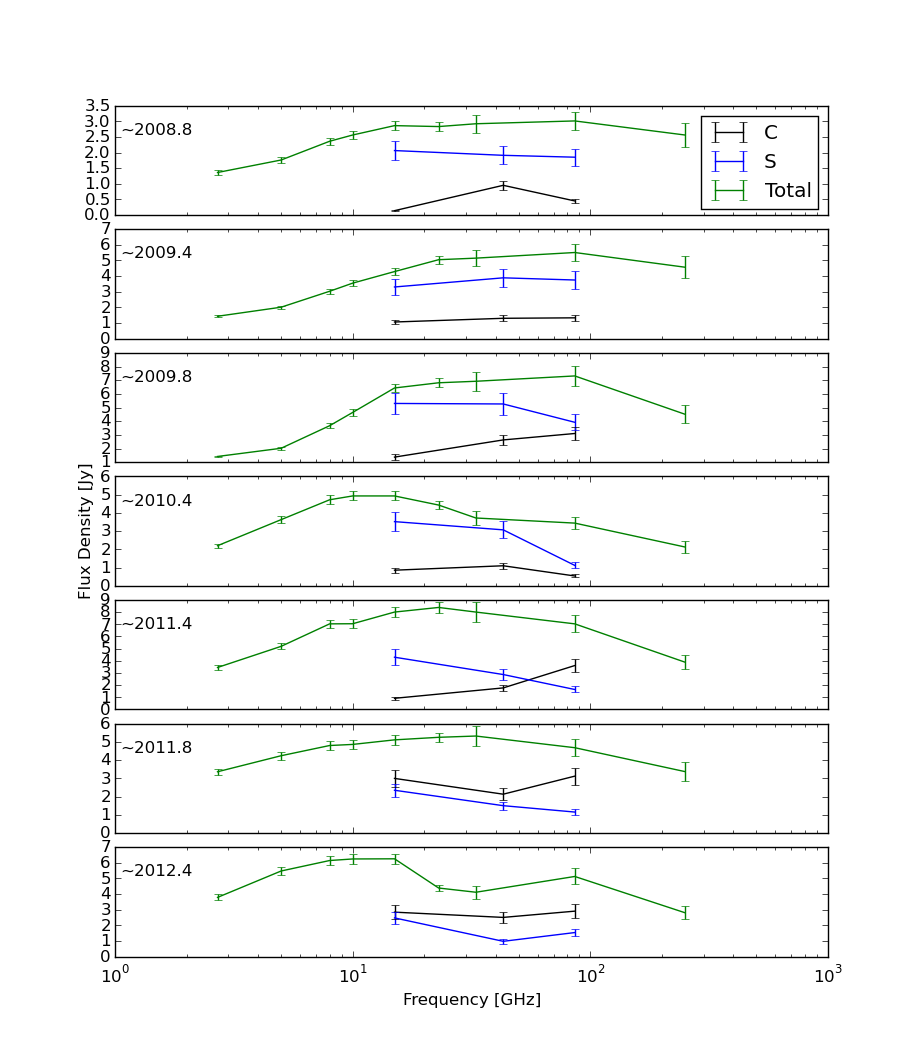}
\caption{Spectral decomposition of OJ 287 for the ``core'' (C, black) and standing feature (S, blue) from Gaussian model-fits of VLBI images. Green lines are total intensity measurements from FGAMMA and SMA flux monitoring programs. 0.85\,mm data were not included due to a lack of near-in-time data at all epochs. }
\label{spec}
\end{figure*}

\begin{figure*}
\includegraphics[width=\linewidth]{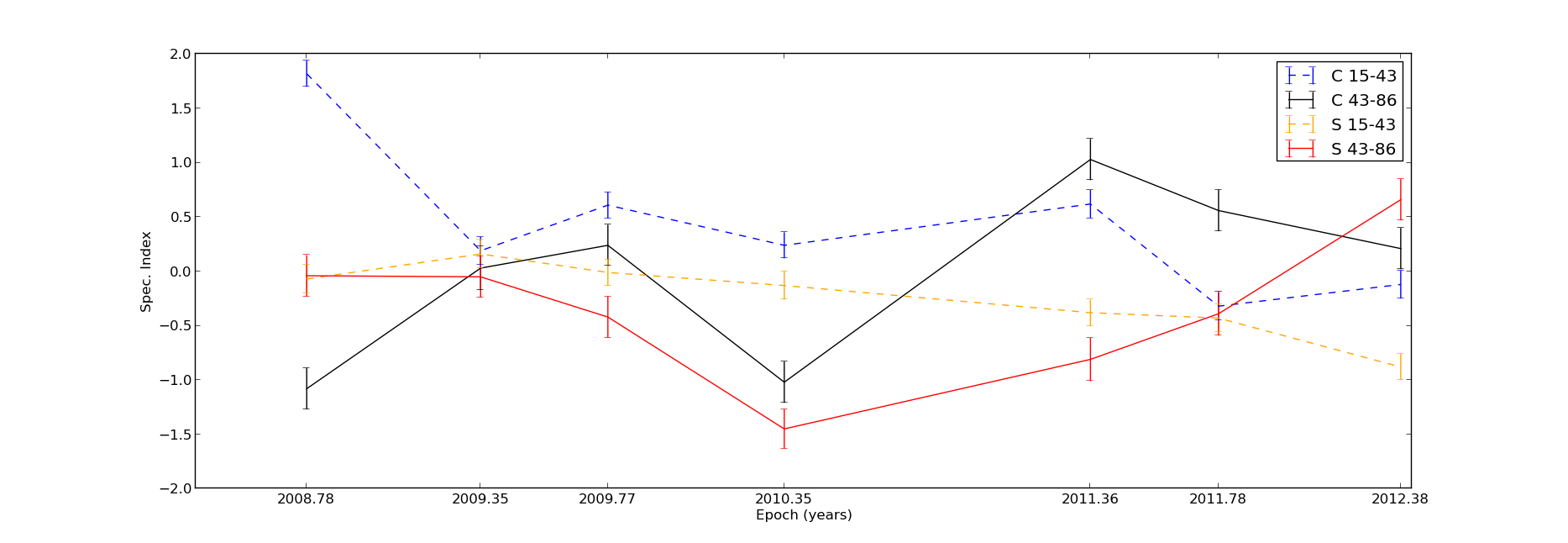}
\caption{Evolution of spectral indices over time of ``core'' and stationary feature.}
\label{spec_ind_plot}
\end{figure*}

\begin{table*}%
\centering
\begin{threeparttable}
\caption{VLBI flux densities with long baselines removed.}
\begin{tabular}{lcccccc}
\hline \hline
Epoch   &   C (86\,GHz)      & S (86\,GHz)   &   C (43\,GHz)      & S (43\,GHz)  &   C (15\,GHz)      & S (15\,GHz)             \\ 
        &   [Jy]            &  [Jy]          &   [Jy]            &  [Jy]         &   [Jy]            &  [Jy]   \\   \hline
$\sim$2008.7 & $0.45 \pm 0.09$   & $1.85 \pm 0.17$ & $0.95 \pm 0.09$   & $1.91 \pm 0.19$ & $0.14 \pm 0.02$   & $2.06 \pm 0.21$ \\
$\sim$2009.4 & $1.33 \pm 0.27$   & $3.75 \pm 0.75$ & $1.30 \pm 0.13$   & $3.89 \pm 0.39$ & $1.06 \pm 0.16$   & $3.30 \pm 0.33$ \\
$\sim$2009.8 & $3.10 \pm 0.62$   & $3.92 \pm 0.75$ & $2.63 \pm 0.26$   & $5.26 \pm 0.52$ & $1.37 \pm 0.14$   & $5.29 \pm 0.53$ \\
$\sim$2010.4 & $0.54 \pm 0.11$   & $1.12 \pm 0.24$ & $1.10 \pm 0.11$   & $3.07 \pm 0.31$ & $0.86 \pm 0.09$   & $3.52 \pm 0.35$ \\
$\sim$2011.4 & $3.61 \pm 0.72$   & $1.64 \pm 0.32$ & $1.77 \pm 0.18$   & $2.87 \pm 0.29$ & $0.91 \pm 0.09$   & $4.29 \pm 0.43$ \\
$\sim$2011.8 & $3.13 \pm 0.62$   & $1.16 \pm 0.23$ & $2.13 \pm 0.21$   & $1.50 \pm 0.15$ & $2.99 \pm 0.30$   & $2.35 \pm 0.23$ \\
$\sim$2012.4 & $2.92 \pm 0.58$   & $1.54 \pm 0.31$ & $2.51 \pm 0.25$   & $0.98 \pm 0.10$ & $2.85 \pm 0.28$   & $2.48 \pm 0.25$ \\ \hline
\end{tabular}
\label{spectra_3mm}
\begin{tablenotes}
\small
\item Long baselines in 3\,mm maps were removed in order to ensure similar resolutions. Flux densities at other frequencies were interpolated to the 3\,mm observation date.
\end{tablenotes}
\end{threeparttable}
\end{table*}

\begin{table*}
\caption{Computed spectral indices for component C and S}
\begin{center}
\begin{tabular}{rrrrr}
\hline \hline
Epoch & $\alpha_{C(15-43)}$ &$\alpha_{C(43-86)}$ & $\alpha_{S(15-43)}$ & $\alpha_{S(43-86)}$ \\ \hline
$\sim$2008.7 & 1.83 $\pm$ 0.12 & -1.08 $\pm$ 0.20 & -0.07 $\pm$ 0.12 & -0.05 $\pm$ 0.19 \\
$\sim$2009.4 & 0.19 $\pm$ 0.13 & 0.03 $\pm$ 0.20 & 0.15 $\pm$ 0.13 & -0.05 $\pm$ 0.19 \\
$\sim$2009.8 & 0.61 $\pm$ 0.13 & 0.24 $\pm$ 0.20 & -0.01 $\pm$ 0.12 & -0.42 $\pm$ 0.19 \\
$\sim$2010.4 & 0.24 $\pm$ 0.13 & -1.03 $\pm$ 0.19 & -0.13 $\pm$ 0.13 & -1.46 $\pm$ 0.20 \\
$\sim$2011.4 & 0.62 $\pm$ 0.13 & 1.04 $\pm$ 0.19 & -0.38 $\pm$ 0.13 & -0.81 $\pm$ 0.19 \\
$\sim$2011.8 & -0.33 $\pm$ 0.13 & 0.56 $\pm$ 0.19 & -0.42 $\pm$ 0.13 & -0.38 $\pm$ 0.19 \\
$\sim$2012.4 & -0.12 $\pm$ 0.13 & 0.22 $\pm$ 0.19 & -0.88 $\pm$ 0.12 & 0.66 $\pm$ 0.19 \\\hline
\end{tabular}
\end{center}
\label{alphas}
\end{table*}

In Fig. \ref{spec}, the total intensity spectrum is over-plotted with the VLBI spectral decomposition for all epochs. In our analysis, we use $S_{\nu} \propto \nu^{\alpha}$. Resolution effects must be taken into consideration when computing the spectral indices. In order to account for this, visibilities over 1500 M$\lambda$ in the 3\,mm maps were removed and maps re-model fitted. The results of this are displayed in Table \ref{spectra_3mm}. The only epoch with significantly different fitted flux densities was $\sim$2009.8. Fitted flux densities at lower frequencies were also interpolated to match the 3\,mm observing date. In Fig. \ref{spec_ind_plot} and Table \ref{alphas}, we see the evolution of the spectral indices ($\alpha$) of the ``core'' and stationary feature.   \\

Usually, the ``core'' is expected to have a flat to slightly inverted spectrum (flux density increasing with frequency), whilst travelling components are expected to have a steep spectrum (flux density increasing with decreasing frequency). In Fig. \ref{spec_ind_plot}, we could see that the ``core'' (black lines) has a flat to inverted spectrum in most epochs, with the exception of $\sim$2008.7 and $\sim$2010.4. The stationary feature also exhibits flat spectra, although steeper on average than the ``core''. Both the ``core'' and stationary feature have similar levels of spectral variability. In two epochs, the turnover frequency was measurable: $\sim$2008.8 and $\sim$2010.4 in the ``core'', as the flux density was highest at 7\,mm than as measured from cross-identified 3\,mm or 2\,cm model-fit components. This indicates that the turnover frequency has been detected as being between 15\,GHz and 86\,GHz at these times. In epochs $\sim$2009.8 and earlier, the standing feature also exhibits a flat to steep 43-86 GHz spectrum. However, in epochs $\sim$2010.4 and later, the 15-43\,GHz and 43-86\,GHz spectral indices indicated an inverted spectrum, consistent with optically thick emission. 

%

\begin{table*}
\centering
\begin{threeparttable}%
\caption{Symbols and units}
\label{units}
\begin{tabular}{ccl}
\hline \hline
Symbol & Unit \\
\hline
$\beta_{\text{app}}$    &   c                   &   Apparent superluminal motion \\
$\beta_{\text{int}}$    &   c                   &   Component source frame velocity \\
$\theta_{0}$            &   Degrees ($^{o}$)    &   Viewing angle to source \\
$\Gamma$                &   -                   &   Lorentz factor \\
$\delta$              &   -                   &   Doppler factor  \\
$\theta_{\text{crit}}$  &   Degrees ($^{o}$)    &   Critical angle, estimated from the highest observed $\beta_{\text{app}}$ \\
$B_{\text{SSA}}$        &   Gauss (G)           &   Magnetic field from SSA \\
$B_{\text{equi}}$       &   Gauss (G)           &   Minimum magnetic field from equipartition \\
$T_{\text{B}}$          &   Kelvin (K)          &   Brightness temperature \\
$\alpha$                &   -                   &   Spectral index \\
$D_{L}$                 &   Gigaparsec (Gpc)   &   Luminosity distance \\
$\nu_{m}$               &   Gigahertz (GHz)     &   Turnover frequency \\
$S_{m}$                 &   Jansky (Jy)         &   Flux density at $\nu_{m}$ \\
$\theta_{m}$            &   Milliarcsecond (mas) &   Angular size of emitting region at $\nu_{m}$ \\
$R_{m}$                 &   Centimeters (cm)    &   Linear Size of emitting region at $\nu_{m}$ \\
$U_{\text{rel}}$        &   erg                 &   Energy due to relativistic particles \\
$U_{\text{mag}}$        &   erg                 &   Energy due to magnetic fields \\
k                       &   -                   &   Scaling value for heavy particles \\
r                       &   Milliarcsecond (mas) &  Separation from jet base \\
R                       &   Milliarcsecond (mas) &  Transverse width of jet \\
$r_{\text{apex}} $      &   Milliarcsecond (mas) &  Separation of ``core'' from jet base \\
$\Delta r_{S}$          &   Milliarcsecond (mas) &  Separation of component S from ``core'' \\ \hline
\end{tabular}
\begin{tablenotes}
\small
\item A list of symbols, units and a brief description used for calculations in the paper.
\end{tablenotes}
\end{threeparttable}
\end{table*}

\section{Analysis}
\label{sec:analysis}
In this section, we describe first how component ejections were related to flaring activity in Section \ref{sec:ejec_relo}. We then in Sections \ref{sec:doppler} and Section \ref{sec:ssa} describe the mathematical framework to derive observational properties of the source and estimate the magnetic field strength. Finally, in Section \ref{sec:bh}, we describe a method to estimate the distance of the ``core'' to the base of the jet. A brief description of symbols and their units was given in Table \ref{units}.

\subsection{Ejection Relations}
\label{sec:ejec_relo}

\begin{figure*}
\includegraphics[width=\linewidth]{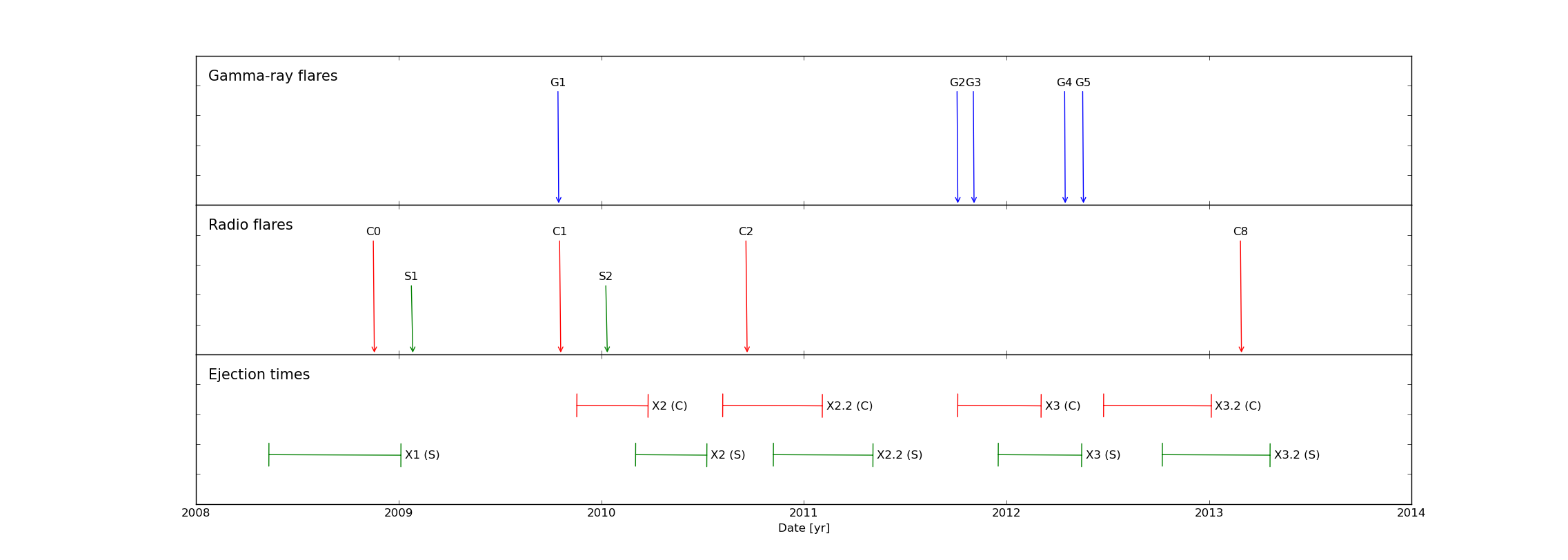}
\caption{A sketch showing the peaks of $\gamma$-ray and significant radio flaring in relation to the estimated ``core'' and quasi-stationary feature passing times. The top panel indicates $\gamma$-ray peaks, the second panel denotes radio flares as detected from VLBI decomposed light-curves at 7\,mm and the bottom panel indicates the estimated ``core'' (C) and quasi-stationary feature (S) passage times. $\gamma$-ray flares are blue, ``core'' flares are red and quasi-stationary feature flares are green. Estimated ejection times for the ``core'' are in red and passage times for the quasi-stationary feature are in green. }
\label{ejec_relo_plot}
\end{figure*}

In order to determine if either $\gamma$-ray or radio flaring could be attributed to structural changes in VLBI maps, estimated component passing times assuming constant motion are plotted against the peaks of flaring activity in Figure \ref{ejec_relo_plot}. The ``core'' ejection times ($C_{0}$) are computed assuming no acceleration. Stationary feature (S) ejection times ($S_{0}$) are estimated by taking the first epoch that a component associated with a travelling component was detected and adopting the error bars computed for ``core'' passage times. A flaring event was associated when any part of the flaring occurred within the estimated component passing times for C and S. \\

We found no radio flaring could be associated with the ``core'' ejection time of component X1 due to the unavailability of LAT $\gamma$-ray data from before 2008. The only association that could be made for $\gamma$-ray flaring was between the G2/G3 pair with the passing of X3 through C and the G4/G5 pair with the passing through S. The only significant radio flaring that could be associated with a component ejection was between C2 and the ``core'' ejection time of component X2.2. \\

\subsection{Component speeds and Doppler factor}
\label{sec:doppler}


The apparent speeds, $\beta_{\text{app}}$, are determined by the intrinsic speed $\beta_{\text{int}}=v/c$ and our viewing angle to the source ($\theta_{0}$) \citep[e.g.,][]{rees66}, with:
\begin{gather}
\beta_{\text{app}} = \frac{\beta_{\text{int}}\sin\theta_{0}}{1-\beta_{\text{int}}\cos\theta_{0}}.
\end{gather}
If we compute the differentiation of this and set the equation to zero, we can calculate the critical angle, $\theta_{\text{crit}}$ at which $\beta_{\text{app}}$ is maximised:
\begin{equation}
\theta_{\text{crit}}=\sin^{-1}(1/\beta_{\text{app}})\text{[degrees]},
\end{equation}
although this is likely to differ from the true viewing angle $\theta_{0}$. We can compute the Lorentz factor,
\begin{equation}
\Gamma = \frac{1}{(1-\beta_{\text{int}}^{2})^{1/2}},
\end{equation}
in two ways. One solves for $\beta_{\text{int}}$ in Eq. 1 with $\theta_{0}=\theta_{\text{crit}}$. The other equates $\Gamma$ to the lowest value that can lead to the observed value of $\beta_{\text{app}}$:
\begin{equation}
\Gamma \approx (1+\beta^{2}_{\text{app}})^{1/2}.
\end{equation}
We can use the derived rough value of $\Gamma$ to estimate the Doppler factor:
\begin{equation}
\delta_{\text{VLBI}} \approx \frac{1}{\Gamma(1-\beta_{\text{int}}\cos\theta_{\text{crit}})}.
\end{equation}
The flux of an optically thin moving components is then boosted by a factor of $\delta^{3-\alpha}$, while that of a stationary feature is boosted by a factor of $\delta^{2-\alpha}$. \\

The results of these computations are displayed in Table \ref{comps2}. Component speeds were computed by first converting from radial coordinates into rectangular coordinates, evaluating the difference between points and smoothing the data. Speeds vary from $\beta_{\text{app}}\approx$3.8\,c to 7.3\,c. As the speeds seen in OJ\,287 are considerably lower than previously reported \citep[e.g.,][]{jor05,agudo11}, the observed $\beta_{\text{app}}$ gives only a lower limit on the Doppler factor, and hence we adopt the bulk Lorentz factor $\Gamma = 16.5 \pm 4.0$, the angle between the axis and the line of sight of $3.2^{\circ} \pm 0.9^{\circ}$ and Doppler factor $\delta = 18.9 \pm 6.4$ from \citet{jor05}, derived from 17 epochs at 7\,mm with the VLBA from 1998 until 2001. This value also agrees well with the variability Doppler factor reported by \citet{hovatta09}.


\begin{table}%
\centering
\begin{threeparttable}
\centering
\caption{Average properties in the ``Core'' and stationary feature.}
\begin{tabular}{lcc}
\hline \hline
Property   & ``Core'' & ``Stationary Feature''     \\ \hline  
$B_{\text{SSA}}$ [G] & $\geq 1.6$  & $\leq 0.4$     \\ 
$B_{\text{equi}}$ [G] & $\geq 1.2$ & $\leq 0.3 $     \\ 
$\delta_{\text{equi}}$ & $\geq 8.7 $ & $\leq 9.9 $      \\ 
$T_{\text{B}}$ [K] & $\geq1.4 \times 10^{11} $ & $ \leq 2.1 \times 10^{11} $    \\ \hline
\end{tabular}
\label{mag_av}
\begin{tablenotes}
\small
\item $B_{\text{SSA}}$: Limits on the magnetic field derived from SSA; $B_{\text{equi}}$ limits on the magnetic field strength assuming equipartition; $\delta_{\text{equi}}$: Limits on the Doppler factor assuming equipartition; $T_{\text{B}}$: Limits on the observer frame brightness temperature assuming equipartition . 
\end{tablenotes}
\end{threeparttable}
\end{table}

\subsection{Magnetic fields}
\label{sec:ssa}

To derive an estimate on the magnetic field from synchrotron self-absorption ($B_{\text{SSA}}$), we required the spectral index of the optically thin emission, ($\alpha$) the turnover frequency ($\nu_{m}$) and turnover flux density ($S_{m}$). A single epoch was analysed in this manner in CTA\,102 by \citet{fromm13_3}, we then applied this method to determine $\nu_{m}$ or limits on it in individual components over time. Using the approach of \citet{LZ99}, we use fitted components to determine the magnetic field of the ``core'' and stationary feature from the expression \citep{mar83,bach05}:
\begin{equation}
B_{\text{SSA}} = 10^{-5}b(\alpha)\frac{\theta_{m}^{4}\nu_{m}^{5}\delta^{2-\alpha}}{S^{2}_{m}(1+z)} \text{ [G]},
\end{equation}
where $ b(\alpha) $ is a parameter between 1.8 and 3.8 for optically thin emission \cite[see:][]{mar83}, $ \theta_{m} $ is 1.8 times the FWHM of the component in mas \citep{mar77}. However, this expression is different than from \citet{mar83}, as stationary features are steady state rather than evolving in time. An additional factor of $(2-\alpha)$ is added to the exponent of $\delta$ to account for this \citep{mar06b}. \\

The minimum magnetic field strength, energies and luminosities can also be computed assuming equipartition between the energy of relativistic particles and the magnetic field, providing an independent check of derived Doppler factors and magnetic fields using the previous estimates of $B_{\text{SSA}}$. An equipartition `critical size' can be calculated based on \citep[e.g.,][]{scott77,daly96}:
\begin{equation}
\theta_{\text{crit,eq}} = \left[\frac{6\pi (1+k) F(\alpha)}{10^{-10}b(\alpha)^{2}D_{L}} \nu^{-17}_{m}\delta^{-(16+2\alpha)}S^{8}_{m} (1+z)^{9}\right]^{1/17} \text{ [mas]},
\end{equation}
where $S_{m}$ and $\nu_{m}$ are the turnover fluxes and frequencies respectively, $\alpha$ is the spectral index, $F(\alpha)$ is a scaling factor from \citet{scott77}, $b(\alpha)$ is a dimensionless parameter from Table 1 in \citet{mar83}, and k is the energy ratio between hadrons and leptons ranging from 0 for an entirely hadronic jet to $\sim$2000 in an entirely leptonic jet. For our calculations, we assume $k\approx100$. If the observed size in the source (scaled by a factor of 1.8 \citep{mar77}) is less than this `critical size' it implies a particle dominated jet and if larger implies a magnetically dominated jet. If we assume equipartition, we can estimate the Doppler factor:
\begin{equation}
\delta_{\text{equi}}= \left( \frac{U_{\text{rel}}}{U_{\text{mag}}} \right)^{1/(7 + 4 \alpha)}
\end{equation}
Where $U_{\text{rel}}$ and $U_{\text{mag}}$ are the energy densities of relativistic particles and magnetic fields respectively:
\begin{equation}
U_{\text{rel}} = F(\alpha)D_{L}^{-1}(\theta_{m}^{-9}\nu_{m}^{-7}S_{m}^{4}(1+z)^{7})
\end{equation}
\begin{equation}
U_{\text{mag}} = \frac{B_{\text{SSA}}^{2}}{8\pi}
\end{equation}
the brightness temperature:
\begin{equation}
T_{\text{B}} = 1.22 \times 10^{12} \frac{S_{m}}{\theta_{obs}^{2}\nu_{m}^{2}} \text{ [K]},
\end{equation}
and an estimate of the minimum magnetic field strength from \citet{bach05}:
\begin{equation}
B_{\text{equi}} = 5.37 \times 10^{12} (S_{m}\nu_{m}D_{L}^{2}R_{m}^{-3})^{-2/7} \text{ [G]},
\end{equation}
where $D_{L}$ is the luminosity distance and $R_{m}$ is the linear radius of the emitting region in cm. The above equation includes a factor of (1+k), where $k\approx100$, which makes assumptions about the matter composition of the jet. However in the case of an entirely leptonic jet (i.e. electron-positrons, $k=0$) the estimated magnetic field strength would be weakened $\approx7$ times. A fully hadronic (i.e. protons) jet with $k\approx2000$, would produce a minimum magnetic energy $\approx150$ times stronger. Averaged quantities are presented in Table \ref{mag_av}. It is important to note that at mm wavelengths, it is possible that the assumption of equipartition could break down, as we could be observing near the acceleration region of the jet, where the jet is thought to be Poynting flux dominated \citep{BZ77}.

Due to the large uncertainties in the magnetic field estimates, values were averaged over all epochs and magnetic fields computed using multiple methods. Table \ref{mag_av} shows magnetic field estimates and limits derived from these averaged quantities. All values are roughly consistent with each other. There was a minimum $\sim75$\% decrease between the ``core'' and the downstream quasi-stationary feature using both methods. The values of the magnetic field strengths derived through both methods are broadly consistent, with the field strengths from SSA being $\sim$20\% higher than when computed from equipartition. The results suggest a magnetic field considerably stronger than previously reported, with \citet{carlosOJ} measuring a magnetic field strength of $\sim$0.2\,G from ``core-shift'' observations.  \\

\subsection{Frequency dependent positions}

Changes in component positions as a function of frequency is more commonly known as ``core-shift'' and is an important parameter in AGN physics. Unfortunately, the uncertainties in positions are too large to obtain reliable results using this method. 

\subsection{Distance to jet apex}
\label{sec:bh}

If we assume as suggested by \citet{daly96} that the magnetic field drops off as:
\begin{equation}
B \propto (1/r)^{n},
\end{equation}
in a smooth jet, where $r$ is the separation from the jet base and n is an exponent for toroidal (n=1) or longitudinal (n=2) magnetic field configurations \citep{LZ99,mgt92}. We can compute an estimate of the distance from the ``core'' to the jet apex $r_{\text{apex}}$ (in $mas$) with the ratio of magnetic fields in component C ($B_{C}$) and S ($B_{S}$) and the separation between the ``core'' and stationary feature $\Delta r_{S}$:
\begin{equation}
\frac{B_{C}}{B_{S}} = \left[ {\frac{r_{\text{apex}} + \Delta r_{S}}{r_{\text{apex}}}} \right]^{n}.
\end{equation}
Rearranging, we get:
\begin{equation}
r_{\text{apex}} = \frac{\Delta r_{S}}{(B_{C}/B_{S})^{1/n}-1}.
\end{equation}

In the previous section (Section \ref{sec:ssa}), we computed the average decrease in the magnetic field strength between the quasi-stationary feature and the ``core''. Recent work by \citet{mb09,gabuzda14} and \citet{koskesh} suggests that toroidal magnetic fields play a more important role in AGN than assumed before. Therefore, if we assume $\Delta r_{s}$=0.2\,mas (the approximate separation of the quasi-stationary feature relative to the ``core''), a toroidal magnetic field (n=1) and equipartition calculations, this would place upper limits on the location of the jet apex at $\lesssim 4$\,pc upstream of the mm-wave ``core''. Using the values derived from SSA calculations, the jet apex is placed $\lesssim 6$\,pc upstream of the mm-wave ``core''.

\section{Discussion and interpretation}

In this Section, we begin by investigating the jet opening angle in Section \ref{open_angle}. We then in Section \ref{sec:ejection} analysed the connection between $\gamma$-ray flaring, mm-wave radio flaring and component ejections, using cross-correlation analysis to find evidence that $\gamma$-rays could originate both within the ``core'' and a downstream quasi-stationary feature. In Section \ref{core_phys}, we discuss the physical nature of the ``core'' and the possible origin of $\gamma$-ray emission. We then estimate an upper limit on the magnetic field strength at the jet base in Section \ref{sec:mag}. Finally, we suggest a possible interpretation for the large PA changes seen in OJ\,287 in Section \ref{pa_theory}. 

\subsection{Jet opening angle}
\label{open_angle}
In Fig. \ref{xy}, we could see two trajectories, with (1) X1 (and X0, not shown) travelling south-westerly and (2) all later components travelling westerly. Additionally, in Fig. \ref{PA}, we could see that X1 was ejected at a PA that differs by $\sim$\,$85^{\circ}$ from a line between C and S, unlike components ejected after 2008. This suggests that X1 was ejected before the stationary feature was established in its current position. Taken alone, the trajectories resemble the ``fanning'' of components reported in BL Lacertae \citep{fanning}. Two interpretations of the jet opening angle are possible: (1) all components lie within the jet cone, which subtends an apparent opening angle of $\approx$\,$85^{\circ}$ or (2) the projected opening angle was smaller ($\approx$\,$30^{\circ}$) at any given time and the jet has changed direction by several degrees to give the appearance in projection of an extremely broad jet. Since there is no known example of a source with an opening angle of $\approx$\,$85^{\circ}$, we found that the second scenario was more likely, in agreement with \citet{agudo11}. \\

\subsection{Flaring-component ejection relations}
\label{sec:ejection}

That radio flares C1 and S2 are superimposed to create one apparently large flare in the total intensity radio light-curves demonstrates the extra consideration must be taken when associating $\gamma$-ray flaring activity with radio flaring using total intensity measurements only. High resolution VLBI was required to spatially resolve the location of flaring within the jet. \\

\begin{figure}
\includegraphics[width=\linewidth]{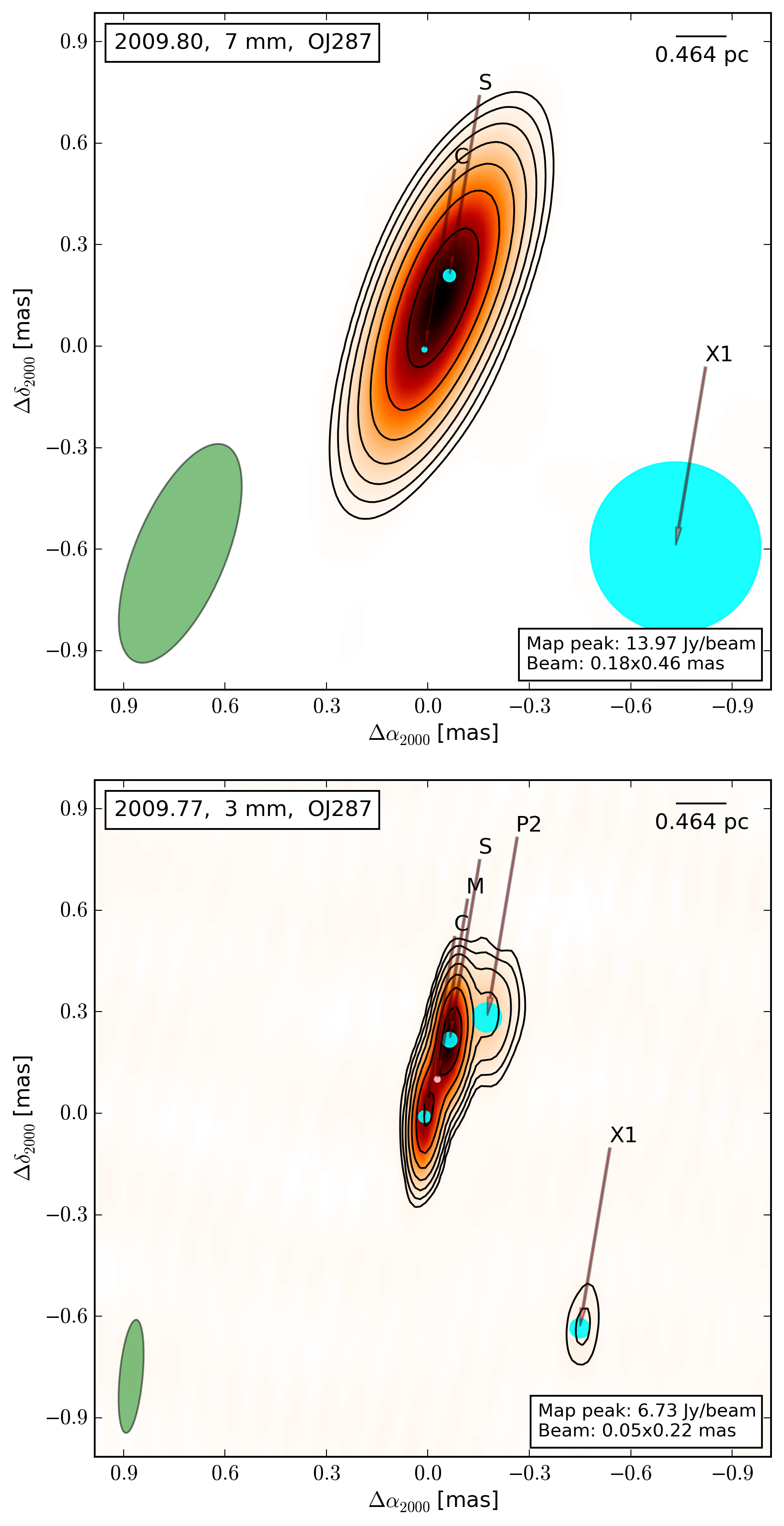}
\caption{Near-in-time 3\,mm and 7\,mm VLBI observations of OJ\,287 in $\sim$2009.8. Contours: -1, 1, 2, 4, 8, 16, 32, 64\% of peak flux density. Other features are described in Fig. \ref{3mm121}.}
\label{3mm092}
\end{figure}

In Section \ref{sec:ejec_relo}, we found that no $\gamma$-ray or radio flaring activity could be associated with the ejection of component X2 (although $\gamma$-ray flaring could be associated for X3). This analysis however assumed constant velocity of the jet components and did not explore the possibility of accelerating jet components, implying that our ejection estimates could be too late. In the case of component X2, structure was resolved around the quasi-stationary feature (labelled P2) as much as a year before a component was ejected. Therefore it was difficult to conclusively associate this activity in the quasi-stationary feature with any component ejection. However, in the 3\,mm map of $\sim$2009.8 (Fig. \ref{3mm092}), we found a component (labelled M) was detected between C and S. Associating this component with X2 allowed us to estimate a new $C_{0}$ time for component X2 of $\sim$2009.0 (with a proper motion of $\sim$0.15\,mas/yr), coinciding well with a large increase in ``core'' flux. If the quasi-stationary feature passage time was also slightly earlier than previously estimated, it would coincide with radio flare S2. 

Similarly, for component X3, if we adopt the same slower speed estimated for component X2 between the ``core'' and the quasi-stationary feature,  $C_{0}$ time for component X3 would be $\sim$2010.8, coinciding with flare C2, which had been previously attributed to X2.2. Similarly, making the quasi-stationary feature passing time slightly earlier would associate the ejection with $\gamma$-ray flares G2 and G3 and the association of $\gamma$-ray flares G1, G2 and G3 with the quasi-stationary feature. A potential explanation for this could be radial acceleration due to the non-linear trajectories seen in Fig \ref{xy} and also in other sources \citep[e.g.,][]{mojave_vii}. The detection of structure between the ``core'' and the quasi-stationary feature in the 3\,mm map leads us to prefer the interpretation that the jet components are accelerating.


\begin{figure*}
\includegraphics[width=\linewidth]{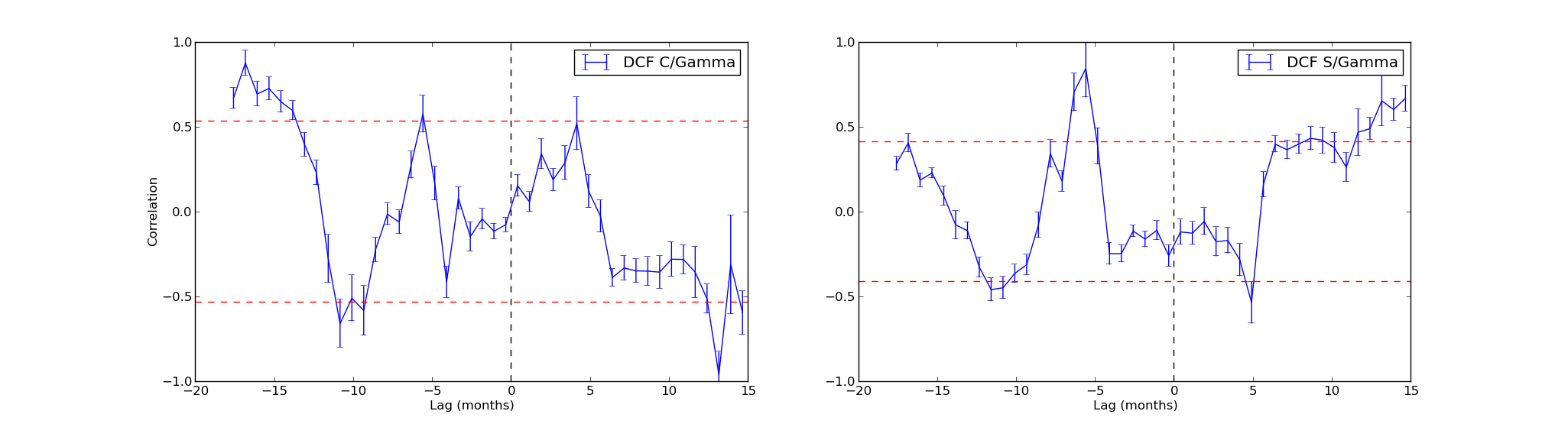}
\includegraphics[width=\linewidth]{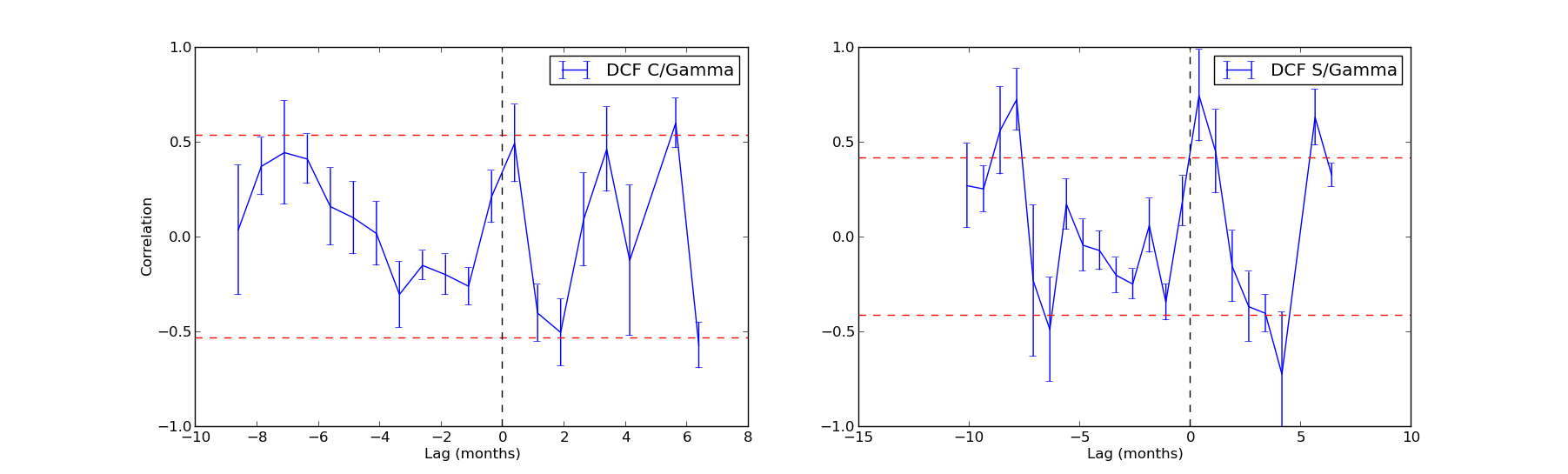}
\includegraphics[width=\linewidth]{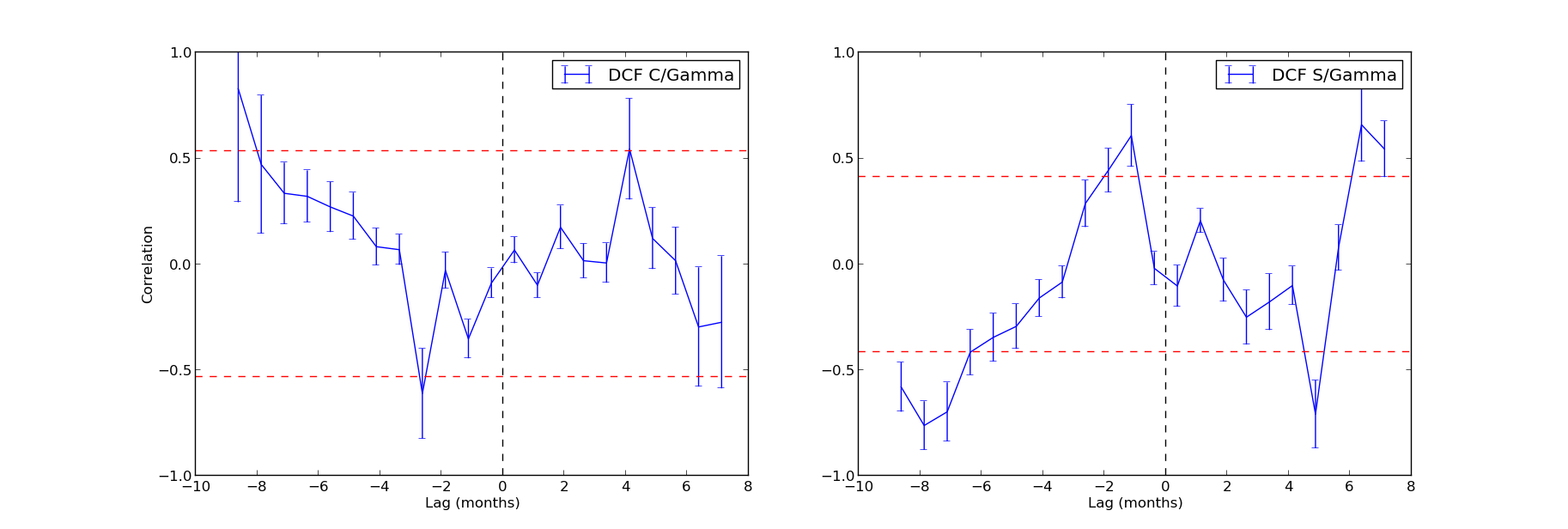}
\caption{Panel 1: Discrete correlation function between VLBI light-curves and $\gamma$-rays. 95\% confidence intervals are denoted with red dashed lines. Left panel between $\gamma$-rays and C, right panel between $\gamma$-rays and S.
Panel 2: Discrete correlation function between VLBI light-curves and $\gamma$-rays before 2011. 95\% confidence intervals are denoted with red dashed lines. Left panel between $\gamma$-rays and C, right panel between $\gamma$-rays and S.
Panel 3: Discrete correlation function between VLBI light-curves and $\gamma$-rays after 2011. 95\% confidence intervals are denoted with red dashed lines. Left panel between $\gamma$-rays and C, right panel between $\gamma$-rays and S. }
\label{ccor_post}
\end{figure*}

\subsubsection{Cross-correlation analysis}

In addition to causal connections between morphological changes and flaring activity, we could also investigate the correlation between VLBI component 
radio flaring and $\gamma$-ray flaring. In order to do this, the VLBI light curves were cross-correlated with the $\gamma$-ray light curves using the discrete
correlation function (DCF) \citep{dcf}. The significance was determined from the simulated light curves described in Section \ref{gamma-LC} and was also extended
to the $\gamma$-ray light curve. From this, 95\% confidence intervals were determined. The results of these correlations are displayed in panel 1 of Fig. \ref{ccor_post}. \\

We found a significant correlation between the $\gamma$-ray light-curves and the VLBI light-curve for both C and S. The radio leads the $\gamma$-rays in both cases by approximately 6 months. The correlation in C was only significant at the 2\,$\sigma$ level, while the correlation in S was much more significant. However, it was likely that the correlations detected here are between flares C1/S2 and G2/G3, due to their relative prominence. These correlations are likely spurious as $\gamma$-ray flaring was thought to correlate with radio flaring on timescales much less than six months \citep[e.g.,][]{fuhrmann14}.  We therefore conclude that the cross-correlation analysis provides inconclusive evidence of a correlation between $\gamma$-ray flaring and VLBI radio flaring in both the ``core'' and the downstream quasi-stationary feature. \\

In order to investigate this further, we split the analysed period in two, i) all data before 2011 (Panel 2 of Fig.  \ref{ccor_post}) and ii) all data after 2011 (Panel 3 of Fig.  \ref{ccor_post}). We found tentative evidence for correlations between the $\gamma$-ray light curves and the decomposed flux densities in the quasi-stationary feature, with a lag consistent with zero before 2011. There was only tentative evidence for a correlation within the ``core'' before 2011. After 2011, there was again a significant correlation between $\gamma$-ray and quasi-stationary feature flux densities, but with $\gamma$-rays leading the quasi-stationary feature flux densities by $\sim$1 month. We found no significant correlation between ``core'' flux densities and $\gamma$-rays after 2011. We conclude that significant $\gamma$-ray activity was correlated with radio flaring in the quasi-stationary feature but we found only tentative evidence in the ``core''.

\subsection{Physical nature of the ``core'' and stationary feature}
\label{core_phys}

Stationary features have been observed previously in this source and in many others \cite[e.g.,][]{jor05,jor13,fromm13_2,frank12}, where they have typically been interpreted as recollimation shocks. A stationary feature could also appear due to maximised Doppler factors in a bent jet \cite[e.g.,][]{alberdi93,alberdi97}, or the base of the jet, but \citet{jor01} suggest that stationary features within 2\,mas of the ``core'' are likely hydrodynamical compressions, while further out, stationary features may be associated with bends in jets. In Section \ref{sec:bh}, we found that the ``core'' must be $\leq6$\,pc downstream of the jet base. This was an upper limit and does not rule out the ``core'' being at or near the jet base, but because total-intensity measurements at 1\,mm and 0.85\,mm are often lower than the 3\,mm flux densities of individual components, we found it it unlikely that we are significantly underestimating the turnover frequency and hence the ``core'' magnetic field. For this reason, we found it unlikely that the ``core'' was the base of the jet, near the black hole. Therefore, any quasi-stationary feature itself must be further downstream. In the quasi-stationary feature of OJ\,287, the high levels of polarisation, spectral variability and downstream proximity to the ``core'' lead us to conclude that component $S$ was probably a recollimation shock or an oblique shock associated with the bend in the jet.  \\

The ``core'' itself could be the $\tau=1$ surface, although this was possibly not the case at mm wavelengths \citep[see][for more details]{Marscher08Rev}. The ``core'' and stationary feature have remarkably similar properties, with both exhibiting high levels of spectral and flux variability and possibly $\gamma$-ray activity. Additionally, we have detected optically thin emission in the ``core'' region on two occasions, consistent with an optically thin travelling feature passing through the ``core''. This suggests that the ``core'' was a recollimation shock \cite[e.g.,][]{DalyMarscher88,Cawthorn06,fromm_sim, mar14}. The stationary feature could itself be a recollimation shock or an oblique shock.  A recollimation shock could be affected by disturbances passing through, resulting in fluctuations of its position, according to numerical simulations by \citet{gomez97}. In Section \ref{sec:PA} and Fig. \ref{xy}, we showed that stationary feature changes position relative to the ``core'', although the ``core'' itself may also shift position on the sky. Multi-epoch phase referencing VLBI observations would be needed to determine this.   \\

Interestingly, in Table \ref{mag_av}, the average size for the ``core'' (0.05 $\pm$ 0.01\,mas) was above the limit for the equipartition critical size of $\geq 0.02$\,mas, consistent with equipartition. The average size for the stationary feature (0.07 $\pm$ 0.01\,mas) was much smaller than the upper limit for the critical size of $\leq 0.12$\,mas, and also consistent with a jet in equipartition.  \\

\subsection{Location of $\gamma$-ray emission}
\label{sec:bh_discuss}

\begin{figure}
\includegraphics[width=\linewidth]{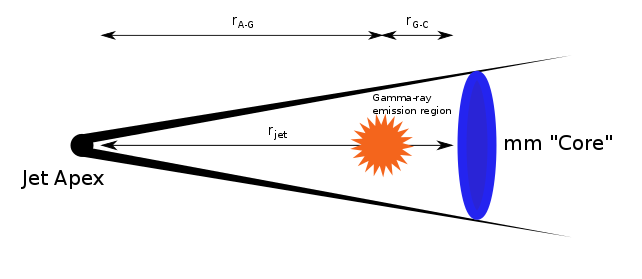}
\caption{A sketch the location of $\gamma$-ray emission region. The $\gamma$-ray emission region is located a distance $R_{A-G}$ from the jet apex.  }
\label{jet_sketch}
\end{figure}

Figure \ref{jet_sketch} depicts a stylised sketch of a relativistic jet. The $\gamma$-ray emission region is located a distance $r_{\text{A-G}}$ from the jet apex. This distance is the difference between the distance to the ``core'' from the jet apex, $r_{\text{apex}}$ and the distance between the ``core'' and the $\gamma$-ray emission region, $r_{G-C}$. We found in Section \ref{sec:ejection} that the $\gamma$-ray emission region, $r_{G-C}$, lies in the region of the mm-wave ``core'' (or if the $\gamma$-ray emission is in the stationary feature, it would be further downstream). The location of the mm-wave ``core'' was determined in Section \ref{sec:bh} to be $\lesssim 4-6$\,pc downstream of the jet apex. Therefore, if the $\gamma$-rays originate in the ``core'' or stationary feature, they are highly likely to be located outside of the BLR, which is considered to be less than a parsec in extent \citep{peterson04,bonnoli11}. This suggests that the synchrotron self-Compton (SSC) mechanism likely dominates $\gamma$-ray production in OJ\,287 \citep{bloom96}. If the interpretation of \citet{agudo11} was correct and $\gamma$-rays are produced in the stationary feature, the $\gamma$-ray emission region would be over 14\,pc (de-projected) further downstream than presented here and would support the same interpretation. If our calculations are correct, this would place the $\gamma$-rays in a region consistent with recollimation shocks \cite[e.g.,][]{mar08}. \\

\subsection{Magnetic field strength in the BLR and at the SMBH}
\label{sec:mag}

The magnetic field strength in the BLR and at the black hole of AGN has been investigated previously by \citet{silan13}, by observing polarisation degrees and position angles of broad H alpha lines. We can extrapolate the magnetic field estimates back and derive an estimate on the magnetic field strength at any arbitrary jet size \cite[e.g.,][]{hodgson15}:
\begin{equation}
\frac{B_{2}}{B_{1}} = \left[\frac{R_{1}}{R_{2}}\right]^{1/n},
\end{equation}
where $B_{1}$ and $B_{2}$ are the magnetic fields at arbitrary \emph{transverse} radii $R_{1}$ and $R_{2}$. Hence if the ``core'' is assumed to be resolved, we can estimate the magnetic field strength at a distance at any arbitrary distance (e.g., $10R_{S}$). We solve for two distances; (i) $B_{\text{0.05}}$, 0.05\,pc from the jet apex (likely within the BLR) and (ii) $B_{\text{apex}}$, at $10 R_{S}$ as an estimate for the size of the jet base. \\

There are many uncertainties in this calculation, as we must assume a constant Doppler factor along the jet, that the magnetic field configuration extends consistently to the jet base and make assumptions about the geometry of the jet (e.g., that the jet is conical). Nevertheless, we could derive upper limits on the magnetic field strength. The computed values are corrected using the Doppler factor from \citet{jor05}. As 1\,mm total intensity flux densities are much lower than at 3\,mm, indicating a small optical depth, these lower limits could be close to the true value. Assuming a toroidal (n=1) configuration and using the magnetic field decrease from SSA calculations, $B_{\text{0.05}} \lesssim$ 600\,G. If we then assume a conservative black hole mass, $m_{\text{BH}} = 4 \times 10^{8} M_{\odot}$, yielding a Schwarzschild Radius $R_{s} \approx 4 \times 10^{-5}$\,pc, we compute a upper limit of $B_{\text{apex}} \lesssim$ 4200\,G at a distance of 10\,$R_{s}$. Under the assumption of equipartition, we compute upper limits of $B_{\text{0.05}} \lesssim$ 500\,G and $B_{\text{apex}} \lesssim$ 3300\,G  The calculated values are broadly consistent with those reported by \citet{silan13} and \citet{marti15}. The presence of such strong magnetic fields at the jet apex would strongly favour the Blandford-Znajek process of jet formation dominating in OJ\,287 \citep{BZ77}, and suggests a magnetic launching mechanism of the VLBI jets, as recently described in magnetically arrested discs and other similar models \citep[e.g.,][]{tchek11,mckinney12,mckinney14,koskesh}. The magnetic field strengths are also near the maximum Eddington magnetic field, making a poloidal (n=2) magnetic field unlikely, as it would imply unphysical field strengths near the jet base \citep[e.g.,][]{rees84}.

\subsubsection{Large PA changes}
\label{pa_theory}

\begin{figure}
\includegraphics[width=\linewidth]{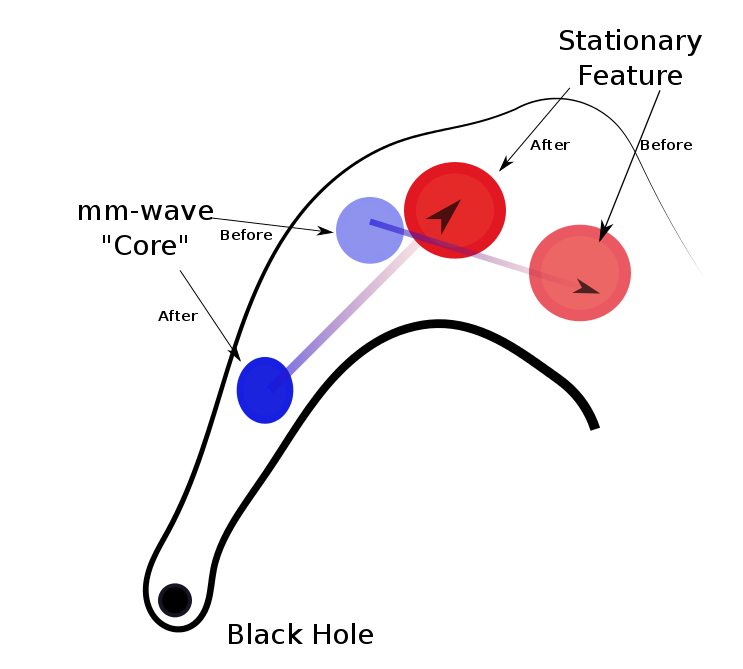}
\caption{A sketch of the cause of the large PA change from 2006-2008. A large disturbance at the jet base causes a change in the pressure of the jet relative to the external medium, changing the location of recollimation shocks within the jet. This sketch shows the \emph{de-projected} jet geometry. Effects such as this would be amplified observationally. }
\label{PA_sketch}
\end{figure}

The large PA change of 2006-2008 was interpreted by \citet{agudo12} as being due to the jet passing through our line-of-sight. This interpretation is plausible, but here we propose an alternative interpretation. If the jet was bent, and if the ``core'' position were to very suddenly change out of our line-of-sight by moving closer to the jet apex, we would observe an apparent large PA change and a coincident drop or increase in Doppler factor. Such a scenario could result from a change in the flow parameters injected at the jet base, perhaps a disrupted star falling into the SMBH, causing significant changes to the pressure ratio between the jet and external medium as shown in Fig. \ref{PA_sketch}. This occurrence would shift the locations of the standing shocks up or downstream. One could also speculate that if the pressure ratios return to their previous ratios in the future, the PA and Doppler factors would also return to their previous directions. While we cannot rule out a binary black hole in this model, it was not necessary to invoke it to explain the ``wobbling''. Assuming a constant external medium and relativistic electron/positron plasma, the `position ($z_{recol}$) of the first recollimation shock was given by \citep{DalyMarscher88}:
\begin{equation}
z_{\text{recol}} \approx 3.3 \Gamma_{0} R_{0} (\rho_{0}/\rho_{ext}),
\end{equation}
where $\Gamma_{0}$ is the Bulk Lorentz factor, $R_{0}$ is the jet opening angle ($\approx 30^{\circ}$), $\rho_{ext}$ is the external pressure ($\approx$ $1.6\times10^{-24}$ Pa) and $\rho_{0}$ is the internal pressure varying from $1.0\times10^{-25}-5\times10^{-28}$ Pa \cite[approximate values from:][]{fromm_thesis}. The results indicate that doubling the internal pressure of the jet results in the $z_{\text{recol}}$ of the first recollimation shock to be approximately twice as far out. This was necessarily very simplistic, but shows that such a scenario should be plausible.

\section{Conclusions}

We have used multi-frequency VLBI data at 2\,cm, 7\,mm and 3\,mm, total intensity radio data at 2\,cm, 7\,mm, 3\,mm, 1\,mm, 0.85\,mm and $\gamma$-ray data from the \emph{Fermi}/LAT space telescope to perform a detailed kinematic and spectral analysis of the highly variable BL Lac OJ\,287. We have used the 2\,cm MOJAVE, 7\,mm VLBA-BU-BLAZAR and 3\,mm GMVA data to determine the spectrum and hence estimates of the magnetic field at multiple locations down the jet. We combine this with kinematics derived from 3\,mm and 7\,mm data to determine the location of radio and $\gamma$-ray flaring events within the jet. We have found:

\begin{enumerate}
\item OJ\,287 exhibits two stationary features, components C (``core'') and S (``stationary feature''), that have very similar properties, with the ``core'' exhibiting an optically thin spectrum on two occasions. Both are interpreted as standing shocks. We postulate that the stationary feature moves on the sky relative to other components and we postulate that the ``core'' could also.

\item The $\approx$\,100$^{\circ}$ PA change reported by \citet{agudo11} resulted in a radically different ejection position angles and trajectories. Recent data suggests that these values may be returning to their pre-2006 values. We suggest an alternate interpretation of this behaviour as due to a large disturbance at the jet base changing the jet pressure causing the location of downstream standing shocks to shift their locations in the jet and changing the viewing angle and hence the Doppler factor.

\item Radio flaring activity was found in both the ``core'' and stationary feature. A large mm-wave radio flare (R2) was found to be a superposition of flares in both the ``core'' and stationary feature. Cross correlation analysis of $\gamma$-ray flaring found that flaring likely correlated with radio flaring in both the ``core'' and stationary feature as flaring coincides with the passage of components through both features. 

\item The magnetic field, as derived from SSA and from equipartition, decreases by $\sim80$ between the ``core'' and stationary feature. This allowed us to derive an estimate of the distance between the mm-wave ``core'' and the jet apex, $r_{\text{apex}}$. We found that $r_{\text{apex}}$, was $\leq 6.0$\,pc upstream of the mm-wave ``core'' and the most upstream site of $\gamma$-ray emission.

\item From SSA calculations, we found magnetic field strengths $B_{\text{SSA}} \geq$ 1.6\,G in the ``core'' and $B_{\text{SSA}} \lesssim$ 0.4\,G in the stationary feature. From equipartition, we estimate $B_{\text{equi}} \geq$ 1.2\,G in the ``core'' and $B_{\text{SSA}} \lesssim$ 0.3\,G in the stationary feature. The results from either method are consistent to within $\sim$20\%.

\item We also extrapolate estimates of the magnetic field strength to within the BLR and at the jet apex. Using SSA and assuming a toroidal magnetic field, this yields $B_{\text{BLR}} \lesssim 600$\,G and $B_{\text{apex}} \lesssim 4200$\,G. Using equipartition, this yields $B_{\text{BLR}} \lesssim 500$\,G and $B_{\text{apex}} \lesssim 3300$\,G.

\end{enumerate}

VLBI at 3\,mm currently lacks sensitivity and cadence, although the recently available 2\,Gbps recording modes and the possible addition of phased ALMA to the array will improve the situation \citep{mmvlbi_whitepaper}. Unfortunately with only 6 month intervals, structural changes may be missed, though high cadence monitoring could performed with the eight 3\,mm equipped stations of the VLBA but without trans-Atlantic baselines. In the future, it would be highly desirable to have monitoring at 3\,mm (and 1\,mm) at least as frequent as 43\,GHz monitoring, but with the longest baselines. In a future paper, we will expand this analysis to other blazars observed at 3\,mm to further investigate the physical conditions and dynamics of jets. We will also include polarisation observations, which may prove important in testing the location of $\gamma$-ray emission.  \\

\begin{acknowledgements}
I would like to thank my collaborators who have helped enormously. Jeffrey Hodgson, V. Karamanavis, I. Nestoras and I. Myserlis were supported in this research with funding from the International Max-Planck Research School (IMPRS) for Astronomy and Astrophysics at the Universities of Bonn and Cologne. I would also like to acknowledge the comments and suggestions from both the journal referee and the internal Fermi collaboration referee Marcello Giroletti. Both have helped greatly in significantly improving the manuscript. This research is partially based on observations performed at the 100\,m Effelsberg Radio Telescope, the IRAM Plateau de Bure Millimetre Interferometer, the IRAM 30\,m Millimeter Telescope, the Onsala 20\,m Radio Telescope, the Mets\"{a}hovi 14\,m Radio Telescope, the Yebes 30\,m Radio Telescope and the Very Long Baseline Array (VLBA). The VLBA is an instrument of the National Radio Astronomy Observatory. The National Radio Observatory is a facility of the National Science Foundation operated under the cooperative agreement by Associated Universities. IRAM is supported by MPG (Germany), INSU/CNRS (France), and IGN (Spain). The GMVA is operated by the MPIfR, IRAM, NRAO, OSO, and MRO. The research at Boston University was supported in part by by NASA through Fermi guest investigator grants s NNX08AV65G, NNX11AQ03G, and NNX12AO90G. The 43\,GHz VLBA data were obtained within the VLBA-BU-BLAZAR program. S.G.J. acknowledges support from Russian RFBR grant 15-02-00949 and St.Petersburg University research grant 6.38.335.2015. This paper made use of data available through the Monitoring Of Jets in Active galactic nuclei with VLBA Experiments (MOJAVE) program. Total intensity data were acquired through the FGAMMA program of the MPIfR and the Submillimeter Array (SMA) flux monitoring programs. The Submillimeter Array is a joint project between the Smithsonian Astrophysical Observatory and the Academia Sinica Institute of Astronomy and Astrophysics and is funded by the Smithsonian Institution and the Academia Sinica. This research has made use of the NASA/IPAC Extragalactic Database (NED) which is operated by the Jet Propulsion Laboratory, California Institute of Technology, under contract with the National Aeronautics and Space Administration. This research made use of Astropy, a community-developed core Python package for Astronomy \citep{astropy}. This research made use of APLpy, an open-source plotting package for Python hosted at http://aplpy.github.com. The {\it Fermi}/LAT Collaboration acknowledges the generous support of a number of agencies and institutes that have supported the {\it Fermi}/LAT Collaboration. These include the National Aeronautics and Space Administration and the Department of Energy in the United States, the Commissariat \`a l'Energie Atomique and the Centre National de la Recherche Scientifique / Institut National de Physique Nucl\'eaire et de Physique des Particules in France, the Agenzia Spaziale Italiana and the Istituto Nazionale di Fisica Nucleare in Italy, the Ministry of Education,Culture, Sports, Science and Technology (MEXT), High Energy Accelerator Research Organization (KEK) and Japan Aerospace Exploration Agency (JAXA) in Japan, and the K.\ A.\ Wallenberg Foundation, the Swedish Research Council and the Swedish National Space Board in Sweden. Additional support for science analysis during the operations phase is gratefully acknowledged from the Istituto Nazionale di Astrofisica in Italy and the Centre National d'\'Etudes Spatiales in France. I would like to thank Tom for helping me articulate my science into simple but inexact metaphors. I would also like to acknowledge the contribution of A. Berterini and the correlation staff at the MPIfR Bonn.
\end{acknowledgements}


\clearpage

\clearpage 
\onecolumn 
\begin{longtable}{ccccccc}


\hline
\hline
Epoch & Freq  & Flux  & Core sep   & PA  & FWHM [mas]  & ID \\ 
      & [GHz] &  [Jy] &  [mas]  & [$^{\circ}$] & \\ \hline
\endhead
\hline 
\caption{Table of model-fit parameters} \\ \hline
\endfoot

2007.45  & 43.2  & 0.49$ \pm $0.05  & 0$ \pm $0  & 0$ \pm $0  & 0.04$ \pm $0.03 & C  \\
2007.45  & 43.2  & 0.55$ \pm $0.06  & 0.13$ \pm $0.02  & -11.7$ \pm $5  & 0.016$ \pm $0.030 & S  \\
2007.45  & 43.2  & 0.13$ \pm $0.01  & 1.00$ \pm $0.15  & -125.9$ \pm $5  & 0.48$ \pm $0.09 & X0  \\
2007.45  & 43.2  & 0.09$ \pm $0.01  & 0.21$ \pm $0.03  & -102.2$ \pm $5  & 0.09$ \pm $0.01 & P1  \\
     &     &     &     &     &    &     \\
2007.66  & 43.2  & 0.52$ \pm $0.05  & 0$ \pm $0  & 0$ \pm $0  & 0.02$ \pm $0.03 & C  \\
2007.66  & 43.2  & 0.79$ \pm $0.08  & 0.15$ \pm $0.02  & -11.5$ \pm $5  & 0.02$ \pm $0.03 & S  \\
2007.66  & 43.2  & 0.12$ \pm $0.01  & 1.01$ \pm $0.15  & -119.6$ \pm $5  & 0.52$ \pm $0.10 & X0  \\
2007.66  & 43.2  & 0.17$ \pm $0.01  & 0.26$ \pm $0.04  & -109.5$ \pm $5  & 0.10$ \pm $0.02 & P1  \\
     &     &     &     &     &    &     \\
2007.74  & 43.2  & 0.68$ \pm $0.07  & 0$ \pm $0  & 0$ \pm $0  & 0.02$ \pm $0.03 & C  \\
2007.74  & 43.2  & 0.93$ \pm $0.10  & 0.15$ \pm $0.02  & -15.4$ \pm $5  & 0.003$ \pm $0.03 & S  \\
2007.74  & 43.2  & 0.12$ \pm $0.01  & 1.04$ \pm $0.15  & -119.6$ \pm $5  & 0.47$ \pm $0.09 & X0  \\
2007.74  & 43.2  & 0.16$ \pm $0.02  & 0.30$ \pm $0.04  & -109.5$ \pm $5  & 0.13$ \pm $0.02 & P1`  \\
     &     &     &     &     &    &     \\
2008.05  & 43.2  & 0.72$ \pm $0.07  & 0$ \pm $0  & 0$ \pm $0  & 0.03$ \pm $0.03 & C    \\
2008.05  & 43.2  & 1.55$ \pm $0.17  & 0.18$ \pm $0.02  & -16.7$ \pm $5  & 0.02$ \pm $0.03 & S   \\
2008.05  & 43.2  & 0.06$ \pm $0.01  & 1.12$ \pm $0.16  & -115.5$ \pm $5  & 0.38$ \pm $0.07 & X0  \\
2008.05  & 43.2  & 0.06$ \pm $0.01  & 0.39$ \pm $0.05  & -119.9$ \pm $5  & 0.17$ \pm $0.03 & P1  \\
     &     &     &     &     &    &     \\
2008.44  & 43.2  & 0.84$ \pm $0.09  & 0$ \pm $0  & 0$ \pm $0  & 0.02$ \pm $0.03 & C  \\
2008.44  & 43.2  & 2.09$ \pm $0.22  & 0.15$ \pm $0.02  & -19.3$ \pm $5  & 0.049$ \pm $0.030 & S  \\
2008.44  & 43.2  & 0.11$ \pm $0.01  & 0.28$ \pm $0.04  & -113.9$ \pm $5  & 0.19$ \pm $0.03 & P1  \\
     &     &     &     &     &    &     \\
2008.62  & 43.2  & 0.57$ \pm $0.06  & 0$ \pm $0  & 0$ \pm $0  & 0.04$ \pm $0.03 & C  \\
2008.62  & 43.2  & 1.99$ \pm $0.21  & 0.20$ \pm $0.03  & -18.1$ \pm $5  & 0.07$ \pm $0.03 & S  \\
2008.62  & 43.2  & 0.20$ \pm $0.02  & 0.30$ \pm $0.04  & -111.2$ \pm $5  & 0.47$ \pm $0.03 & P1  \\
     &     &     &     &     &    &     \\
2008.69  & 43.2  & 0.60$ \pm $0.06  & 0$ \pm $0  & 0$ \pm $0  & 0.05$ \pm $0.05 & C   \\
2008.69  & 43.2  & 1.22$ \pm $0.15  & 0.20$ \pm $0.03  & -15.0$ \pm $5  & 0.02$ \pm $0.03 & S  \\
2008.69  & 43.2  & 0.57$ \pm $0.06  & 0.18$ \pm $0.03  & -31.9$\pm  $5  & 0.15$ \pm $0.03 & P2 \\
2008.69  & 43.2  & 0.10$ \pm $0.01  & 0.45$ \pm $0.06  & -122.6$ \pm $5  & 0.35$ \pm $0.07 & X1  \\
     &     &     &     &     &    &     \\
2008.78  & 86.2  & 0.53$ \pm $0.05  & 0$ \pm $0  & 0$ \pm $0  & 0.01$ \pm $0.01 & C   \\
2008.78  & 86.2  & 1.82$ \pm $0.20  & 0.19$ \pm $0.02  & -12.2$ \pm $5  & 0.01$ \pm $0.01 & S  \\
2008.78  & 86.2  & 0.13$ \pm $0.01  & 0.17$ \pm $0.02  & -59.4$ \pm $5  & 0.11$ \pm $0.02 & U  \\
2008.78  & 86.2  & 0.42$ \pm $0.04  & 0.33$ \pm $0.04  & -10.8$ \pm $5  & 0.05$ \pm $0.01 & P2  \\
2008.78  & 86.2  & 0.07$ \pm $0.01  & 0.47$ \pm $0.07  & -98.4$ \pm $5  & 0.39$ \pm $0.07 & X1  \\
     &     &     &     &     &    &     \\
2008.88  & 43.2  & 1.34$ \pm $0.14  & 0$ \pm $0  & 0$ \pm $0  & 0.02$ \pm $0.03 & C  \\
2008.88  & 43.2  & 2.42$ \pm $0.26  & 0.22$ \pm $0.03  & -12.4$ \pm $5  & 0.02$ \pm $0.03 & S  \\
2008.88  & 43.2  & 0.11$ \pm $0.01  & 0.39$ \pm $0.05  & -127.5$ \pm $5  & 0.11$ \pm $0.02 & X1  \\
     &     &     &     &     &    &     \\
2008.97  & 43.2  & 1.27$ \pm $0.13  & 0$ \pm $0  & 0$ \pm $0  & 0.04$ \pm $0.03 & C  \\
2008.97  & 43.2  & 2.08$ \pm $0.22  & 0.22$ \pm $0.03  & -11.3$ \pm $5  & 0.04$ \pm $0.03 & S  \\
2008.97  & 43.2  & 0.13$ \pm $0.01  & 0.45$ \pm $0.06  & -115.1$ \pm $5  & 0.40$ \pm $0.08 & X1  \\
     &     &     &     &     &    &     \\
2009.07  & 43.2  & 1.39$ \pm $0.15  & 0$ \pm $0  & 0$ \pm $0  & 0.028$ \pm $0.03 & C  \\
2009.07  & 43.2  & 3.74$ \pm $0.38  & 0.22$ \pm $0.03  & -10.6$ \pm $5  & 0.03$ \pm $0.03 & S  \\
2009.07  & 43.2  & 1.04$ \pm $0.10  & 0.29$ \pm $0.03  & -13.5$ \pm $5  & 0.04$ \pm $0.03 & P2  \\
2009.07  & 43.2  & 0.10$ \pm $0.01  & 0.45$ \pm $0.06  & -116.6$ \pm $5  & 0.22$ \pm $0.04 & X1  \\
     &     &     &     &     &    &     \\
2009.15  & 43.2  & 1.33$ \pm $0.14  & 0$ \pm $0  & 0$ \pm $0  & 0.042$ \pm $0.03 & C  \\
2009.15  & 43.2  & 4.17$ \pm $0.45  & 0.23$ \pm $0.03  & -12.1$ \pm $5  & 0.04$ \pm $0.03 & S  \\
2009.15  & 43.2  & 0.18$ \pm $0.02  & 0.46$ \pm $0.07  & -121.0$ \pm $5  & 0.55$ \pm $0.11 & X1  \\
     &     &     &     &     &    &     \\
2009.35  & 86.2  & 1.45$ \pm $0.15  & 0$ \pm $0  & 0$ \pm $0  & 0.021$ \pm $0.014 & C  \\
2009.35  & 86.2  & 3.68$ \pm $0.40  & 0.26$ \pm $0.03  & -13.0$ \pm $5  & 0.04$ \pm $0.02 & S  \\
2009.35  & 86.2  & 0.24$ \pm $0.02  & 0.45$ \pm $0.06  & -129.3$ \pm $5  & 0.13$ \pm $0.026 & X1  \\
     &     &     &     &     &    &     \\
2009.41  & 43.2  & 1.21$ \pm $0.13  & 0$ \pm $0  & 0$ \pm $0  & 0.035$ \pm $0.03 & C  \\
2009.41  & 43.2  & 1.52$ \pm $0.23  & 0.27$ \pm $0.03  & -16.9$ \pm $5  & 0.04$ \pm $0.03 & S \\
2009.41  & 43.2  & 1.84$ \pm $0.27  & 0.23$ \pm $0.03  & -15.0$ \pm $5  & 0.03$ \pm $0.03 & P2  \\
2009.41  & 43.2  & 0.06$ \pm $0.02  & 0.30$ \pm $0.03  & -102.5$ \pm $5  & 0.10$ \pm $0.03 & U \\
2009.41  & 43.2  & 0.18$ \pm $0.02  & 0.61$ \pm $0.09  & -122.0$ \pm $5  & 0.52$ \pm $0.10 & X1  \\
     &     &     &     &     &    &     \\
2009.47  & 43.2  & 1.26$ \pm $0.13  & 0$ \pm $0  & 0$ \pm $0  & 0.033$ \pm $0.03 & C  \\
2009.47  & 43.2  & 1.47$ \pm $0.15  & 0.20$ \pm $0.03  & -13.3$ \pm $5  & 0.04$ \pm $0.03 & S  \\
2009.47  & 43.2  & 1.45$ \pm $0.15  & 0.30$ \pm $0.03  & -13.3$ \pm $5  & 0.06$ \pm $0.03 & P2 \\
2009.47  & 43.2  & 0.19$ \pm $0.02  & 0.56$ \pm $0.08  & -118.7$ \pm $5  & 0.55$ \pm $0.11 & X1  \\
     &     &     &     &     &    &     \\
2009.57  & 43.2  & 1.43$ \pm $0.15  & 0$ \pm $0  & 0$ \pm $0  & 0.031$ \pm $0.03 & C  \\
2009.57  & 43.2  & 3.01$ \pm $0.30  & 0.21$ \pm $0.03  & -14.8$ \pm $5  & 0.03$ \pm $0.03 & S  \\
2009.57  & 43.2  & 0.68$ \pm $0.07  & 0.33$ \pm $0.03  & -16.2$ \pm $5  & 0.07$ \pm $0.03 & P2  \\
2009.57  & 43.2  & 0.18$ \pm $0.01  & 0.55$ \pm $0.08  & -119.1$ \pm $5  & 0.68$ \pm $0.13 & X1  \\
     &     &     &     &     &    &     \\
2009.62  & 43.2  & 1.42$ \pm $0.15  & 0$ \pm $0  & 0$ \pm $0  & 0.032$ \pm $0.03 & C  \\
2009.62  & 43.2  & 4.62$ \pm $0.50  & 0.23$ \pm $0.03  & -16.4$ \pm $5  & 0.03$ \pm $0.03 & S  \\
2009.62  & 43.2  & 0.16$ \pm $0.01  & 0.73$ \pm $0.11  & -124.8$ \pm $5  & 0.61$ \pm $0.12 & X1   \\
     &     &     &     &     &    &     \\
2009.71  & 43.2  & 2.02$ \pm $0.22  & 0$ \pm $0  & 0$ \pm $0  & 0.003$ \pm $ $\leq 0.01$ & C  \\
2009.71  & 43.2  & 1.75$ \pm $0.18  & 0.19$ \pm $0.03  & -16.0$ \pm $5  & $\leq 0.01$ & S  \\
2009.71  & 43.2  & 2.96$ \pm $0.30  & 0.25$ \pm $0.03  & -19.3$ \pm $5  & 0.04$ \pm $0.03 & P2  \\
2009.71  & 43.2  & 0.19$ \pm $0.02  & 0.82$ \pm $0.12  & -122.8$ \pm $5  & 0.61$ \pm $0.12 & X1  \\
     &     &     &     &     &    &     \\
2009.77  & 86.2  & 2.95$ \pm $0.32  & 0$ \pm $0  & 0$ \pm $0  & 0.036$ \pm $0.03 & C   \\
2009.77  & 86.2  & 0.51$ \pm $0.05  & 0.11$ \pm $0.02  & -19.0$ \pm $5  & 0.02$ \pm $0.01 & M  \\
2009.77  & 86.2  & 4.73$ \pm $0.52  & 0.23$ \pm $0.03  & -18.1$ \pm $5  & 0.04$ \pm $0.03 & S  \\
2009.77  & 86.2  & 0.81$ \pm $0.08  & 0.34$ \pm $0.05  & -32.2$ \pm $5  & 0.09$ \pm $0.03 & P2  \\
2009.77  & 86.2  & 0.13$ \pm $0.01  & 0.77$ \pm $0.11  & -143.7$ \pm $5  & 0.06$ \pm $0.01 & X1  \\
     &     &     &     &     &    &     \\
2009.79  & 43.2  & 2.83$ \pm $0.31  & 0$ \pm $0  & 0$ \pm $0  & 0.014$ \pm $0.03 & C   \\
2009.79  & 43.2  & 5.44$ \pm $0.59  & 0.23$ \pm $0.03  & -18.6$ \pm $5  & 0.03$ \pm $0.03 & S  \\
2009.79  & 43.2  & 0.17$ \pm $0.01  & 0.94$ \pm $0.14  & -128.2$ \pm $5  & 0.55$ \pm $0.11 & X1   \\
     &     &     &     &     &    &     \\
2009.80  & 43.2  & 2.92$ \pm $0.32  & 0$ \pm $0  & 0$ \pm $0  & $\leq 0.01$ & C  \\
2009.80  & 43.2  & 5.42$ \pm $0.59  & 0.23$ \pm $0.03  & -18.9$ \pm $5  & 0.03$ \pm $0.03 & S  \\
2009.80  & 43.2  & 0.18$ \pm $0.02  & 0.94$ \pm $0.14  & -123.3$ \pm $5  & 0.51$ \pm $0.10 & X1  \\
     &     &     &     &     &    &     \\
2009.82  & 43.2  & 2.11$ \pm $0.23  & 0$ \pm $0  & 0$ \pm $0  & 0.042$ \pm $0.03 & C  \\
2009.82  & 43.2  & 0.91$ \pm $0.10  & 0.14$ \pm $0.03  & -26.3$ \pm $5  & 0.12$ \pm $0.03 & P2  \\
2009.82  & 43.2  & 3.86$ \pm $0.39  & 0.22$ \pm $0.03  & -19.3$ \pm $5  & 0.03$ \pm $0.03 & S  \\
2009.82  & 43.2  & 0.25$ \pm $0.02  & 0.74$ \pm $0.11  & -124.6$ \pm $5  & 0.64$ \pm $0.12 & X1  \\
     &     &     &     &     &    &     \\
2009.91  & 43.2  & 2.83$ \pm $0.31  & 0$ \pm $0  & 0$ \pm $0  & 0.066$ \pm $0.03 & C  \\
2009.91  & 43.2  & 6.78$ \pm $0.74  & 0.22$ \pm $0.03  & -21.6$ \pm $5  & 0.05$ \pm $0.03 & S  \\
2009.91  & 43.2  & 0.15$ \pm $0.01  & 0.99$ \pm $0.14  & -125.2$ \pm $5  & 0.58$ \pm $0.11 & X1  \\
     &     &     &     &     &    &     \\
2010.03  & 43.2  & 1.68$ \pm $0.18  & 0$ \pm $0  & 0$ \pm $0  & 0.039$ \pm $0.03 & C  \\
2010.03  & 43.2  & 1.41$ \pm $0.14  & 0.19$ \pm $0.03  & -23.8$ \pm $5  & $\leq 0.01$ & S  \\
2010.03  & 43.2  & 6.82$ \pm $0.68  & 0.24$ \pm $0.03  & -25.6$ \pm $5  & 0.06$ \pm $0.03 & P2  \\
2010.03  & 43.2  & 0.17$ \pm $0.01  & 0.91$ \pm $0.13  & -124.2$ \pm $5  & 0.62$ \pm $0.12 & X1  \\
     &     &     &     &     &    &     \\
2010.11  & 43.2  & 0.94$ \pm $0.10  & 0$ \pm $0  & 0$ \pm $0  & 0.044$ \pm $0.03 & C  \\
2010.11  & 43.2  & 3.27$ \pm $0.35  & 0.23$ \pm $0.03  & -25.7$ \pm $5  & 0.06$ \pm $0.03 & S  \\
2010.11  & 43.2  & 0.13$ \pm $0.01  & 0.89$ \pm $0.13  & -121.7$ \pm $5  & 0.63$ \pm $0.12 & X1  \\
     &     &     &     &     &    &     \\
2010.18  & 43.2  & 1.25$ \pm $0.13  & 0$ \pm $0  & 0$ \pm $0  & 0.059$ \pm $0.03 & C  \\
2010.18  & 43.2  & 5.94$ \pm $0.65  & 0.23$ \pm $0.03  & -27.8$ \pm $5  & 0.06$ \pm $0.03 & S  \\
2010.18  & 43.2  & 0.18$ \pm $0.02  & 1.05$ \pm $0.15  & -130.5$ \pm $5  & 0.53$ \pm $0.10 & X1  \\
     &     &     &     &     &    &     \\
2010.35  & 86.2  & 0.87$ \pm $0.09  & 0$ \pm $0  & 0$ \pm $0  & 0.023$ \pm $0.014 & C  \\
2010.35  & 86.2  & 0.43$ \pm $0.04  & 0.18$ \pm $0.02  & -36.6$ \pm $5  & 0.03$ \pm $0.01 & S  \\
2010.35  & 86.2  & 0.84$ \pm $0.09  & 0.23$ \pm $0.03  & -43.9$ \pm $5  & 0.08$ \pm $0.01 & P2  \\
2010.35  & 86.2  & 0.07$ \pm $0.01  & 1.09$ \pm $0.16  & -126.8$ \pm $5  & 0.41$ \pm $0.08 & X1  \\
     &     &     &     &     &    &     \\
2010.45  & 43.2  & 0.88$ \pm $0.09  & 0$ \pm $0  & 0$ \pm $0  & 0.04$ \pm $0.03 & C   \\
2010.45  & 43.2  & 1.38$ \pm $0.15  & 0.20$ \pm $0.03  & -35.5$ \pm $5  & 0.04$ \pm $0.03 & S  \\
2010.45  & 43.2  & 0.57$ \pm $0.06  & 0.27$ \pm $0.04  & -46.4$ \pm $5  & 0.12$ \pm $0.02 & P2  \\
2010.45  & 43.2  & 0.13$ \pm $0.01  & 1.11$ \pm $0.16  & -129.1$ \pm $5  & 0.46$ \pm $0.09 & X1  \\
     &     &     &     &     &    &     \\
2010.58  & 43.2  & 0.82$ \pm $0.09  & 0$ \pm $0  & 0$ \pm $0  & 0.036$ \pm $0.030 & C   \\
2010.58  & 43.2  & 0.90$ \pm $0.09  & 0.21$ \pm $0.03  & -40.1$ \pm $5  & 0.05$ \pm $0.03 & S  \\
2010.58  & 43.2  & 0.85$ \pm $0.09  & 0.31$ \pm $0.04  & -47.3$ \pm $5  & 0.11$ \pm $0.02 & P2  \\
2010.58  & 43.2  & 0.17$ \pm $0.02  & 1.07$ \pm $0.16  & -130.5$ \pm $5  & 0.63$ \pm $0.12 & X1  \\
     &     &     &     &     &    &     \\
2010.72  & 43.2  & 2.82$ \pm $0.31  & 0$ \pm $0  & 0$ \pm $0  & 0.038$ \pm $0.030 & C  \\
2010.72  & 43.2  & 1.37$ \pm $0.15  & 0.15$ \pm $0.02  & -39.9$ \pm $5  & 0.08$ \pm $0.03 & S  \\
2010.72  & 43.2  & 0.89$ \pm $0.09  & 0.31$ \pm $0.04  & -58.6$ \pm $5  & 0.23$ \pm $0.04 & P2  \\
2010.72  & 43.2  & 0.33$ \pm $0.03  & 1.09$ \pm $0.16  & -130.9$ \pm $5  & 0.63$ \pm $0.12 & X1  \\
     &     &     &     &     &    &     \\
2010.81  & 43.2  & 1.72$ \pm $0.18  & 0$ \pm $0  & 0$ \pm $0  & 0.023$ \pm $0.030 & C  \\
2010.81  & 43.2  & 2.02$ \pm $0.22  & 0.18$ \pm $0.02  & -40.1$ \pm $5  & 0.06$ \pm $0.03 & S  \\
2010.81  & 43.2  & 0.43$ \pm $0.04  & 0.31$ \pm $0.04  & -54.9$ \pm $5  & 0.19$ \pm $0.03 & P2  \\
2010.81  & 43.2  & 0.14$ \pm $0.01  & 1.15$ \pm $0.17  & -127.5$ \pm $5  & 0.58$ \pm $0.11 & X1  \\
     &     &     &     &     &    &     \\
2010.84  & 43.2  & 2.64$ \pm $0.29  & 0$ \pm $0  & 0$ \pm $0  & 0.035$ \pm $0.030 & C  \\
2010.84  & 43.2  & 2.21$ \pm $0.24  & 0.19$ \pm $0.02  & -41.2$ \pm $5  & 0.09$ \pm $0.03 & S  \\
2010.84  & 43.2  & 0.62$ \pm $0.06  & 0.31$ \pm $0.04  & -55.6$ \pm $5  & 0.24$ \pm $0.04 & P2  \\
2010.84  & 43.2  & 0.16$ \pm $0.01  & 1.18$ \pm $0.17  & -128.0$ \pm $5  & 0.52$ \pm $0.10 & X1  \\
     &     &     &     &     &    &     \\
2010.85  & 43.2  & 2.61$ \pm $0.28  & 0$ \pm $0  & 0$ \pm $0  & 0.035$ \pm $0.030 & C  \\
2010.85  & 43.2  & 2.36$ \pm $0.25  & 0.19$ \pm $0.02  & -42.4$ \pm $5  & 0.08$ \pm $0.03 & S  \\
2010.85  & 43.2  & 0.59$ \pm $0.06  & 0.31$ \pm $0.04  & -60.5$ \pm $5  & 0.20$ \pm $0.04 & P2  \\
2010.85  & 43.2  & 0.19$ \pm $0.02  & 1.18$ \pm $0.17  & -131.4$ \pm $5  & 0.56$ \pm $0.11 & X1  \\
     &     &     &     &     &    &     \\
2010.87  & 43.2  & 2.25$ \pm $0.24  & 0$ \pm $0  & 0$ \pm $0  & 0.012$ \pm $0.030 & C  \\
2010.87  & 43.2  & 2.22$ \pm $0.24  & 0.21$ \pm $0.03  & -44.0$ \pm $5  & 0.07$ \pm $0.03 & S  \\
2010.87  & 43.2  & 0.52$ \pm $0.05  & 0.32$ \pm $0.04  & -64.0$ \pm $5  & 0.20$ \pm $0.04 & P2  \\
2010.87  & 43.2  & 0.15$ \pm $0.01  & 1.20$ \pm $0.18  & -131.2$ \pm $5  & 0.64$ \pm $0.12 & X1  \\
     &     &     &     &     &    &     \\
2010.93  & 43.2  & 2.20$ \pm $0.24  & 0$ \pm $0  & 0$ \pm $0  & 0.04$ \pm $0.03 & C  \\
2010.93  & 43.2  & 3.19$ \pm $0.35  & 0.22$ \pm $0.03  & -46.9$ \pm $5  & 0.10$ \pm $0.03 & S  \\
2010.93  & 43.2  & 0.42$ \pm $0.04  & 0.35$ \pm $0.05  & -64.8$ \pm $5  & 0.15$ \pm $0.03 & P2  \\
2010.93  & 43.2  & 0.13$ \pm $0.01  & 1.24$ \pm $0.18  & -130.0$ \pm $5  & 0.50$ \pm $0.10 & X1  \\
     &     &     &     &     &    &     \\
2011.01  & 43.2  & 0.27$ \pm $0.03  & 0$ \pm $0  & 0$ \pm $0  & $\leq 0.01$ & C  \\
2011.01  & 43.2  & 2.09$ \pm $0.22  & 0.12$ \pm $0.01  & -22.0$ \pm $5  & 0.01$ \pm $0.03 & M  \\
2011.01  & 43.2  & 2.19$ \pm $0.24  & 0.32$ \pm $0.04  & -35.6$ \pm $5  & 0.11$ \pm $0.03 & S  \\
2011.01  & 43.2  & 1.07$ \pm $0.11  & 0.43$ \pm $0.06  & -54.4$ \pm $5  & 0.17$ \pm $0.03 & P2  \\
2011.01  & 43.2  & 0.14$ \pm $0.01  & 1.20$ \pm $0.18  & -124.8$ \pm $5  & 0.50$ \pm $0.10 & X1  \\
     &     &     &     &     &    &     \\
2011.10  & 43.2  & 2.19$ \pm $0.24  & 0$ \pm $0  & 0$ \pm $0  & 0.036$ \pm $0.030 & C  \\
2011.10  & 43.2  & 1.79$ \pm $0.19  & 0.19$ \pm $0.02  & -26.8$ \pm $5  & 0.05$ \pm $0.03 & S  \\
2011.10  & 43.2  & 1.40$ \pm $0.15  & 0.27$ \pm $0.04  & -41.6$ \pm $5  & 0.12$ \pm $0.02 & P2.2   \\
2011.10  & 43.2  & 0.92$ \pm $0.10  & 0.40$ \pm $0.06  & -70.8$ \pm $5  & 0.21$ \pm $0.04 & X2  \\
2011.10  & 43.2  & 0.17$ \pm $0.02  & 1.17$ \pm $0.17  & -131.4$ \pm $5  & 0.58$ \pm $0.11 & X1   \\
     &     &     &     &     &    &     \\
2011.16  & 43.2  & 2.25$ \pm $0.24  & 0$ \pm $0  & 0$ \pm $0  & 0.014$ \pm $0.030 & C  \\
2011.16  & 43.2  & 1.97$ \pm $0.21  & 0.18$ \pm $0.02  & -28.4$ \pm $5  & 0.04$ \pm $0.03 & S  \\
2011.16  & 43.2  & 0.94$ \pm $0.10  & 0.28$ \pm $0.04  & -39.0$ \pm $5  & 0.12$ \pm $0.02 & P2.2  \\
2011.16  & 43.2  & 0.79$ \pm $0.08  & 0.42$ \pm $0.06  & -74.5$ \pm $5  & 0.22$ \pm $0.04 & X2  \\
2011.16  & 43.2  & 0.14$ \pm $0.01  & 1.23$ \pm $0.18  & -131.3$ \pm $5  & 0.48$ \pm $0.09 & X1  \\
     &     &     &     &     &    &     \\
2011.30  & 43.2  & 1.40$ \pm $0.15  & 0$ \pm $0  & 0$ \pm $0  & 0.035$ \pm $0.03 & C  \\
2011.30  & 43.2  & 2.52$ \pm $0.27  & 0.20$ \pm $0.03  & -23.0$ \pm $5  & 0.04$ \pm $0.03 & S  \\
2011.30  & 43.2  & 1.48$ \pm $0.16  & 0.26$ \pm $0.04  & -33.9$ \pm $5  & 0.08$ \pm $0.03 & P2.2  \\
2011.30  & 43.2  & 0.34$ \pm $0.03  & 0.34$ \pm $0.05  & -50.3$ \pm $5  & 0.09$ \pm $0.03 & U  \\
2011.30  & 43.2  & 0.49$ \pm $0.05  & 0.49$ \pm $0.07  & -77.3$ \pm $5  & 0.19$ \pm $0.03 & X2  \\
2011.30  & 43.2  & 0.18$ \pm $0.02  & 1.24$ \pm $0.18  & -131.8$ \pm $5  & 0.56$ \pm $0.11 & X1  \\
     &     &     &     &     &    &     \\
2011.36  & 86.2  & 3.71$ \pm $0.40  & 0$ \pm $0  & 0$ \pm $0  & 0.043$ \pm $0.014 & C  \\
2011.36  & 86.2  & 1.79$ \pm $0.19  & 0.19$ \pm $0.02  & -43.8$ \pm $5  & 0.11$ \pm $0.01 & S  \\
2011.36  & 86.2  & 0.74$ \pm $0.08  & 0.45$ \pm $0.06  & -66.0$ \pm $5  & 0.15$ \pm $0.03 & X2  \\
     &     &     &     &     &    &     \\
2011.39  & 43.2  & 1.95$ \pm $0.21  & 0$ \pm $0  & 0$ \pm $0  & 0.053$ \pm $0.030 & C  \\
2011.39  & 43.2  & 3.05$ \pm $0.33  & 0.22$ \pm $0.03  & -29.3$ \pm $5  & 0.05$ \pm $0.03 & S  \\
2011.39  & 43.2  & 2.55$ \pm $0.28  & 0.28$ \pm $0.04  & -38.6$ \pm $5  & 0.16$ \pm $0.03 & P2.2   \\
2011.39  & 43.2  & 0.73$ \pm $0.08  & 0.49$ \pm $0.07  & -81.2$ \pm $5  & 0.26$ \pm $0.05 & X2  \\
2011.39  & 43.2  & 0.15$ \pm $0.01  & 1.30$ \pm $0.19  & -128.7$ \pm $5  & 0.21$ \pm $0.04 & X1  \\
     &     &     &     &     &    &     \\
2011.45  & 43.2  & 1.68$ \pm $0.18  & 0$ \pm $0  & 0$ \pm $0  & 0.016$ \pm $0.030 & C  \\
2011.45  & 43.2  & 1.65$ \pm $0.18  & 0.21$ \pm $0.03  & -30.3$ \pm $5  & 0.03$ \pm $0.03 & S  \\
2011.45  & 43.2  & 1.61$ \pm $0.17  & 0.31$ \pm $0.04  & -37.7$ \pm $5  & 0.09$ \pm $0.01 & P2.2  \\
2011.45  & 43.2  & 0.68$ \pm $0.07  & 0.49$ \pm $0.07  & -75.3$ \pm $5  & 0.26$ \pm $0.05 & X2  \\
2011.45  & 43.2  & 0.18$ \pm $0.02  & 1.29$ \pm $0.19  & -139.5$ \pm $5  & 0.44$ \pm $0.08 & X1  \\
     &     &     &     &     &    &     \\
2011.55  & 43.2  & 1.00$ \pm $0.11  & 0$ \pm $0  & 0$ \pm $0  & 0.019$ \pm $0.030 & C  \\
2011.55  & 43.2  & 1.45$ \pm $0.15  & 0.19$ \pm $0.02  & -31.5$ \pm $5  & 0.03$ \pm $0.03 & S  \\
2011.55  & 43.2  & 1.24$ \pm $0.13  & 0.28$ \pm $0.04  & -33.1$ \pm $5  & 0.04$ \pm $0.01 & P2.2  \\
2011.55  & 43.2  & 0.40$ \pm $0.04  & 0.36$ \pm $0.05  & -45.2$ \pm $5  & 0.14$ \pm $0.02 & U  \\
2011.55  & 43.2  & 0.48$ \pm $0.05  & 0.56$ \pm $0.08  & -81.0$ \pm $5  & 0.24$ \pm $0.04 & X2  \\
2011.55  & 43.2  & 0.17$ \pm $0.02  & 1.34$ \pm $0.20  & -134.6$ \pm $5  & 0.46$ \pm $0.09 & X1  \\
     &     &     &     &     &    &     \\
2011.64  & 43.2  & 2.07$ \pm $0.22  & 0$ \pm $0  & 0$ \pm $0  & 0.084$ \pm $0.030 & C  \\
2011.64  & 43.2  & 1.64$ \pm $0.18  & 0.22$ \pm $0.03  & -34.3$ \pm $5  & 0.05$ \pm $0.03 & S  \\
2011.64  & 43.2  & 1.10$ \pm $0.12  & 0.33$ \pm $0.04  & -39.1$ \pm $5  & 0.12$ \pm $0.02 & X2.2  \\
2011.64  & 43.2  & 0.46$ \pm $0.05  & 0.58$ \pm $0.08  & -83.7$ \pm $5  & 0.28$ \pm $0.05 & X2  \\
2011.64  & 43.2  & 0.19$ \pm $0.02  & 1.37$ \pm $0.20  & -135.0$ \pm $5  & 0.48$ \pm $0.09 & X1  \\
     &     &     &     &     &    &     \\
2011.71  & 43.2  & 1.62$ \pm $0.17  & 0$ \pm $0  & 0$ \pm $0  & 0.017$ \pm $0.030 & C  \\
2011.71  & 43.2  & 0.55$ \pm $0.06  & 0.13$ \pm $0.02  & -32.2$ \pm $5  & 0.02$ \pm $0.03 & M  \\
2011.71  & 43.2  & 0.78$ \pm $0.08  & 0.28$ \pm $0.04  & -38.0$ \pm $5  & 0.05$ \pm $0.03 & S  \\
2011.71  & 43.2  & 0.36$ \pm $0.03  & 0.38$ \pm $0.05  & -46.1$ \pm $5  & 0.14$ \pm $0.02 & X2.2  \\
2011.71  & 43.2  & 0.28$ \pm $0.03  & 0.63$ \pm $0.09  & -85.6$ \pm $5  & 0.30$ \pm $0.06 & X2  \\
2011.71  & 43.2  & 0.16$ \pm $0.01  & 1.38$ \pm $0.20  & -134.3$ \pm $5  & 0.45$ \pm $0.09 & X1  \\
     &     &     &     &     &    &     \\
2011.77  & 86.2  & 3.12$ \pm $0.34  & 0$ \pm $0  & 0$ \pm $0  & 0.052$ \pm $0.030 & C  \\
2011.77  & 86.2  & 1.15$ \pm $0.12  & 0.18$ \pm $0.02  & -33.2$ \pm $5  & 0.05$ \pm $0.03 & S  \\
2011.77  & 86.2  & 0.29$ \pm $0.03  & 0.46$ \pm $0.06  & -74.4$ \pm $5  & 0.06$ \pm $0.01 & U  \\
     &     &     &     &     &    &     \\
2011.79  & 43.2  & 2.30$ \pm $0.25  & 0$ \pm $0  & 0$ \pm $0  & 0.029$ \pm $0.030 & C  \\
2011.79  & 43.2  & 0.78$ \pm $0.08  & 0.09$ \pm $0.01  & -28.9$ \pm $5  & 0.05$ \pm $0.03 & M  \\
2011.79  & 43.2  & 1.82$ \pm $0.20  & 0.21$ \pm $0.03  & -37.0$ \pm $5  & 0.06$ \pm $0.03 & S  \\
2011.79  & 43.2  & 0.48$ \pm $0.05  & 0.36$ \pm $0.05  & -45.5$ \pm $5  & 0.18$ \pm $0.03 & X2.2  \\
2011.79  & 43.2  & 0.26$ \pm $0.02  & 0.67$ \pm $0.10  & -87.8$ \pm $5  & 0.33$ \pm $0.06 & X2  \\
2011.79  & 43.2  & 0.21$ \pm $0.02  & 1.41$ \pm $0.21  & -133.9$ \pm $5  & 0.50$ \pm $0.10 & X1  \\
     &     &     &     &     &    &     \\
2011.92  & 43.2  & 2.18$ \pm $0.23  & 0$ \pm $0  & 0$ \pm $0  & 0.019$ \pm $0.030 & C  \\
2011.92  & 43.2  & 0.87$ \pm $0.09  & 0.17$ \pm $0.02  & -33.7$ \pm $5  & 0.04$ \pm $0.03 & S  \\
2011.92  & 43.2  & 1.21$ \pm $0.13  & 0.27$ \pm $0.04  & -38.4$ \pm $5  & 0.06$ \pm $0.03 & P3  \\
2011.92  & 43.2  & 0.59$ \pm $0.06  & 0.35$ \pm $0.05  & -51.7$ \pm $5  & 0.14$ \pm $0.02 & X2.2  \\
2011.92  & 43.2  & 0.18$ \pm $0.02  & 0.72$ \pm $0.10  & -88.9$ \pm $5  & 0.41$ \pm $0.08 & X2  \\
2011.92  & 43.2  & 0.17$ \pm $0.02  & 1.48$ \pm $0.22  & -133.9$ \pm $5  & 0.50$ \pm $0.10 & X1  \\
     &     &     &     &     &    &     \\
2012.17  & 43.2  & 1.76$ \pm $0.19  & 0$ \pm $0  & 0$ \pm $0  & 0.036$ \pm $0.030 & C  \\
2012.17  & 43.2  & 1.68$ \pm $0.18  & 0.19$ \pm $0.02  & -26.0$ \pm $5  & 0.01$ \pm $0.03 & S  \\
2012.17  & 43.2  & 0.97$ \pm $0.10  & 0.30$ \pm $0.04  & -33.1$ \pm $5  & 0.09$ \pm $0.03 & P3  \\
2012.17  & 43.2  & 0.42$ \pm $0.04  & 0.41$ \pm $0.06  & -48.4$ \pm $5  & 0.12$ \pm $0.02 & U  \\
2012.17  & 43.2  & 0.28$ \pm $0.03  & 0.55$ \pm $0.08  & -66.2$ \pm $5  & 0.23$ \pm $0.04 & X2.2  \\
2012.17  & 43.2  & 0.11$ \pm $0.01  & 1.00$ \pm $0.15  & -98.3$ \pm $5  & 0.24$ \pm $0.04 & X2  \\
2012.17  & 43.2  & 0.17$ \pm $0.02  & 1.52$ \pm $0.22  & -131.0$ \pm $5  & 0.56$ \pm $0.11 & X1  \\
     &     &     &     &     &    &     \\
2012.40  & 43.2  & 2.09$ \pm $0.22  & 0$ \pm $0  & 0$ \pm $0  & 0.013$ \pm $0.030 & C  \\
2012.40  & 43.2  & 1.52$ \pm $0.16  & 0.16$ \pm $0.02  & -31.1$ \pm $5  & 0.04$ \pm $0.03 & S  \\
2012.40  & 43.2  & 1.87$ \pm $0.20  & 0.30$ \pm $0.04  & -31.2$ \pm $5  & 0.03$ \pm $0.03 & P3  \\
2012.40  & 43.2  & 0.79$ \pm $0.08  & 0.43$ \pm $0.06  & -42.6$ \pm $5  & 0.17$ \pm $0.03 & U  \\
2012.40  & 43.2  & 0.27$ \pm $0.02  & 0.59$ \pm $0.08  & -67.4$ \pm $5  & 0.24$ \pm $0.04 & X2.2  \\
2012.40  & 43.2  & 0.12$ \pm $0.01  & 1.05$ \pm $0.15  & -96.5$ \pm $5  & 0.23$ \pm $0.04 & X2  \\
2012.40  & 43.2  & 0.23$ \pm $0.03  & 1.50$ \pm $0.22  & -128.1$ \pm $5  & 0.71$ \pm $0.14 & X1  \\
     &     &     &     &     &    &     \\
2012.38  & 86.2  & 2.91$ \pm $0.32  & 0$ \pm $0  & 0$ \pm $0  & 0.057$ \pm $0.014 & C  \\
2012.38  & 86.2  & 1.46$ \pm $0.16  & 0.26$ \pm $0.03  & -37.6$ \pm $5  & 0.05$ \pm $0.014 & S   \\
2012.38  & 86.2  & 0.28$ \pm $0.03  & 0.42$ \pm $0.06  & -61.3$ \pm $5  & 0.19$ \pm $0.038 & X3   \\
2012.38  & 86.2  & 0.48$ \pm $0.05  & 0.62$ \pm $0.09  & -64.6$ \pm $5  & 0.24$ \pm $0.048 & X2.2  \\
     &     &     &     &     &    &     \\
2012.40  & 43.2  & 2.58$ \pm $0.28  & 0$ \pm $0  & 0$ \pm $0  & 0.034$ \pm $0.030 & C  \\
2012.40  & 43.2  & 0.90$ \pm $0.09  & 0.19$ \pm $0.02  & -40.3$ \pm $5  & 0.04$ \pm $0.03 & S  \\
2012.40  & 43.2  & 1.30$ \pm $0.14  & 0.36$ \pm $0.05  & -40.2$ \pm $5  & 0.13$ \pm $0.02 & P3  \\
2012.40  & 43.2  & 0.26$ \pm $0.02  & 0.57$ \pm $0.08  & -70.2$ \pm $5  & 0.22$ \pm $0.04 & X2.2  \\
2012.40  & 43.2  & 0.12$ \pm $0.01  & 0.96$ \pm $0.14  & -96.8$ \pm $5  & 0.32$ \pm $0.06 & X2  \\
2012.40  & 43.2  & 0.22$ \pm $0.02  & 1.58$ \pm $0.23  & -131.3$ \pm $5  & 0.68$ \pm $0.13 & X1  \\
     &     &     &     &     &    &     \\
2012.62  & 43.2  & 1.92$ \pm $0.21  & 0$ \pm $0  & 0$ \pm $0  & 0.02$ \pm $0.030 & C  \\
2012.62  & 43.2  & 0.67$ \pm $0.07  & 0.19$ \pm $0.02  & -40.1$ \pm $5  & 0.05$ \pm $0.03 & S  \\
2012.62  & 43.2  & 0.45$ \pm $0.04  & 0.37$ \pm $0.05  & -41.4$ \pm $5  & 0.07$ \pm $0.03 & X3  \\
2012.62  & 43.2  & 0.26$ \pm $0.02  & 0.46$ \pm $0.06  & -51.7$ \pm $5  & 0.13$ \pm $0.02 & U  \\
2012.62  & 43.2  & 0.12$ \pm $0.01  & 0.62$ \pm $0.09  & -70.1$ \pm $5  & 0.34$ \pm $0.06 & X2.2  \\
2012.62  & 43.2  & 0.09$ \pm $0.00  & 1.00$ \pm $0.15  & -94.5$ \pm $5  & 0.41$ \pm $0.08 & X2  \\
2012.62  & 43.2  & 0.14$ \pm $0.01  & 1.53$ \pm $0.23  & -127.7$ \pm $5  & 0.72$ \pm $0.14 & X1  \\
     &     &     &     &     &    &     \\
2012.82  & 43.2  & 2.36$ \pm $0.25  & 0$ \pm $0  & 0$ \pm $0  & 0.049$ \pm $0.030 & C  \\
2012.82  & 43.2  & 1.37$ \pm $0.15  & 0.19$ \pm $0.02  & -43.5$ \pm $5  & 0.09$ \pm $0.03 & S  \\
2012.82  & 43.2  & 0.88$ \pm $0.09  & 0.47$ \pm $0.07  & -48.6$ \pm $5  & 0.20$ \pm $0.04 & X3  \\
2012.82  & 43.2  & 0.09$ \pm $0.01  & 0.62$ \pm $0.09  & -69.3$ \pm $5  & 0.13$ \pm $0.02 & X2.2  \\
2012.82  & 43.2  & 0.07$ \pm $0.01  & 0.98$ \pm $0.14  & -94.4$ \pm $5  & 0.19$ \pm $0.03 & X2  \\
2012.82  & 43.2  & 0.20$ \pm $0.02  & 1.58$ \pm $0.23  & -123.9$ \pm $5  & 0.87$ \pm $0.17 & X1  \\
     &     &     &     &     &    &     \\
2013.04  & 43.2  & 2.47$ \pm $0.27  & 0$ \pm $0  & 0$ \pm $0  & 0.007$ \pm $0.030 & C  \\
2013.04  & 43.2  & 0.68$ \pm $0.07  & 0.14$ \pm $0.02  & -34.7$ \pm $5  & 0.05$ \pm $0.03 & S  \\
2013.04  & 43.2  & 0.20$ \pm $0.02  & 0.31$ \pm $0.04  & -47.4$ \pm $5  & 0.09$ \pm $0.03 & X3.2  \\
2013.04  & 43.2  & 0.14$ \pm $0.01  & 0.52$ \pm $0.07  & -54.6$ \pm $5  & 0.10$ \pm $0.02 & X3  \\
2013.04  & 43.2  & 0.14$ \pm $0.01  & 0.66$ \pm $0.09  & -61.7$ \pm $5  & 0.22$ \pm $0.04 & X2.2  \\
2013.04  & 43.2  & 0.09$ \pm $0.01  & 1.00$ \pm $0.15  & -93.7$ \pm $5  & 0.46$ \pm $0.09 & X2  \\
2013.04  & 43.2  & 0.11$ \pm $0.01  & 1.73$ \pm $0.26  & -127.7$ \pm $5  & 0.82$ \pm $0.16 & X1  \\
     &     &     &     &     &    &     \\
2013.16  & 43.2  & 3.22$ \pm $0.35  & 0$ \pm $0  & 0$ \pm $0  & 0.05$ \pm $0.030 & C  \\
2013.16  & 43.2  & 1.41$ \pm $0.15  & 0.13$ \pm $0.02  & -41.8$ \pm $5  & 0.12$ \pm $0.03 & S  \\
2013.16  & 43.2  & 0.12$ \pm $0.01  & 0.43$ \pm $0.06  & -61.0$ \pm $5  & 0.18$ \pm $0.03 & X3.2  \\
2013.16  & 43.2  & 0.15$ \pm $0.01  & 0.62$ \pm $0.09  & -61.1$ \pm $5  & 0.22$ \pm $0.04 & X3  \\
2013.16  & 43.2  & 0.04$ \pm $0.01  & 0.77$ \pm $0.11  & -76.6$ \pm $5  & 0.23$ \pm $0.04 & X2.2  \\
2013.16  & 43.2  & 0.08$ \pm $0.01  & 0.94$ \pm $0.14  & -91.3$ \pm $5  & 0.41$ \pm $0.08 & X2  \\
2013.16  & 43.2  & 0.12$ \pm $0.01  & 1.66$ \pm $0.25  & -126.3$ \pm $5  & 0.91$ \pm $0.19 & X1  \\
     &     &     &     &     &    &     \\
2013.29  & 43.2  & 1.58$ \pm $0.17  & 0$ \pm $0  & 0$ \pm $0  & 0.02$ \pm $0.030 & C  \\
2013.29  & 43.2  & 1.18$ \pm $0.12  & 0.15$ \pm $0.02  & -36.3$ \pm $5  & 0.04$ \pm $0.03 & S  \\
2013.29  & 43.2  & 0.99$ \pm $0.10  & 0.22$ \pm $0.03  & -48.1$ \pm $5  & 0.12$ \pm $0.02 & P4  \\
2013.29  & 43.2  & 0.08$ \pm $0.01  & 0.47$ \pm $0.07  & -61.6$ \pm $5  & 0.12$ \pm $0.02 & X3.2  \\
2013.29  & 43.2  & 0.11$ \pm $0.01  & 0.63$ \pm $0.09  & -65.6$ \pm $5  & 0.22$ \pm $0.04 & X3  \\
2013.29  & 43.2  & 0.08$ \pm $0.01  & 0.90$ \pm $0.13  & -89.9$ \pm $5  & 0.68$ \pm $0.13 & X2  \\
2013.29  & 43.2  & 0.10$ \pm $0.01  & 1.72$ \pm $0.25  & -126.1$ \pm $5  & 0.87$ \pm $0.17 & X1  \\
     &     &     &     &     &    &     \\
2013.41  & 43.2  & 1.60$ \pm $0.17  & 0$ \pm $0  & 0$ \pm $0  & 0.015$ \pm $0.030 & C  \\
2013.41  & 43.2  & 0.50$ \pm $0.05  & 0.15$ \pm $0.02  & -39.6$ \pm $5  & 0.06$ \pm $0.03 & S  \\
2013.41  & 43.2  & 0.60$ \pm $0.06  & 0.24$ \pm $0.03  & -43.3$ \pm $5  & 0.11$ \pm $0.02 & P4  \\
2013.41  & 43.2  & 0.22$ \pm $0.02  & 0.55$ \pm $0.08  & -67.4$ \pm $5  & 0.20$ \pm $0.04 & X3  \\
2013.41  & 43.2  & 0.13$ \pm $0.01  & 1.11$ \pm $0.16  & -110.9$ \pm $5  & 1.00$ \pm $0.20 & X2  \\
     &     &     &     &     &    &     \\
2013.49  & 43.2  & 2.50$ \pm $0.27  & 0$ \pm $0  & 0$ \pm $0  & 0.024$ \pm $0.030 & C  \\
2013.49  & 43.2  & 1.05$ \pm $0.11  & 0.13$ \pm $0.02  & -42.1$ \pm $5  & 0.05$ \pm $0.03 & S  \\
2013.49  & 43.2  & 0.44$ \pm $0.04  & 0.26$ \pm $0.03  & -48.5$ \pm $5  & 0.12$ \pm $0.02 & P4  \\
2013.49  & 43.2  & 0.19$ \pm $0.02  & 0.62$ \pm $0.09  & -70.5$ \pm $5  & 0.20$ \pm $0.04 & X3  \\
2013.49  & 43.2  & 0.07$ \pm $0.01  & 1.02$ \pm $0.15  & -91.4$ \pm $5  & 0.61$ \pm $0.12 & X2  \\
2013.49  & 43.2  & 0.05$ \pm $0.01  & 1.88$ \pm $0.28  & -130.1$ \pm $5  & 0.86$ \pm $0.17 & X1  \\
     &     &     &     &     &    &     \\
2013.57  & 43.2  & 2.45$ \pm $0.26  & 0$ \pm $0  & 0$ \pm $0  & 0.029$ \pm $0.030 & C  \\
2013.57  & 43.2  & 1.11$ \pm $0.12  & 0.17$ \pm $0.02  & -48.8$ \pm $5  & 0.05$ \pm $0.03 & S  \\
2013.57  & 43.2  & 0.51$ \pm $0.05  & 0.27$ \pm $0.04  & -49.8$ \pm $5  & 0.10$ \pm $0.02 & P4  \\
2013.57  & 43.2  & 0.17$ \pm $0.02  & 0.65$ \pm $0.09  & -72.0$ \pm $5  & 0.27$ \pm $0.05 & X3  \\
2013.57  & 43.2  & 0.10$ \pm $0.01  & 1.27$ \pm $0.19  & -114.1$ \pm $5  & 1.26$ \pm $0.25 & X2 \\ \hline
\label{comp_info}

\end{longtable}



\end{document}